\documentclass[10pt]{article}
\usepackage[utf8]{inputenc}
\usepackage{cite}
\usepackage{hyperref}
\hypersetup{colorlinks=true,linkcolor=bluecyan,citecolor=orange3,urlcolor=green3,pdfencoding=auto,linktocpage}
\usepackage{amsmath,amssymb,amsbsy,amstext,amsthm,simplewick,amsfonts}
\usepackage{mathrsfs}
\usepackage{graphicx}
\usepackage{wrapfig}
\usepackage{upgreek}
\usepackage{bm} 
\usepackage{framed}
\usepackage{bbm}
\usepackage{textcomp}
\usepackage{adjustbox}
\usepackage{makecell}
\usepackage{tcolorbox}
\usepackage{empheq}
\usepackage[normalem]{ulem}
\usepackage{enumitem}
\usepackage{braket} 
\usepackage{array}
\usepackage{dsfont}
\usepackage{physics}
\usepackage{ulem}
\usepackage{xcolor}
\usepackage{caption}
\usepackage{subcaption}
\usepackage{tocloft}
\usepackage{tikz}
\usepackage{float}
\usepackage[compat=1.1.0]{tikz-feynman}
\tikzfeynmanset{
      every blob={/tikz/fill=gray!20}
    }
\usepackage{appendix}
\usepackage[framemethod=default]{mdframed}
\usepackage[makeroom]{cancel}
\usepackage{diagbox}
\usepackage{tabularray}
\usepackage{multirow}
\usepackage{bigfoot}
\usepackage{flowchart}
\usetikzlibrary{shapes.geometric, arrows}

\usetikzlibrary {arrows.meta}
\usetikzlibrary{decorations.markings}
\usetikzlibrary{calc,patterns,angles,quotes}

\tikzstyle{failure} = [rectangle, rounded corners, minimum width=2cm, minimum height=1cm,text centered, draw=black, fill=red!30]
\tikzstyle{good} = [rectangle, rounded corners, minimum width=1cm, minimum height=1cm,text centered, draw=black, fill=green!30]
\tikzstyle{process} = [rectangle, rounded corners, minimum width=2cm, minimum height=1cm, text centered, draw=black, fill=gray!30]
\tikzstyle{arrow} = [thick,->,>=stealth]
\tikzstyle{auxiliaryline} = [line width=0mm]

\newmdenv[backgroundcolor=gray!15,%
skipabove=5pt,%
skipbelow=5pt,%
leftmargin=2pt,%
rightmargin=2pt,%
innertopmargin=-6pt,%
innerbottommargin=5pt,%
innerleftmargin=5pt,%
innerrightmargin=5pt,%
splittopskip=0pt,%
splitbottomskip=0pt,%
linewidth=0pt,%
nobreak=true]%
{keyeqn}

\newmdenv[backgroundcolor=gray!15,%
skipabove=5pt,%
skipbelow=5pt,%
leftmargin=2pt,%
rightmargin=2pt,%
innertopmargin=-2pt,%
innerbottommargin=5pt,%
innerleftmargin=5pt,%
innerrightmargin=5pt,%
splittopskip=0pt,%
splitbottomskip=0pt,%
linewidth=0pt,%
nobreak=true]%
{keythrm}

\graphicspath{{Fig/}}

\definecolor{lightgreen}{cmyk}{0.2, 0, 0.2, 0.2}
\definecolor{lightgray}{cmyk}{0.1,0.2,0,0.1}
\definecolor{lightgray2}{cmyk}{0.1,0.1,0,0.1}
\definecolor{bluecyan}{RGB}{0, 100, 200}
\definecolor{blue3}{RGB}{31,119,180}
\definecolor{red3}{RGB}{214,39,40}
\definecolor{orange3}{RGB}{255,127,14}
\definecolor{green3}{RGB}{44,160,44}

\definecolor{red2}{RGB}{255,0,0}
\definecolor{green2}{RGB}{0,170,0}
\definecolor{blue2}{RGB}{0,128,255}
\definecolor{magenta2}{RGB}{191,64,191}
\definecolor{purple2}{RGB}{112,48,160}
\definecolor{orange2}{RGB}{255,192,0}

\def\bfk{\textbf{k}}
\def\bfx{\textbf{x}}

\numberwithin{equation}{section}

\allowdisplaybreaks[1]
\begin{document}

\begin{titlepage}
	\setcounter{page}{1} \baselineskip=15.5pt 
	\thispagestyle{empty}

    \begin{center}
        {\centering {{\fontsize{18}{18}\bf{General Spin-1 Dynamics During Inflation:}} \\ \vspace{0.2cm}  {\fontsize{15}{15}{\it Microcausality Constraints and Cosmological Collider Signals }}}}
	\end{center}
 
	\vskip 18pt
	\begin{center}
		\noindent
		{\fontsize{12}{18}\selectfont Trevor Cheung\footnote{\tt kaiheitrevor.cheung@nottingham.ac.uk}$^{,a}$, Carlos Duaso Pueyo\footnote{\tt carlos.duasopueyo@sns.it}$^{,b,c}$, David Stefanyszyn\footnote{\tt david.stefanyszyn@nottingham.ac.uk}$^{,a}$, Dong-Gang Wang\footnote{\tt wdgang@ust.hk}$^{,d,e}$}
	\end{center}
	
	\begin{center}
		\vskip 8pt
		$a$ \textit{School of Physics and Astronomy,
			University of Nottingham, University Park, \\ Nottingham, NG7 2RD, United Kingdom} \\
            $b$ \textit{Scuola Normale Superiore, Piazza dei Cavalieri 7, 56126 Pisa, Italy} \\
            $c$ \textit{INFN, Sezione di Pisa, Largo Bruno Pontecorvo 3, 56127 Pisa, Italy} \\
            $d$ \textit{Institute for Advanced Study and Department of Physics,\\ Hong Kong University of Science and Technology, Clear Water Bay, HK, China} \\
            $e$ \textit{Department of Applied Mathematics and Theoretical Physics\\ University of Cambridge,
Wilberforce Road, Cambridge, CB3 0WA, United Kingdom} \\
	\end{center}
	
	
	\noindent\rule{\textwidth}{0.4pt}
	\noindent \textbf{Abstract} ~~ We construct the general theory of a spin-$1$ field propagating on an inflationary background consistent with scale invariance, rotations, Bunch-Davies conditions and microcausality. Our \textit{Boostless Proca} theory generalises the usual massive Proca theory to a Lagrangian with 5 free parameters: 4 corresponding to parity-even terms and a single parity-odd term. We impose three degrees of freedom from the outset and find that microcausality, the statement that field operators vanish outside the lightcone, requires the kinetic terms to be gauge invariant. We couple this spin-$1$ state to the inflaton and study the signatures it imprints on cosmological correlators. We find no new signals for the bispectrum meaning that the known signals are the most general thereby providing a very clean model-independent probe of new physics in this leading form of non-Gaussianity. For the trispectrum and higher-point functions, the boostless Proca theory produces new shapes with all 5 parameters in principle measurable.   
    
	\noindent\rule{\textwidth}{0.4pt}
	
	
\end{titlepage} 


\newpage
\setcounter{page}{2}
{
	\tableofcontents
}

\setcounter{footnote}{0}


\section{Introduction}\label{sec:intro}
The leading paradigm of cosmology involves a period of exponential expansion before the hot big bang, known as inflation. The energy scale during inflation could be as high as $10^{13}$ to $10^{14}$ GeV, around 10 orders of magnitude higher than what can be achieved in terrestrial colliders, inspiring the ``cosmological collider'' programme
\cite{Chen:2009we,Chen:2009zp,Noumi:2012vr,Arkani-Hamed:2015bza}. Due to the expanding spacetime and the associated high energies, heavy particles, possibly with spin, can be spontaneously produced during inflation, and if they decay into massless particles like the inflaton $\pi$ and graviton $\gamma_{ij}$, they leave distinctive imprints in end-of-inflation cosmological correlators $\langle \varphi \varphi \ldots \varphi \rangle$ where $\varphi$ collectively denotes $\pi$ and $\gamma$. These correlators are a statistical description of the structure of the universe at the end of inflation and provide the initial seeds from which the Cosmic Microwave Background (CMB) and Large Scale Structures (LSS) evolved. 

The two-point functions are fixed by symmetry up to an overall normalisation, so they lack the complexity to provide evidence of new particles. Higher-point functions on the other hand, namely non-Gaussianities, possess enough kinematic freedom to open up new discovery channels. As is now well-known, massive spinning particles leave distinct \textit{oscillatory signals} in certain kinematic limits of such higher-point functions with the mass, spin, and speed of sound of new particles encoded in the frequency of the oscillations, angular dependence, and resonance features respectively\cite{Arkani-Hamed:2015bza, Noumi:2012vr,Pimentel:2022fsc,Jazayeri:2022kjy,Jazayeri:2023xcj}. This opportunity to do particle physics spectroscopy at extremely high energies with cosmological observations has led to an intense search for such signals in the cosmological data\cite{Cabass:2024wob,Sohn:2024xzd,Suman:2025tpv,Suman:2025vuf,Cabass:2022oap,Philcox:2026bfa,Kumar:2026dih,Philcox:2026njr,Cassem:2026ygh}.

We can use one of many formalisms to compute these cosmological correlators, including the conventional in-in/Schwinger-Keldysh formalism \cite{Weinberg:2005vy,Chen:2017ryl}, the wavefunction of the universe formalism \cite{Maldacena:2011nz}, and the more recent in-out formalism \cite{Donath:2024utn}. However, the computations are tricky in practice due to the presence of nested time integrals that compute such correlators and the propagators of new heavy particles being built out of Hankel functions (or Whittaker functions in the presence of parity violation). In recent years, there have been many approaches developed to tackle or even bypass these computations, including analytic approaches\cite{Qin:2022fbv,Qin:2023ejc,Qin:2023nhv,Xianyu:2025lbk}, bootstrap techniques\cite{CosmoBootstrap1,Pimentel:2022fsc,Wang:2022eop,CosmoBootstrap2,CosmoBootstrap3,MLT,BBBB,Hogervorst:2021uvp,DiPietro:2021sjt,Cabass:2021fnw,Cabass:2022jda,Wang:2025qww,Jazayeri:2025vlv}, differential equations\cite{Werth:2023pfl,Arkani-Hamed:2023kig}, convex geometry\cite{Arkani-Hamed:2017fdk,Baumann:2025qjx}, the use of alternative variables\cite{Sleight:2019mgd,Sleight:2019hfp,Baumann:2024ttn,Arundine:2026fbr,Belrhali:2026ktb,Belrhali:2026ygh,CarrilloGonzalez:2025qjk}, and spectral representations \cite{Sleight:2020obc,Werth:2024mjg,Grafe:2026avi}. This is far from a complete list and we refer the reader to \cite{Baumann:2022jpr,Benincasa:2022gtd,Albayrak:2026auu} for reviews. The result is an excellent understanding of how new particles manifest themselves in cosmological observables, given a particular description for how such particles evolve during inflation and for how they interact with the inflaton and graviton.

For massive spinning particles in de Sitter (inflationary) spacetime, there are currently two formulations on the market: {\it cosmological collider (CC) physics} considered in \cite{Lee:2016vti,Arkani-Hamed:2015bza,Pimentel:2022fsc} and {\it cosmological condensed matter (CCM) physics} considered in \cite{Bordin:2018pca}. In the former case, new massive spinning fields are described as irreps of the full de Sitter (dS) isometries with the free theories arising from covariant and minimal couplings to gravity. The fields are $\sigma_{\mu \nu \ldots}$ where $\mu,\nu, \ldots = 0,1,2,3$. The dS boost symmetry is then broken when these fields couple to inflaton perturbations. In the latter case, new spinning fields are described as irreps of SO(3), and hence boosts are already broken in the free theory and remain so in the coupling to inflaton perturbations. In this case the fields are $\sigma_{ij\ldots}$ where $i,j,\dots = 1,2,3$. The logic for this case is that fluctuations on an FRW background should be organised according to the way they transform under the unbroken symmetries---that is, rotations---without worrying about how the theory is extended to other backgrounds. 

An important difference between the two scenarios is the existence of temporal modes in the CC case. Although these modes are not dynamical, once integrated out they produce equations of motion for the $2S+1$ dynamical modes that differ from those of the CCM case. So even though in both set-ups $2S+1$ degrees of freedom propagate, they do so in different ways. In \cite{Cheung:2025dmc} the resulting cosmological correlators were compared and it was concluded that the collider signals in the inflaton bispectrum are degenerate in the two set-ups but this degeneracy is broken for higher-point functions such as the trispectrum, where the two set-ups could in principle be distinguished from each other. 

One potential downside of the CC set-up described above is that the breaking of dS boost symmetry at the level of the interactions and not the free theory seems somewhat artificial. At the very least, we would expect some breaking to be induced in the free theory from symmetry-breaking loops. So in this paper we consider the phenomenology of an ``intermediate'' set-up where the free theory of a massive spinning field is described by a field $\sigma_{\mu\nu\ldots}$ but we allow for boost-breaking terms while maintaining $2S+1$ degrees of freedom. In other words, we aim to answer the following questions:
\begin{enumerate}
    \item \textit{Is there a general description of massive spinning fields during inflation that encompasses both the CC and CCM scenarios and any intermediate theories?} 
    \item \textit{If so, does this more general description lead to new cosmological collider signals in inflationary observables?}
\end{enumerate}
In this work we answer these questions for a spin-$1$ field.\footnote{This problem was also considered in \cite{Kehagias:2017cym} working at the level of the equations of motion where they imposed the constraint $\nabla^{\mu} \sigma_{\mu} = 0$. Here we work with an action instead and don't impose this extra constraint by hand.} To do so, in Section \ref{lagrangiansec} we write down the most general free action for a massive spin-1 field consistent with rotations and scale invariance while maintaining three propagating degrees of freedom. We treat $\sigma_0$ and $\sigma_i$ separately thereby allowing for the breaking of dS boosts. Not all such terms are physically admissible and to find the allowed subset we impose $i)$ Bunch-Davies conditions on the corresponding mode function solutions and $ii)$ consistency of the corresponding retarded propagator with \textit{microcausality} which requires any local operators in the theory to commute outside the lightcone \cite{Dubovsky:2007ac,Hui:2025aja,Creminelli:2025rxj}. Consistency with microcausality turns out to be very non-trivial due to the non-dynamical temporal mode which introduces non-localities in the action for the dynamical modes once integrated out. Such non-localities are a feature of the dS-invariant massive Proca theory (henceforth referred to as ``Proca theory") so are not an issue per se, but we need to check if they are consistent with causal propagation. We review these physical principles in Sections \ref{BDsec} and \ref{sec:microcausality}, and use them to constrain the free theory, first in flat space in Section \ref{sec3}; then for inflation in Section \ref{sec:dS}. At first glance it is not obvious that microcausality yields constraints on inflationary theories but two important points make it so \cite{Hui:2025aja}. Firstly, the statement that local operators commute outside the lightcone holds on \textit{all states} of a theory, even symmetry-breaking ones like inflation. Secondly, working in conformal time and without dynamical gravity, we have a well-defined lightcone.\footnote{With dynamical gravity the situation is more subtle given our freedom to perform field redefinitions on the metric, see e.g.\cite{deRham:2020zyh}.} Constraints from causality on cosmological observables have been considered in \cite{Tong:2021wai,Ema:2024hkj,Chowdhury:2026upp,Benincasa:2020aoj,Jazayeri:2025vlv,Agui-Salcedo:2023wlq,CarrilloGonzalez:2025fqq}.

In both flat-space and inflation we find that microcausality is very powerful and leaves little freedom in how we can describe the propagation of a new massive spin-$1$ field. Most notably, we find that the general theory can be split into two distinct cases: the CCM case described above where $\sigma_0$ and $\sigma_i$ decouple, and a new class of theories we refer to as {\it boostless Proca} where $\sigma_0$ and $\sigma_i$ do not decouple. This theory contains the Proca theory as a subset, but generalises it in a number of ways:
\begin{enumerate}
    \item The kinetic terms are forced to be gauge invariant but two structures are allowed $F_{\mu\nu}^2$ and $F_{0 \mu}^2$ meaning that the longitudinal and transverse modes can propagate with different speeds. We do not assume gauge invariant kinetic terms, rather microcausality forces them upon us.
    \item The temporal and spatial modes are not required to have the same mass since both $\sigma_\mu^2$ and $\sigma_0^2$ terms are allowed. Notably, the Higuchi-like bound we find requires only the mass squared of the temporal mode to be positive. If the masses are fixed to be the same by symmetry, this yields the usual bound \cite{Higuchi:1986py}.
    \item A one-derivative term $\sigma_0 \nabla^{\mu} \sigma_{\mu}$ is allowed. 
    \item A single parity-odd term $\epsilon^{0 \mu\nu \rho} \sigma_{\mu} \nabla_{\nu} \sigma_{\rho}$ is allowed. 
\end{enumerate}
The Lagrangian for the boostless Proca theory in de Sitter space can then be expressed as 
\begin{align}
    2\mathcal{L} = \underbrace{- F_{0i}F^{0i} - \frac {c_T^2}2 F_{ij}F^{ij}}_{\text{gauge-invariant kinetic terms}} + \underbrace{m^2a^2(\eta) \sigma_0^2 - M^2a^2(\eta) \sigma_i^2}_{\text{mass terms}} + \underbrace{2\beta a(\eta) \sigma_0 \partial_i \sigma_i}_{\text{1-derivative term}} + \underbrace{2\kappa a(\eta) \epsilon_{ijk} \sigma_i \partial_j \sigma_k}_{\text{parity-violating term}}\,,
\end{align}
where $a(\eta) = -1/(H \eta)$ is the scale factor in conformal time. This theory therefore shows that the first question above is answered in the affirmative. As we will discuss in detail, any deviations from this allowed set of terms must be treated as perturbative corrections so that microcausality is obeyed within a finite regime of validity. 

\begin{figure}[H]
\begin{subfigure}[h]{0.45\linewidth}
\includegraphics[width=\linewidth]{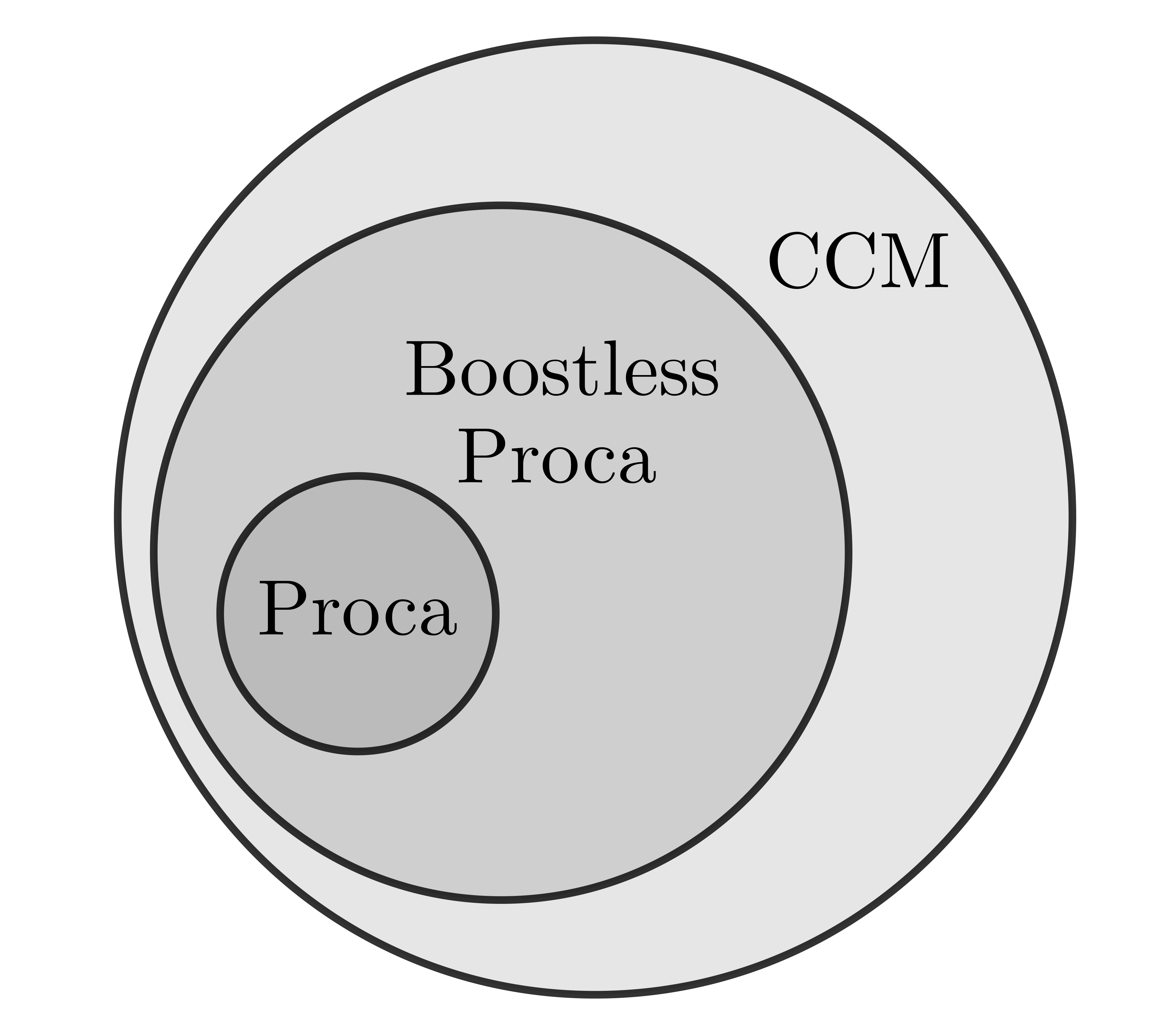}
\caption{Bispectrum}
\label{bispectrumvenndiagram}
\end{subfigure}
\hfill
\begin{subfigure}[h]{0.45\linewidth}
\includegraphics[width=\linewidth]{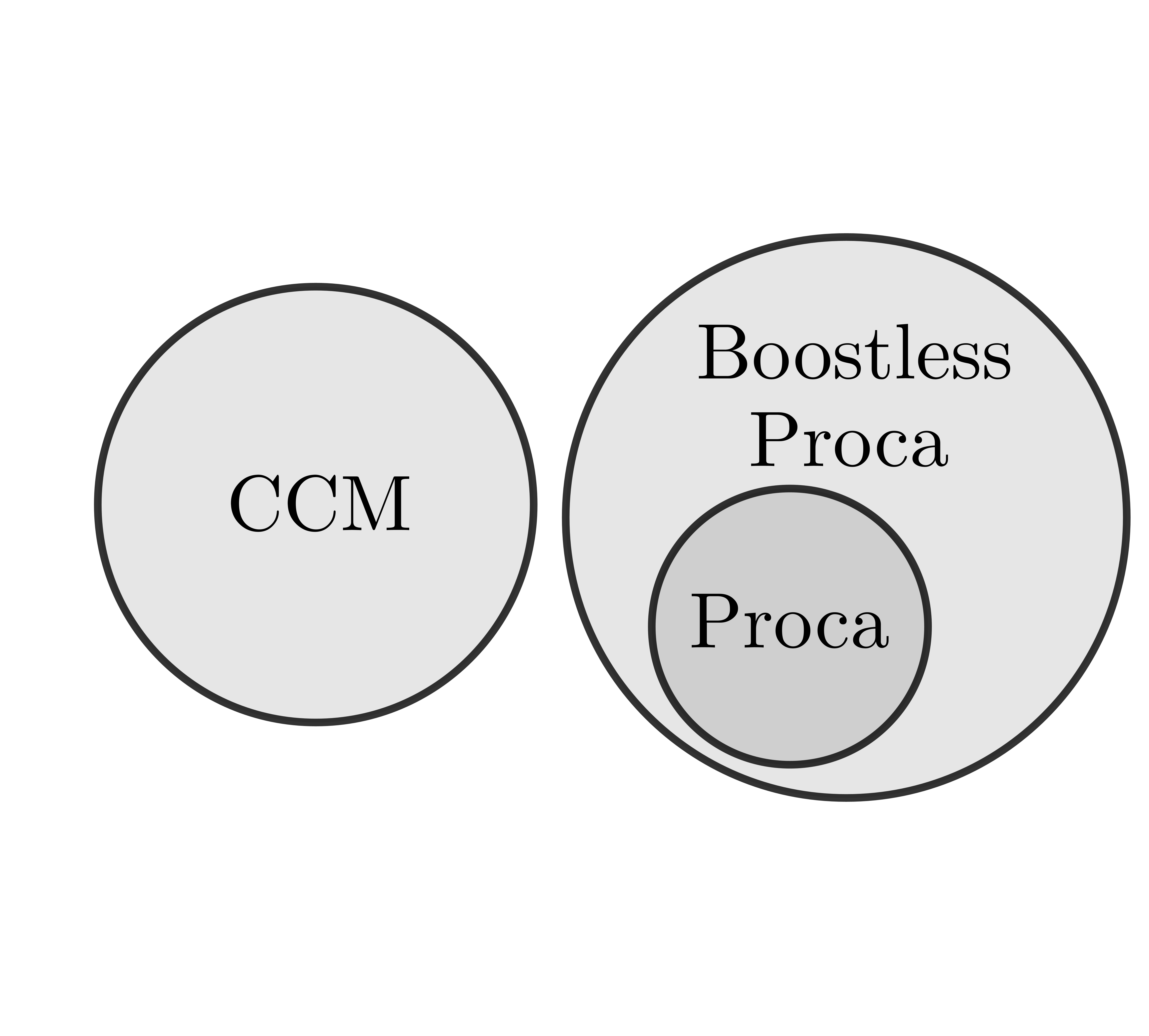}
\caption{\centering Trispectrum}
\label{trispectrumvenndiagram}
\end{subfigure}%
\caption{\centering a) An illustration that scalar bispectra produced by local interactions between the spinning field and the inflaton in the CCM theory contain all bispectra of the Boostless Proca theory which in turn contain all bispectra of the dS-invariant Proca theory. b) An illustration that at the level of the scalar trispectra (and higher-point functions) CCM and boostless Proca yield distinct signals.}
\label{Venndiagramhierarchy}
\end{figure}

In Section \ref{sec:phenomenology}, we answer the second question by coupling the boostless Proca field to the inflaton according to the rules of the Effective Field Theory of Inflation \cite{Cheung:2007st} followed by computing the corresponding correlators.\footnote{Since there is only a single parity-odd term whose signature in inflationary trispectra has already been computed \cite{Cabass:2022rhr,Stefanyszyn:2023qov}, we focus on parity-even interactions.} Similar to the conclusions of \cite{Cheung:2025dmc}, in Section \ref{sec:bispectrum} we show that all bispectra arising in the boostless Proca case are contained within the CCM case meaning that relative to CCM no new collider signals arise. The relationship among predictions from CCM, boostless Proca, and Proca is illustrated in Figure \ref{Venndiagramhierarchy}. Our results therefore show that the oscillations in the bispectrum are a very general and clean probe of new physics that makes no assumptions on the underlying theory describing the new field other than rotations, scale invariance, Bunch-Davies initial conditions and microcausality. Once we have access to the trispectrum, however, we are able to distinguish between the three cases. The key difference here is that all modes of the spin-$1$ exchanged field contribute whereas only helicity-$0$ modes contribute to the bispectrum. In Section \ref{sec:trispectrum}, we show that in the collapsed limit of the trispectrum, where one sees the characteristic oscillations, one can cleanly read off various parameters of the boostless Proca theory and distinguish it from the Proca theory. Therefore, the answer to the second question is a no for bispectra, but yes for trispectra (and any other higher-point functions).
One ``elephant in the room" that we haven't yet mentioned is that the collider signals for spin-$1$ exchange are suppressed in the relevant kinematic limits relative to non-oscillatory contributions \cite{Bordin:2018pca,Pimentel:2022fsc,Qin:2022lva} meaning that spin-$1$ exchange processes do not yield the most exciting prospect for new discoveries. However, for us the motivation to start with spin-$1$ was to show that our techniques can at least in principle open up new possibilities for cosmological collider physics and our positive results in this regard motivate us to continue this exploration with higher spins. We discuss this, and other avenues for future work when we conclude in Section \ref{Conclusion}. We also include three appendices. Appendix \ref{sec:contourintegration} provides more details on the conditions of microcausality at the level of retarded propagators making use of contour integration when going from momentum space to position space, Appendix \ref{appBswitchingpicture} provides an alternative perspective on microcausality and connects it with having a local basis for the equations of motion, and finally Appendix \ref{app:Seed} provides technical details required to compute exchange trispectra from known ``seeds" \cite{Jazayeri:2022kjy,CosmoBootstrap1}.


\paragraph{Notations and conventions}We use the following convention and notation for Fourier transforms:
\begin{align}
    \int_{\bfk} = \int \frac {d^3\bfk}{(2\pi)^3}\qquad \int_{\bfx - {\bf y}} = \int d^3(\bfx - {\bf y}) \,.
\end{align}The polarisation 3-vectors $e^{(h)}_{i}$, with $h=0, \pm 1$, are required to satisfy
\begin{align}
    \left [e^{(h)}_{i}(\bfk)\right ]^* & = e^{(h)}_{i}(-\bfk)\tag{reality}\\
    e^{(h)}_{i}(\bfk) e^{(h')}_{i}(-\bfk) & = \delta_{hh'} \tag{normalisation} \label{norm}  \\
    k_{i} e^{(\pm 1)}_{i}(\bfk) & = 0 \,. \tag{transverse}
    \end{align}
These can be fulfilled by taking
\begin{equation} \label{eq:polarisvecs}
    e^{(0)}_{i}(\bfk) = i\hat{k}_i, \qquad e^{(\pm 1)}_i (\bfk) = \frac {\hat{n}_i - \hat{n}_j \hat{k}_j\hat{k}_i \pm i \epsilon_{ijk} \hat{k}_j \hat{n}_k}{\sqrt{2(1- (\hat{n}_i \hat{k}_i)^2)}}\,,
\end{equation}
for any unit vector $\hat{\bf n}$ not parallel to $\bfk$. When computing cosmological correlators, we will make use of the following properties of the polarisation vectors when summed over helicities: 
\begin{align}
    \sum_{h=\pm 1} e^{(h)}_i (\bfk) e^{(h)}_j (-\bfk) & = \delta_{ij} - \hat{k}_i \hat{k}_j \,, \label{sumpolarisation}\\
    \sum_{h=\pm 1} h e^{(h)}_i (\bfk) e^{(h)}_j (-\bfk) & = -i\epsilon_{ijk} \hat{k}_{k} \,. \label{differencepolarisation}
\end{align}
Throughout the paper, we model the inflationary spacetime as de Sitter space and we work in the Poincar\'{e} patch with metric
\begin{align} \label{dSMetric}
ds^2 = a^2(\eta)(- d \eta^2 + d\bfx^2), \qquad a(\eta) = -\frac{1}{H \eta} \,,
\end{align}
where $H$ is the approximately constant Hubble parameter. 


\section{Checklist of constraints} \label{sec2}
In this section, we review the different physical constraints that we will impose on a QFT of a massive vector field that is dynamical during inflation. 
\subsection{Rotational and scale invariance} \label{lagrangiansec}
In the UV completion of inflationary perturbations, we have in mind a four-component vector $\sigma^{\mu}$ coupled to a scalar field $\phi$ that is responsible for slow-roll inflation.\footnote{This is just to justify the field content of our Lagrangian, in particular the fact that it contains only one $SO(3)$ scalar $\sigma^0$. However, the Lagrangian captures more general situations in which $\sigma^0$ and $\sigma^i$ do not descend from a UV four vector. \label{fn:4vector}} At the level of perturbations and working in the unitary gauge, the free theory for $\sigma^{\mu}$ will obey the unbroken $SO(3)$ symmetry and scale invariance of the quasi-de Sitter background, but not necessarily de Sitter boosts which could be strongly broken. This means the components $\sigma^{0}$ and $\sigma^{i}$ are not required to come as a package deal thanks to our freedom to use upper $0$ indices when building theories of inflationary perturbations \cite{Cheung:2007st}. With these symmetries in mind, the most general unitary gauge Lagrangian for a massive vector during inflation, up to second order in derivatives, is given by 
    \begin{align}
        2 \frac{\mathcal{L}}{\sqrt{-g}} = & \,\, \hat{\alpha}_1 (\nabla^{0} \sigma^{0})^2 + \hat{\alpha}_2 \nabla_\mu \sigma^{0} \nabla^{\mu} \sigma^{0} + \hat{\alpha}_3 (\sigma^{0})^2 \nonumber \\
       & + \hat{\beta}_1 (\nabla_{\mu} \sigma^{\mu})^2 + \hat{\beta}_2 \nabla_{\mu}\sigma_{\nu} \nabla^{\mu}\sigma^{\nu} + \hat{\beta}_3 \sigma_{\mu}\sigma^{\mu} + \hat{\beta}_4 \nabla^{0} \sigma^{\mu} \nabla^{0} \sigma_{\mu} + \hat{\beta}_5\epsilon^{0 \mu \nu \rho} \sigma_{\mu} \nabla_{\nu} \sigma_{\rho}\nonumber \\
    &   + \hat{\gamma}_1 \nabla^{0} \sigma^{0} \nabla_{\mu}\sigma^{\mu}+  \hat{\gamma}_2 \sigma^{0} \nabla_{\mu}\sigma^{\mu} \,,
    \end{align}
which in $3+1$ component form, and converting to conformal time, can be written as 
\begin{align}
2 \mathcal{L} &= \alpha_1 \partial_i \sigma_0 \partial_i \sigma_0 + \alpha_2 a^2(\eta) (\sigma_0)^2 \nonumber \\
& \quad  + \sigma_i' \sigma_i' + \beta_1 \partial_i\sigma_i \partial_j \sigma_j +\beta_2  \partial_i \sigma_j \partial_i \sigma_j + \beta_3 a^2(\eta) (\sigma_i)^2 + 2\beta_4 a(\eta) \epsilon_{ijk} \sigma_i \partial_j \sigma_k \nonumber \\
& \quad + 2\gamma_1 \partial_i  \sigma_0 \sigma_i' + 2 \gamma_2 a(\eta) \sigma_0 \partial_i \sigma_i \,,
\label{spin1generalfreewithparityodd}
\end{align}
where spatial indices are raised and lowered with $\delta_{ij}$, and a prime denotes a conformal time derivative. We emphasise (see Footnote~\ref{fn:4vector}) that it is not necessary to assume that $\sigma_i$ and $\sigma_0$ combine into a four-vector in the UV and indeed $\sigma_0$ could be a fundamental scalar with the spin-$1$ perturbations arising in a different manner as in e.g. Solid Inflation \cite{SolidInflation}. It is for this reason that we do not assume the kinetic structures are gauge invariant as that is not the case for Solids and Framids \cite{Zoology}. In flat-space, Lorentz invariance plus the requirement of three propagating degrees of freedom necessitates gauge invariant kinetic terms, but in a theory without Lorentz boosts there is no such requirement. The parameters in the two Lagrangians are related by 
    \begin{align}
        \hat{\alpha}_1  &= \alpha_1 + 1 + \beta_1 + \beta_2 + 2 \gamma_1, \quad 
        \hat{\alpha}_2  = \alpha_1 + \beta_2, \quad 
        \hat{\alpha_3}  = \alpha_2 + \beta_3 - (4\alpha_1 + 6\gamma_1 + 2)H^2 + 3\gamma_2 H,  \\
        \hat{\beta}_1 & = \beta_1,\quad 
        \hat{\beta}_2  = \beta_2, \quad
        \hat{\beta}_3  = \beta_3 - (\alpha_1 + 2)H^2, \quad
        \hat{\beta}_4  = 1 + \beta_2, \quad
        \hat{\beta}_5  = 2\beta_4, \\
        \hat{\gamma}_1 & = 2(\beta_1 + \gamma_1), \quad
        \hat{\gamma}_2  = 2H(\alpha_1 - \beta_2 - 3\beta_1 -\gamma_1) - 2\gamma_2 \,.
    \end{align}
There are a few comments in order:
\begin{itemize}
    \item We have not included a kinetic term for the temporal mode i.e. $(\sigma'_0)^2$ in \eqref{spin1generalfreewithparityodd} in order to maintain three degrees of freedom, which can be achieved by imposing the linear dependence of the original parameters: $\hat{\alpha}_1 - \hat{\alpha}_2 + \hat{\beta}_1 + \hat{\beta}_2  - \hat{\beta}_4 - \hat{\gamma}_1 = 0$.     One might wonder if we could instead make the longitudinal mode of $\sigma_i$ non-dynamical but this is not possible while keeping the transverse modes dynamical and maintaining $SO(3)$ symmetries. 
    \item The above Lagrangian \eqref{spin1generalfreewithparityodd} is general up to integration by parts meaning that we have dropped boundary terms. Such boundary terms could alter the correlators of the spinning field e.g. its power spectrum, but will not alter inflaton correlators coming from exchanging this field which is what we are ultimately interested in.
    \item The two-derivative parity-odd term $\epsilon_{ijk} \sigma'_{i} \partial_j \sigma_k$ can be removed using integration by parts, up to the $\beta_4$ term and boundary terms. 
    \item The CCM case \cite{Bordin:2018pca} corresponds to $\alpha_1=\alpha_2=\gamma_1=\gamma_2=\beta_4 = 0$ (or just $\gamma_1=\gamma_2=\beta_4 = 0$ so that $\sigma^0$ and $\sigma^i$ decouple). 
    \item The Proca theory with action
    \begin{align} \label{MassiveProca}
        S = \int d^4x \sqrt{-g} \left(-\frac{1}{4}F^2_{\mu\nu} - \frac{1}{2} m^2 \sigma^2_{\mu} \right) \,,
    \end{align}
    with $F_{\mu\nu} = \nabla_{\mu}\sigma_{\nu} - \nabla_{\nu}\sigma_{\mu}$,
    corresponds to $\alpha_1=\beta_1=-\beta_2=-\gamma_1=1, \alpha_2 = - \beta_3 = m^2, \gamma_2=\beta_4=0$. The parity-odd $\beta_4$ term arises from $\phi F^{\mu\nu} \tilde{F}_{\mu\nu}$ expanded around the inflationary background.
\end{itemize}

\subsection{Bunch-Davies vacuum} \label{BDsec}
When computing cosmological correlators, we need to quantise the fields and choose a vacuum state from which we evolve the universe from in the far past to the conformal boundary at $\eta_0 \rightarrow 0$. Throughout we will work with the Bunch-Davies choice for the vacuum which requires the mode functions of our massive vector to satisfy
\begin{align}
    \sigma_{\mu} (\eta,k) \sim \exp (-i\omega_k \eta), \qquad \text{as }\eta \to -\infty (1-i\epsilon)\,, \label{BunchDavies}
\end{align}
where $\omega_k^2 > 0$ for all $k> 0$. This means that at very early times the modes fluctuate as if they are in flat space and means that in the path integral we integrate over fields that vanish in the far past. In Section \ref{sec3} we will work in flat space, in which case this condition will be replaced by
\begin{align}
    \sigma_{\mu} (t,k) \propto \exp (-i\omega_k t)\,, \label{positivedispersionflatspace}
\end{align}
where $\omega_k^2 > 0$ for all $k>0$ to ensure that the mode function is a stable plane wave at all times.
\subsection{Microcausality} \label{sec:microcausality}
We have in mind a theory that is ultimately relativistic, but we are considering a state of the theory (inflation) for which some of the relativistic symmetries are spontaneously broken. The statement of {\it microcausality}, namely that if $x$ and $y$ are two spacetime points that are spacelike separated, local observables at these two points commute, i.e.
\begin{align}
    [\mathcal{O}_1(x), \mathcal{O}_2(y)]=0 \qquad \text{for }(x-y)^2 > 0 \,,
\end{align}
holds on all states of the theory since it is an operator statement \cite{Dubovsky:2007ac,Hui:2025aja}. What is particularly important for us, is that the expectation value of this commutator, calculated using the Bunch-Davies state, has to vanish outside the lightcone. We take the local operators to be the components of our vector field, and we require 
\begin{align} \label{VanishingCommutator}
    \langle [\hat{\sigma}_{i} (x), \hat{\sigma}_{j} (y)]\rangle = 0 \qquad \text{for }(x-y)^2 > 0\,.
\end{align}
We concentrate on the commutator of $\hat{\sigma}_i$ since these components of the full $\hat{\sigma}_\mu$ capture the dynamical modes of the theory, but we will see that any commutator involving $\hat{\sigma}_0$ also vanishes outside of the lightcone once we have satisfied \eqref{VanishingCommutator}. This is important as we will allow both $\sigma_i$ and $\sigma_0$ to couple locally to other degrees of freedom. Due to spatial translation invariance, the dependence on the spatial coordinates enters only via $\bfx - {\bf y}$ so we can Fourier transform with respect to the variable $\bfx - {\bf y}$ to define the mixed time-momentum representation Green's function\footnote{With the appropriate $\theta$-functions this is the retarded/advanced propagator. The $\theta$-functions won't play any role in our discussion since we only need to check analyticity in $\vec{k}$-space. We therefore drop any $\theta$-functions and just work with the commutator. We nevertheless still use the \textit{Green's function} terminology.}
\begin{align}
G_{ij}(x^0,y^0,\bfk) \equiv \int_{\bfx - {\bf y}}\langle [\hat{\sigma}_{i}(x^0,\bfx), \hat{\sigma}_{j}(y^0,{\bf y})]\rangle e^{-i\bfk \cdot (\bfx - {\bf y})}\,.
\end{align}
The Paley-Wiener-Schwartz theorem then implies that the necessary and sufficient conditions for the Green's function $G_{ij}(x^0,y^0,\bfk)$ to be consistent with microcausality are \cite{Hui:2025aja}:
\begin{enumerate}[label=(\Alph*)]
    \item \label{analyticity}{\bf A}nalyticity: $G_{ij}(x^0,y^0,\bfk)$ is analytic in complex $\bfk$ for all $x^0$ and $y^0$.
    \item \label{bound} {\bf B}ound: there exists some positive constants $C$, $D$, and $N$ such that, for all $i$ and $j$, we have
    \begin{align}
        |G_{ij}(x^0,y^0,\bfk)| < C(D + \|\bfk \|)^{N} \exp (\| \text{Im}(\bfk)\| \cdot |x^0 - y^0|) \,, \label{boundeqn}
    \end{align}
    for all $x^0$, $y^0$, and complex $\bfk$. 
\end{enumerate}
When we are working in de Sitter space, $x^0$ and $y^0$ are in conformal time, so that $|\bfx - {\bf y}| = |x^0 - y^0|$ defines the lightcone, as in flat space. Spacelike separations have $|\bfx - {\bf y}| > |x^0 - y^0|$.

For more details on these conditions of microcausality we refer the reader to \cite{Hui:2025aja}. To impose microcausality on  our spin-$1$ theory, we must first compute the Green's function $G_{ij}(x^0,y^0,\bfk)$ followed by demanding analyticity and \eqref{boundeqn}. As shown in \cite{Hui:2025aja}, familiar theories in flat and de Sitter space satisfy these conditions in a non-trivial way.

\paragraph{Computing the Green's function} To compute the Green's function $G_{ij}(x^0, y^0, \bfk)$, we work in momentum space and decompose the fields in terms of the helicities:
\begin{align}
\sigma_0(x^0,\bfx) & = \int_{\bfk} e^{i\bfk \cdot \bfx} \sigma_0(x^0,\bfk)   \,,\\
\sigma_{i}(x^0, \bfx)& = \int_{\bfk} e^{i\bfk \cdot \bfx} \sum_{h=-1}^1 e^{(h)}_i(\bfk) \sigma^{(h)}(x^0,\bfk) \,.
\end{align}
We then canonically quantise via
\begin{align}
\hat{\sigma}_0(x^0,\bfk) & = \sigma_0(x^0,k)  \hat{a}^{(0)}_{\bfk} +\sigma^{*}_0(x^0,k)  \hat{a}^{(0)\dagger}_{-\bfk}  \,,\\
\hat{\sigma}^{(h)}(x^0,\bfk) & = \sigma^{(h)}(x^0,k)  \hat{a}^{(h)}_{\bfk} +\sigma^{(h)^*}(x^0,k)  \hat{a}^{(h)\dagger}_{-\bfk} \,,
\end{align}
where the unhatted $\sigma_{0}(x^0, k)$ and $\sigma^{(h)}(x^0,k)$ are the helicity mode functions, only depending on the magnitude $k$ due to the $SO(3)$ symmetry of the theory. We will often denote the longitudinal mode of $\sigma_i$ as $\sigma_L \equiv \sigma^{(0)}$ to avoid confusion with the temporal mode $\sigma_0$. The creation and annihilation operators satisfy the usual commutation relations
\begin{align} \label{CandACommutators}
    [\hat{a}^{(h)}_{\bfk}, \hat{a}_{\bfk'}^{(h') \dagger}] = (2\pi)^3 \delta^{hh'} \delta^{(3)}(\bfk - \bfk')\,,
\end{align}
and in our mode function solutions we will impose the Wronskian conditions to satisfy the canonical commutation relations. Note that $\sigma_0$ and $\sigma_L$ share the same creation and annihilation operators as dictated by the three degrees of freedom of the theory \cite{Higuchi:1986py}. The Green's function is then
\begin{align}
    G_{ij}(x^0, y^0,\bfk ) &= \frac 1{(2\pi)^3}\langle [\hat{\sigma}_{i}(x^0,\bfk), \hat{\sigma}_{j}(y^0,-{\bfk})]\rangle \\
     & = G^{(0)}\hat{k}_i\hat{k}_j + \sum_{h=\pm 1} G^{(h)}e^{(h)}_i(\bfk) e^{(h)}_j(-\bfk) \\
&= G^{(0)} \hat{k}_i \hat{k}_j + \frac 12 (G^{(+1)}+ G^{(-1)}) (\delta_{ij} - \hat{k}_i \hat{k}_j) + \frac 12 (G^{(+1)} - G^{(-1)}) (-i\epsilon_{ijk} \hat{k}_k) \,, \label{ijcommutator}
\end{align}
where we introduced the Green's functions for individual helicity modes:
\begin{align} \label{HelicityGreensFunctions}
G^{(h)} (x^0,y^0,k) & = \sigma^{(h)}(x^0,k)\sigma^{(h)^*}(y^0,k) - \text{c.c.}\,.
\end{align}
In arriving at these expressions, we used \eqref{sumpolarisation} and \eqref{differencepolarisation}. Since we have included parity-violating terms in the free theory, $G^{(+1)}$ and $G^{(-1)}$ are not necessarily equal. With these expressions, once we compute the mode functions by solving the free equations of motion, we can easily read off Green's function $G_{ij}(x^0,y^0,\bfk)$. 
\paragraph{Analyticity} By considering the different tensor structures in \eqref{ijcommutator}, the analyticity condition \ref{analyticity} can be broken down into:
\begin{enumerate} [label = (A\arabic*)]
    \item \label{analyticitycondition1} $\delta_{ij}$ coefficient: $\frac 12(G^{(+1)}+ G^{(-1)})$ is analytic in $k^2$. 
    \item \label{analyticitycondition2}$\hat{k}_k$ coefficient: $\frac 12 (G^{(+1)} - G^{(-1)})$ is $k$ times an analytic function of $k^2$.
    \item   $\hat{k}_i \hat{k}_j$ coefficient: $G^{(0)} - \frac 12(G^{(+1)}+ G^{(-1)})$ is $k^2$ times an analytic function of $k^2$.
\end{enumerate}
We can combine the conditions coming from the $\delta_{ij}$ and $\hat{k}_i\hat{k}_j$ structures to modify the last condition to:
\begin{enumerate} [start=3,label = (A\arabic*)]
    \item \label{analyticitycondition3}  $\hat{k}_i \hat{k}_j$ coefficient: $G^{(0)}$ is analytic in $k^2$, and $G^{(0)}$ and $\frac 12 (G^{(+1)}+G^{(-1)})$ tend to the same limit as $k\to 0$.
\end{enumerate}
Note that in the final condition \ref{analyticitycondition3}, the factor of $1$ in front of $G^{(0)}$ and the factor $\frac{1}{2}$ in front of $(G^{(+1)}+G^{(-1)})$ will be crucial in checking that they tend to the same limit as $k \rightarrow 0$. Given that these coefficients ultimately come from our choice of normalisation for the polarisation vectors $e_i^{(h)}$, one might think these relative factors are not physical and can be altered abitrarily. However, our choice of normalisation will again become important when we solve for the mode functions subject to satisfying the canonical commutation relations and therefore any changes in how we normalise the polarisation vectors will be canceled by changes in the mode functions and therefore changes in the Green's functions $G^{(h)}$. Any collaboration between $G^{(0)}$ and  $(G^{(+1)}+G^{(-1)})$ in satisfying analyticity is therefore physically meaningful.  

In practice, to check a function's analyticity in $k^2$, we can check its analyticity in $k$ and its parity in complex $k$: if $f(k)$ is analytic and $f(-k) = f(k)$, it is analytic in $k^2$; if $f(k)$ is analytic and $f(-k) = -f(k)$, it is $k$ times an analytic function in $k^2$. Note that the complex conjugated mode functions $\sigma^{(h)^*}(y^0,k)$ appearing in the Green's functions are the complex conjugate of $\sigma^{(h)}(y^0,k)$ when $k\in\mathbb{R}$. When we analytically continue to $k \in \mathbb{C}$, we define the complex conjugate mode functions by $\sigma^{(h)^*}(y^0,k) \equiv \left [\sigma^{(h)}(y^0,k^*)\right ]^*$, such that they are the complex conjugate when $k\in \mathbb{R}$, and are analytic wherever $\sigma^{(h)}$ is analytic.
\paragraph{Bound} The bound \eqref{boundeqn} is only useful in the limit of large momenta, otherwise it can always be satisfied thanks to the freedom in the polynomial prefactor. We will therefore first assume that the bound applies for all momenta, including arbitrarily large momenta, in which case our resulting theory will obey microcausality in the same manner as e.g. a free scalar field with $c_s \leq 1$ or the Proca theory. We will then relax this restriction and work within a finite regime of validity where we will use the bound to ascertain which operators should be treated perturbatively. At large momenta where the polynomial prefactor can be ignored, it was shown in \cite{Hui:2025aja} that the bound can be formulated in terms of the magnitude $k$ as
\begin{align} \label{HelicityBound}
   | G^{(h)}(x^0, y^0, k) | \lesssim e^{|\text{Im}(k)| \cdot |x^0 -y^0|} \,,
\end{align}
which will be a useful relation for us.

In appendix \ref{sec:contourintegration} we use contour integration to show that once we Fourier transform $G_{ij}(x^0, y^0,\bfk)$ back to position space, our analyticity conditions and the bound are sufficient for the Green's function to vanish outside of the lightcone. We will close the contour using large arcs at infinity which is why the exponential boundedness is really a condition that needs to be satisfied as $|k| \rightarrow \infty$. The observation that such a large momentum limit is important for checking if a theory obeys causality goes back to the work of Sommerfeld and Brillouin who, using contour integration, realised that the \textit{front velocity} (which is the $k \rightarrow \infty$ limit of the phase velocity) is the quantity that must not exceed the speed of light in vacuum rather than the phase or group velocities which could be superluminal \cite{Sommerfeld,Brillouin} (see \cite{Brillouin:1960tos} for English translations of these papers).

\paragraph{On the importance of coupling to sources} So far our discussion has been within the context of operators and QFT but we will be computing the commutator at tree-level and so it makes sense to draw a connection with classical propagation. This discussion will also highlight the importance of coupling to sources when checking if a theory obeys causality. 

To this end, consider a scalar field (this argument goes through for a vector too just with extra indices) coupled to a local external source with equation of motion
\begin{align} \label{SourcedEoM}
\hat{\mathcal{O}}(t, \bfx) \phi(t, \bfx) = J(t, \bfx) \,,
\end{align}
where the differential operator $\hat{\mathcal{O}}(t, \bfx)$ comes from the free theory. Using the Green's function method, we can write a formal solution to this equation of the form 
\begin{align}
\phi(t, \bfx) = \phi_{\text{free}}(t, \bfx) + i \int dt' \int d^3 \bfx' \, G(t,t',\bfx - \bfx') \, J(t', \bfx')\,, 
\end{align}
where $\phi_{\text{free}}(t, \bfx)$ solves \eqref{SourcedEoM} in the absence of the source and $G$ is a Green's function that by spatial translation invariance depends on the difference $\bfx - \bfx'$, but not necessarily on the difference $t - t'$ because we are not assuming invariance under time translations, and solves the equation
\begin{align}
 \hat{\mathcal{O}}(t, \bfx)  G(t,t',\bfx - \bfx') = - i \delta(t-t') \delta^{(3)}(\bfx -\bfx') \,.  
\end{align}
The appropriate Green's function is chosen based on the physical problem we are solving. If we are asking how this scalar field responds to the presence of a source, then the Green's function must satisfy
\begin{align}  \label{RetardedGreensFunction}
G(t, t', \bfx -\bfx') = 0 ~~~~~ \text{for} ~~ t < t' \,,
\end{align}
which simply means the field cannot respond to the source before it is switched on. A Green's function satisfying such a boundary condition is unique and is the \textit{retarded Green's function}. Additionally, if our theory has a well-defined lightcone, the field cannot respond to the source if they are spacelike separated. This further imposes  
\begin{align} \label{Microcausality2}
G(t, t', \bfx -\bfx') = 0 ~~~~~ \text{for} ~~ |t-t'| < |\bfx - \bfx'| \,.
\end{align}
This final condition is that of \textit{relativistic causality} and ensures that no signal can travel faster than the speed of light in vacuum. 

In the mixed domain we are working with in this paper, the retarded Green's function obeys the equation
\begin{align} \label{MomentumSpaceGreensFunction}
    \hat{\mathcal{O}}(t, k) G(t, t', k) = - i \delta(t-t')\,,
\end{align}
with invariance under rotations ensuring the Green's function depends only on the two times and the magnitude of the spatial momentum. If two solutions to the homogeneous equation are $\phi^{+}(t, k)$ and $\phi^{-}(t,k)$, then we can attempt to find a solution to \eqref{MomentumSpaceGreensFunction} of the form
\begin{align} \label{GreensFunctionSolution}
    G(t, t', k) = (\phi^{+}(t, k) \phi^{-}(t', k)-\phi^{+}(t', k) \phi^{-}(t, k)) \theta(t-t') \,,
\end{align}
where the $\theta$-function ensures we obey the boundary condition \eqref{RetardedGreensFunction}. If we take $\hat{\mathcal{O}}(t,k)$ to be of the form we will encounter in this paper, i.e. a second-order differential operator with time-dependent and momentum-dependent coefficients, we can write 
\begin{align}
 \hat{\mathcal{O}}(t, k) = A(t, k) \frac{\partial^2}{\partial t^2} + B(t, k) \frac{\partial}{\partial t} + C(t, k)\,,
\end{align}
and then \eqref{GreensFunctionSolution} solves \eqref{MomentumSpaceGreensFunction} if the homogeneous solutions we are using in \eqref{GreensFunctionSolution} satisfy  
\begin{align}
A(t, k) \left( \frac{\partial \phi^{+}(t,k)}{\partial t} \phi^{-}(t,k) - \phi^{+}(t,k) \frac{\partial \phi^{-}(t,k)}{\partial t}  \right) = - i \,.
\end{align}
This condition, which we recognise as a non-vanishing Wronskian, tells us that $\phi^{+}$ and $\phi^{-}$ are two linearly-independent solutions to the homogeneous equation. For the cases of interest in this paper, we can take two linearly-independent solutions to be conjugate pairs such that $\phi^{+}(t,k) = (\phi^{-}(t,k))^{*} \equiv \phi(t,k)$.\footnote{In the far past we will impose Bunch-Davies conditions meaning that our equations of motion reduce to wave equations with stable plane wave solutions for which linearly-independent solutions are conjugate pairs. Linearly-independent solutions remain conjugate pairs away from this limit since any linearly-dependent solutions are related by time-independent coefficients.} We then have  
\begin{align} \label{GreensFunctionSolutionNew}
    G(t, t', k) = (\phi(t, k) \phi^{*}(t', k)-\phi(t', k) \phi^{*}(t, k)) \theta(t-t') \,,
\end{align}
and the Wronskian condition is 
\begin{align} \label{WronskianNew}
A(t, k) \left(  \phi(t,k) \frac{\partial \phi^{*}(t,k)}{\partial t} - \frac{\partial \phi(t,k)}{\partial t} \phi^{*}(t,k)   \right) =  i \,,
\end{align}
which normalises the homogeneous solutions to give us the correct Green's function. Expressing the Green's function in this way allows us to draw a connection with the QFT discussion above since in canonical quantisation \eqref{GreensFunctionSolutionNew} is simply the commutator
\begin{align}
G(t, t', k) = [\hat{\phi}(t,k), \hat{\phi}(t',k)]  \theta(t-t') \,,
\end{align}
computed in the free theory i.e it is the tree-level result, and \eqref{WronskianNew} arises from imposing the canonical commutator relation with the factor of $A(t,k)$ arising from computing the conjugate momentum. 

This discussion highlights the important role that the coupling to a source has played. Indeed, as emphasised in \cite{Dubovsky:2007ac}, the free equation of motion is not enough to ascertain if a theory obeys microcausality since there only the ratios $B/A$ and $C/A$ are important, whereas in computing the Green's function we need to know $A(t,k)$ which is a form of off-shell data.  
\paragraph{Commutators involving $\sigma^0$} Finally, we check that the commutators involving the temporal mode obey analyticity and the bound. The relevant Green's functions are
\begin{align}
    G_{0i}(x^0, y^0, \bfk) &= \frac 1{(2\pi)^3}\langle [\hat{\sigma}_{0}(x^0, \bfk), \hat{\sigma}_i(y^0, -\bfk)] \rangle \\
    & = -i \hat{k}_i (\sigma_0(x^0, k) \sigma^{*}_L(y^0, k) - \text{c.c.}) \\
    & = - i \hat{k}_i G_{\text{mixed}}(x^0, y^0, k)\,,
\end{align}
and 
\begin{align}
    G_{00}(x^0, y^0, \bfk) &= \frac 1{(2\pi)^3}\langle [\hat{\sigma}_{0}(x^0, \bfk), \hat{\sigma}_{0}(y^0, -\bfk)] \rangle \\
    & = \sigma_0(x^0, k) \sigma^{*}_0(y^0, k) - \text{c.c.} = G_{\text{temporal}}(x^0, y^0, k) \,,
\end{align}
where again we defined the helicity mode functions. Note that the Green's function $G_{0i}$ mixes the temporal mode and the longitudinal mode but does not depend on the transverse modes due to \eqref{CandACommutators}. For analyticity, we require $G_{\text{mixed}}$ to be $k$ times an analytic function of $k^2$ and $G_{\text{temporal}}$ to be an analytic function of $k^2$. For the bound, we impose \eqref{HelicityBound} for each Green's function. 

\paragraph{Commutators involving the conjugate momentum} The conjugate momentum of the dynamical modes can be straightforwardly computed form \eqref{spin1generalfreewithparityodd}:
\begin{align}
\Pi_i &= \frac{\partial \mathcal{L}}{\partial \sigma'_i} \\
 &= \sigma'_i + \gamma_1 \partial_i \sigma_0 \,. \label{ConjugateMomentum}
\end{align}
Given that $\Pi_i$ is related to the field operators in a local way, once the commutators $[\hat{\sigma}_\mu, \hat{\sigma}_\nu]$ vanish outside of the lightcone, any commutators involving the conjugate momentum will vanish too.\footnote{We thank Zhong-Zhi Xianyu for raising this point.} Similarly, any new fields we might want to use after a field redefinition that are related to $\sigma_{\mu}$ in a local manner will also lead to commutators that vanish outside of the lightcone. 
\subsection{Checklist summary}
In summary, we can obtain constraints on the parameters in \eqref{spin1generalfreewithparityodd} by the following procedure: 
\begin{mdframed}
\underline{\bf Procedure for obtaining constraints on the free theory}
\begin{enumerate} [label=(\Roman*)]
    \item \label{BDchecklist} Solve the equations of motion subject to the Bunch-Davies condition \eqref{BunchDavies}. In flat space, we check the positive dispersion relation in the solutions \eqref{positivedispersionflatspace}.
    \item \label{computeGreenchecklist} Compute the Green's functions for different helicity modes $G^{(h)}(x^0,y^0,k)$.
    \item \label{analyticitychecklist} Impose the analyticity conditions \ref{analyticitycondition1}, \ref{analyticitycondition2}, and \ref{analyticitycondition3}. When working with a finite regime of validity, impose analyticity only for momenta below the cut-off. In the complex plane of physical momenta $k$, this means check analyticity within a circle centered at the origin of fixed finite radius $\Lambda$. 
    \item \label{boundchecklist} Check that the bound \ref{bound} holds. When working with a finite regime of validity, use the bound to ascertain which operators should be treated perturbatively.
    \item \label{temporalchecklist} Check that the commutators involving the temporal mode $\sigma_0$ are analytic and obey the bound. 
\end{enumerate}
\end{mdframed}

\section{Flat space}\label{sec3}
As a warm up, we begin by working in flat space where the absence of explicit time dependence in the equations of motion makes things far simpler. Even so, as we will see, microcausality is still very powerful. Working with the various helicity modes, the equations of motion are 
\begin{align}
\frac{\delta S}{\delta \sigma_{0}} &= 0 \implies (\alpha_1 k^2 + \alpha_2)\sigma_0 + \gamma_1 k \partial_t \sigma_L  - \gamma_2 k \sigma_L  = 0 \,, \label{eomsigma0} \\
\frac{\delta S}{\delta \sigma_{L}} &= 0 \implies -\partial_t^2 \sigma_L + [(\beta_1+\beta_2)k^2 + \beta_3] \sigma_L  - \gamma_1 k \partial_t \sigma_0 - \gamma_2 k\sigma_0  = 0 \,, \label{eomsigmaL}\\
\frac{\delta S}{\delta \sigma^{(\pm 1)}} &= 0 \implies -\partial_t^2 \sigma^{(\pm 1)} + (\beta_2 k^2 + \beta_3 \pm 2\beta_4 k)\sigma^{(\pm 1)}  = 0\,. \label{eomsigmaT}
\end{align}
In general, we can solve \eqref{eomsigma0} for $\sigma_0$ in terms of the longitudinal mode $\sigma_L$ and substitute the solution into \eqref{eomsigmaL} to find a dynamical equation for $\sigma_L$. However, if $\alpha_1 = \alpha_2 = 0$, this is not possible. We therefore consider these special cases separately in Section \ref{specialcases} followed by considering the case when at least one of $\alpha_1$ and $\alpha_2$ is non-zero in Section \ref{genericcases}. In both of these sections, we impose the microcausality conditions for all momenta meaning that we assume the free theory is valid for arbitrarily high momenta. In Section \ref{EFTflatspace}, we impose a cut-off on the momenta and rederive the conditions of microcausality assuming that the free theory has a finite regime of validity.
\subsection{Special cases} \label{specialcases}
If $\alpha_1 = \alpha_2 = 0$, \eqref{eomsigma0} is not a constraint equation for $\sigma_0$. Depending on whether $\gamma_1$ and/or $\gamma_2$ also vanish, the analysis is slightly different so we work with three subcases: (1) $\alpha_1 = \alpha_2 = \gamma_1=\gamma_2 = 0$, (2) $\alpha_1= \alpha_2 = \gamma_1=0$ but $\gamma_2 \neq 0$, and (3) $\alpha_1=\alpha_2=0$ but $\gamma_1 \neq 0$. We follow the procedure outlined above to obtain constraints on the parameters in \eqref{spin1generalfreewithparityodd} in each case.
\subsubsection*{Case 1: $\alpha_1 = \alpha_2 = \gamma_1 = \gamma_2 = 0$} In this case, the Lagrangian \eqref{spin1generalfreewithparityodd} has no mixing between the temporal mode and the longitudinal mode, and the $\sigma_0$ and its equation of motion \eqref{eomsigma0} does not exist. In inflation, this would be the CCM case \cite{Bordin:2018pca} and here we are working with the flat space counterpart. Note that if $\beta_3 = \beta_4 = 0$ i.e. the transverse and longitudinal modes are massless and there is no parity violation, this theory describes the phonons of a Solid/Framid corresponding to the Goldstones associated with the spontaneous breaking of some Poincar\'{e} symmetries \cite{Nicolis:2015sra}. If we further take $\beta_1=-\beta_2=1$ we recover Maxwell theory in temporal gauge, whereas $\beta_2=-1$ and $\beta_1=1-c_L^2$ yields the Material Frame theory of~\cite{Serra:2026qnp}. 
\paragraph{\ref{BDchecklist} Check positive dispersion relation} In flat space, we check that all helicity modes have positive dispersion relations for all $k > 0$. The equation of motion \eqref{eomsigmaL} yields the dispersion relation for the longitudinal mode:
    \begin{align}
        (\omega^{(0)}_k)^2 = -(\beta_1 + \beta_2)k^2 - \beta_3\,. \label{dispersionlongitudinalCCM}
    \end{align}
   The requirement of positive dispersion relation implies $\beta_1 + \beta_2 < 0, \; \beta_3 < 0$. For the transverse modes, the equation of motion \eqref{eomsigmaT} yields the dispersion relation:
\begin{align}
    (\omega_k^{(\pm 1)})^2 = -\beta_2 k^2 - \beta_3 \mp 2\beta_4k\,. \label{transversedispersion}
\end{align}
To obtain a positive dispersion relation, we require $\beta_2 < 0, \; \beta_3 < 0, \; \beta_4^2 < \beta_2 \beta_3$.
\paragraph{\ref{computeGreenchecklist} Compute the Green's functions} Now we compute the Green's functions for the different helicity modes. To fix the overall normalisation of the mode functions, we impose the canonical commutation relations
\begin{align}
    [\hat{\sigma}^{(h)}(t, \bfk), \hat{\Pi}^{(h')}(t,  \bfk')] = i (2\pi)^3 \delta^{hh'} \delta^{(3)}(\bfk + \bfk')\,,
\end{align}
where $\Pi$ are the conjugate momenta. Given the normalisation conditions we have imposed on the polarisation vectors, each mode has a canonical kinetic term and therefore the conjugate momenta are 
\begin{align}
 \hat{\Pi}^{(h)}(t, \bfk) =  \partial_t \sigma^{(h)}(t,k) \hat{a}^{(h)}_{\bfk} +\partial_t  \sigma^{(h)^*}(t,k)  \hat{a}^{(h)\dagger}_{-\bfk} \,,
\end{align}
leading to the Wronskian conditions
    \begin{align}
        \sigma^{(h)}\partial_t \sigma^{(h)^*} - \sigma^{(h)^*}\partial_t \sigma^{(h)} & = i\,. \label{CCMWronskian}
    \end{align}
The appropriately normalised mode function solutions are therefore
\begin{align}
    \sigma^{(h)}(t,k) & = \frac {1}{\sqrt{2\omega_k^{(h)}}} \exp \left (-i \omega_k^{(h)} t\right ) \label{flatspacetransversemodefn} \,,
\end{align}
and the Green's functions \eqref{HelicityGreensFunctions} are 
  \begin{align}  
    G^{(h)}(t_1,t_2,k) & = - \frac {i\sin (\omega_k^{(h)} (t_1-t_2))}{\omega_k^{(h)}}\,. \label{flatspacecommutators}
\end{align}
\paragraph{\ref{analyticitychecklist} Check the analyticity conditions} We now impose the analyticity conditions \ref{analyticitycondition1}, \ref{analyticitycondition2}, and \ref{analyticitycondition3}.
\begin{itemize}
    \item To check \ref{analyticitycondition1}, we note that $(\sin z)/z$ is analytic in $z^2$, and hence \eqref{flatspacecommutators} suggests that $G^{(\pm 1)}$ are analytic in $(\omega_k^{(\pm 1)})^2$, and hence analytic in $k$ given \eqref{transversedispersion} (in the absence of any parity-violation $G^{(\pm 1)}$ would already be analytic in $k^2$). To check the analyticity properties in $k^2$, it suffices to check that $\frac 12 (G^{(+1)} + G^{(-1)})$ is even in complex $k$. To this end, note that
\begin{align}
    G^{(\pm 1)}(t_1,t_2,-k) = G^{(\mp 1)}(t_1,t_2,k)\,, \label{TCparity}
\end{align}
and therefore
\begin{align}
    \frac 12 (G^{(+1)}(t_1,t_2,-k) + G^{(-1)}(t_1,t_2,-k)) & = \frac 12 (G^{(-1)}(t_1,t_2,k) + G^{(+1)}(t_1,t_2,k)) \,,
\end{align}
which is even in complex $k$ and hence analytic in $k^2$, satisfying \ref{analyticitycondition1}.
\item To check \ref{analyticitycondition2}, since $G^{(\pm 1)}$ are analytic in $k$, it suffices to check that $\frac 12 (G^{(+1)} - G^{(-1)})$ is odd in complex $k$. Using \eqref{TCparity}, we have  
\begin{align}
    \frac 12 (G^{(+1)}(t_1,t_2,-k) - G^{(-1)}(t_1,t_2,-k)) & = \frac 12(G^{(-1)}(t_1,t_2,k)- G^{(+1)}(t_1,t_2,k))\,,
\end{align}
which is odd in complex $k$ and is hence $k$ times an analytic function in $k^2$, satisfying \ref{analyticitycondition2}.
\item To check \ref{analyticitycondition3}, we note that $G^{(0)}$ is analytic in $(\omega_k^{(0)})^2$ for the same reason as above, but now it is also directly analytic in $k^2$ given \eqref{dispersionlongitudinalCCM}. The difference here compared to the $(\pm 1)$ modes is that there is no parity-violating term for the longitudinal mode. In addition, we have $\lim_{k \rightarrow 0}(\omega_k^{(h)})^2 = -\beta_3$ for all helicities $h$, and hence $G^{(0)}$ and $\frac 12 (G^{(+1)} + G^{(-1)})$ both tend to the same limit, satisfying \ref{analyticitycondition3}.
\end{itemize}
Therefore, analyticity does not impose any extra constraints on the free theory in this case. Let us emphasise that satisfying analyticity is non-trivial: it requires a collaboration between the transverse and longitudinal modes to cancel the non-analyticity in the polarisation sums and requires these modes to have the same mass as forced by $SO(3)$ invariance. Furthermore, for each helicity, the mode function is non-analytic due to the factor of $1 / \sqrt{\omega_k^{(h)}}$ forced by the Wronskian condition, but this disappears once we take the imaginary part as required to compute the commutator. The massless and parity-even case was discussed in \cite{Hui:2025aja} and here we see that adding a mass and the parity-odd correction also passes the analyticity test.

\paragraph{\ref{boundchecklist} Check the bound} Now we check that the bound condition \ref{bound} holds, which requires us to check that \eqref{HelicityBound} holds for each helicity. For complex $\bfk$ and therefore complex $\omega_k^{(h)}$, from \eqref{flatspacecommutators} we have
\begin{align}
| G^{(h)}(t_1, t_2, k) | \lesssim e^{|\text{Im}(\omega_k^{(h)})| \cdot |t_1 -t_2|}\,,
\end{align}
and therefore the bound \eqref{HelicityBound} requires
\begin{align}
    |\operatorname{Im} \omega_k^{(h)}| \leq |\operatorname{Im} k| \,,\label{dispersionbound}
\end{align}
for all helicities $h$. Recall that we are working in the large momentum limit here so that we can ignore the polynomial prefactors. It is then clear that to satisfy this condition we require the longitudinal and transverse speeds of sound to be less than unity: $-(\beta_1 + \beta_2) \leq 1$ and $-\beta_2 \leq 1$. 

\paragraph{\ref{temporalchecklist} Check the temporal commutators} In this case $\sigma_0$ decouples and is therefore not part of the free theory of the vector field. There is therefore no need to check its commutators. 

\paragraph{Summary} In conclusion, for this case stability and microcausality imposes:
\begin{keyeqn}
    \begin{equation} \label{CCMflatspaceconstraints}
    \begin{split}
       \text{Case 1:} ~~~
        -1 \leq \beta_1+\beta_2 < 0, \qquad -1 \leq \beta_2 < 0, \qquad \beta_3 < 0, \qquad \beta_4^2 < 4\beta_2 \beta_3\,.
        \end{split}
    \end{equation}
    \end{keyeqn}
Physically, the first two inequalities refer to subluminal propagation of the longitudinal and transverse modes respectively, the third condition refers to positive mass squared, and the last one is a constraint on the size of parity violation.
\subsubsection*{Case 2: $\alpha_1 = \alpha_2 = \gamma_1= 0$ but $\gamma_2 \neq 0$} In this case, \eqref{eomsigma0} imposes $\sigma_L = 0$ i.e. $\sigma_0$ is a Lagrange multiplier enforcing the longitudinal mode to vanish on-shell which in turn implies $G^{(0)}(t_1,t_2,k) = 0$. The transverse modes equation of motion remains the same as for case 1 so for $h= \pm 1$ the Green's function is given by \eqref{flatspacecommutators}. The analyticity conditions \ref{analyticitycondition1} and \ref{analyticitycondition2} are therefore satisfied since they only depend on $G^{(\pm 1)}$, while \ref{analyticitycondition3} now requires 
\begin{align}
    \lim_{k \rightarrow 0} (G^{(+1)} + G^{(-1)}) = 0 \,.
\end{align}
We have $\lim_{k \rightarrow 0} (\omega_k^{(\pm 1)})^2 = -\beta_3$ and therefore
\begin{align}
    \lim_{k \rightarrow 0} (G^{(+1)} + G^{(-1)}) = -\frac {2i\sin (\sqrt{-\beta_3} (t_1 -t_2))}{\sqrt{-\beta_3}} \neq 0  \,.
\end{align}
This violation of analyticity holds for both zero and non-zero $\beta_3$ respectively corresponding to the transverse modes being massless or massive. Therefore, the analyticity condition \ref{analyticitycondition3} is violated, and case 2 is not admissible. In words, it is not possible to have the propagating transverse modes without also having the propagating longitudinal mode. This makes sense given that the projection onto longitudinal or transverse modes is non-local with only the sum of all helicities being a local object.
\subsubsection*{Case 3: $\alpha_1 = \alpha_2 = 0$ but $\gamma_1 \neq 0$} In this case, \eqref{eomsigma0} is a first-order differential equation for the longitudinal mode, which does not yield a plane wave solution. Therefore, this case is ruled out.
\subsection{Generic cases} \label{genericcases}
We now turn to the generic case where at least one of $\alpha_1$ and $\alpha_2$ is non-zero. In this case, \eqref{eomsigma0} is a constraint equation for $\sigma_0$, which has the solution
\begin{align}
    \sigma_0 & = -\frac {k}{\alpha_1 k^2 + \alpha_2} (\gamma_1 \partial_t - \gamma_2) \sigma_L\,.
\end{align}
This can be substituted back into \eqref{spin1generalfreewithparityodd} to obtain the Lagrangian for the longitudinal mode
\begin{align} \label{genericcaseLagrangianintegratedoutflat}
2\mathcal{L}& \supset A \partial_t \sigma_L \partial_t \sigma_L + B \sigma_L^2 \,,
\end{align}
where 
\begin{align}
    A = \frac {(\alpha_1 - \gamma_1^2) k^2 + \alpha_2}{\alpha_1 k^2 + \alpha_2}, \qquad \qquad B = (\beta_1+\beta_2)k^2+\beta_3 - \frac {\gamma_2^2k^2}{\alpha_1 k^2+\alpha_2}\,, \label{ABabbreviation}
\end{align}
and we dropped a total derivative.\footnote{When computing the conjugate momentum to match \eqref{ConjugateMomentum} this total derivative would need to be included.} Note the apparent non-locality that has arisen in the kinetic term of the longitudinal mode. We emphasise that this is also a property of the Proca theory \eqref{MassiveProca} where we have \cite{Himmetoglu:2008zp}
\begin{align}
    A_{\text{CC}} = \frac {m^2}{\alpha_1 k^2 +m^2}, \qquad \qquad B_{\text{CC}} = -m^2 \,.
\end{align}
The longitudinal mode equation of motion then reads
\begin{align}
-A\partial_t^2 \sigma_L + B\sigma_L & = 0\,. \label{eomsigmaLsubbed}
\end{align}
Since the constraint does not interfere with the transverse modes, the dispersion relations, mode functions, and Green's functions for the transverse modes are the same as for case 1 of the special cases and given by \eqref{transversedispersion}, \eqref{flatspacetransversemodefn}, and \eqref{flatspacecommutators} (for $h = \pm 1$). 
\paragraph{\ref{BDchecklist} Check positive dispersion relation} For real momenta, we don't want $A$ to vanish or diverge since these would respectively correspond to infinitely strong or weak coupling of the longitudinal mode. This is possible if e.g. $\alpha_1$ and $\alpha_2$ have opposite signs and $\gamma_1 \neq 0$. With this assumption, we can safely divide through by $A$ to find the following dispersion relation for the longitudinal mode: 
\begin{align} \label{GenericLongDR}
    (\omega_k^{(0)})^2 & = -\frac BA = -\frac {\alpha_1 (\beta_1 + \beta_2)k^4 + (\alpha_1\beta_3 + \alpha_2(\beta_1 + \beta_2) - \gamma_2^2)k^2 + \alpha_2 \beta_3}{(\alpha_1 - \gamma_1^2)k^2 + \alpha_2} \,.
\end{align}
We will derive the necessary conditions for $A$ to not vanish or diverge after we have imposed the conditions from microcausality. Let us emphasise that we are solving the equations of motion and therefore finding the Green's functions for real momenta and then analytically continuing into the complex plane afterwards. In this sense, our commutator does not have to be a Green's function in the full complex plane as there can be values of $k$ away from the real line for which the equation of motion becomes ill-defined. Note that in this dispersion relation we have a $k^4$ term even though we started with a Lagrangian with at most two derivatives. This has arisen upon integrating out the temporal mode. There are too many cases to check by directly imposing $(\omega_k^{(0)})^2 > 0$ for all $k>0$, so we will use the microcausality conditions first and impose positive dispersion relation once this relation has been simplified. For the transverse modes, we obtain the same conditions as for case 1 of the special cases: $\beta_2 < 0, \; \beta_3 < 0, \; \beta_4^2 < \beta_2 \beta_3$. 
\paragraph{\ref{computeGreenchecklist} Compute the Green's functions} As can be seen from the Lagrangian \eqref{genericcaseLagrangianintegratedoutflat}, the conjugate momentum for the longitudinal mode has changed and is now given by 
\begin{align}
 \hat{\Pi}_{L}(t, \bfk) = A( \partial_t \sigma_{L}(t,k) \hat{a}^{(0)}_{\bfk} +\partial_t  \sigma^*_L(t,k)  \hat{a}^{(0)\dagger}_{-\bfk} ) \,,
\end{align}
so the new Wronskian condition is 
\begin{align} \label{GenericWronskian}
    A(\sigma_L \partial_t \sigma^*_L -  \sigma^*_L \partial_t \sigma_L)& = i\,.
\end{align}
Using this to solve \eqref{eomsigmaLsubbed}, we find the longitudinal mode function and Green's function to be
\begin{align}
    \sigma_L(t,k) & = \frac {1}{\sqrt{2A\omega_k^{(0)}}} \exp \left (-i \omega_k^{(0)} t\right ) \,,\label{sigmaLmodefunction} \\
   G^{(0)}(t_1,t_2,k) & =  -\frac {i\sin (\omega_k^{(0)} (t_1-t_2))}{A\omega_k^{(0)}}\,.  \label{flatspaceLC}
\end{align}
The transverse mode functions and Green's functions have been computed in \eqref{flatspacetransversemodefn} and \eqref{flatspacecommutators} for $h = \pm 1$.
\paragraph{\ref{analyticitychecklist} Check the analyticity conditions} Since the transverse modes' Green's functions are the same as for case 1 of the special cases, the conditions \ref{analyticitycondition1} and \ref{analyticitycondition2} are automatically satisfied. Therefore, we focus on \ref{analyticitycondition3}, where we require:
\begin{itemize}
    \item $G^{(0)}$ is analytic in $k^2$: note that if $\alpha_1 \neq \gamma_1^2, \; \alpha_2 = 0$ or $\alpha_1 = \gamma_1^2, \; \alpha_2 \neq 0$, both $1/A$ and $(\omega_k^{(0)})^2$ are analytic in $k^2$. Since $(\sin z)/z$ is analytic in $z^2$, the full $G^{(0)}$ is analytic in $k^2$ in these cases. The case where both $\alpha_1 = \gamma_1^2$ and $\alpha_2 = 0$ yields a vanishing kinetic term and therefore the longitudinal mode is infinitely strongly coupled so we can rule this case out. The remaining case is where both $\alpha_1 \neq \gamma_1^2$ and $\alpha_2 \neq 0$. Under this assumption, note that
    \begin{align} \label{AExpression}
        \frac 1A & = \frac {\alpha_1}{\alpha_1 - \gamma_1^2} - \frac {\alpha_2\gamma_1^2}{(\alpha_1-\gamma_1^2)[(\alpha_1 - \gamma_1^2)k^2 + \alpha_2]}\,,
    \end{align}
    and
    \begin{align}
        (\omega_k^{(0)})^2 & = \text{analytic in }k^2 + \frac {\alpha_2\left [-\alpha_2\gamma_1^2(\beta_1+\beta_2) + (\beta_3 \gamma_1^2-\gamma_2^2) (\alpha_1 - \gamma_1^2)\right ]}{(\alpha_1-\gamma_1^2)^2[(\alpha_1 - \gamma_1^2)k^2 + \alpha_2]}\,. 
    \end{align}
    If $ -\alpha_2 \gamma_1^2 (\beta_1 + \beta_2) + (\beta_3 \gamma_1^2-\gamma_2^2)(\alpha_1-\gamma_1^2) \neq 0$, then $(\omega_k^{(0)})^2$ blows up at $k_{\pm} = \pm \sqrt{-\frac {\alpha_2}{\alpha_1-\gamma_1^2}}$. In particular, as $k\to k_{\pm}$,
    \begin{align}
        (\omega_k^{(0)})^2 & = \frac {\alpha_2\left [-\alpha_2\gamma_1^2(\beta_1+\beta_2) + (\beta_3 \gamma_1^2-\gamma_2^2) (\alpha_1 - \gamma_1^2)\right ]}{2k_{\pm}(\alpha_1-\gamma_1^2)^3(k-k_{\pm})} + \mathcal{O}(1)\,.
    \end{align}
    Depending on the values of $\alpha_1, \alpha_2$ and $\gamma_1^2$, this singularity in $(\omega_k^{(0)})^2$ could arise for purely real or purely imaginary $k$. The former is not allowed by our assumption that $A$ does not vanish for real momenta, while for the latter $\sin(\omega_k^{(0)}(t_2-t_1))$ grows exponentially and therefore dominates over $A\omega_k^{(0)}$ meaning that $G^{(0)}$ is not analytic in $k^2$. We therefore require
    \begin{align}
        -\alpha_2 \gamma_1^2 (\beta_1 + \beta_2) + (\beta_3 \gamma_1^2-\gamma_2^2)(\alpha_1-\gamma_1^2) = 0 \,, \label{omegaanalytic}
    \end{align}
    which makes $\omega_{k}^{(0)}$ analytic in $k^2$. Therefore, at any finite $k$,
    we can use the infinite radius of convergence of $\sin(z)$ to expand
    \begin{align}
        G^{(0)}(t_1,t_2,k) & = -\frac {i(t_1-t_2)}A + \mathcal{O}((t_1-t_2)^3)\,.
    \end{align}
    For $G^{(0)}(t_1,t_2,k)$ to be analytic in $k^2$ for all $t_1$ and $t_2$, we therefore require $\frac{1}{A}$ to be analytic. Since we are under the assumption that both $\alpha_1 \neq \gamma_1^2$ and $\alpha_2 \neq 0$, this forces $\gamma_1 = 0$ c.f. \eqref{AExpression}. Substituting $\gamma_1 = 0$ into the requirement \eqref{omegaanalytic}, yields $\gamma_1 = \gamma_2 = 0$. From the Lagrangian \eqref{spin1generalfreewithparityodd}, this corresponds to no mixing between the temporal and spatial modes and therefore reduces us back to case 1 of the special cases worked through in Section \ref{specialcases}. This is simply the flat-space version of the CCM theory so is nothing new. At this stage we therefore require $\alpha_1 \neq \gamma_1^2, \; \alpha_2 = 0$ or $\alpha_1 = \gamma_1^2, \; \alpha_2 \neq 0$.       
    \item $G^{(0)}$ and $\frac 12 (G^{(+1)} + G^{(-1)})$ tend to the same limit as $k\to 0$: we can separate the discussion into the two cases identified above:
    \begin{itemize}
        \item $\alpha_1 \neq \gamma_1^2, \; \alpha_2 = 0$: as $k\to 0$, we have 
        \begin{align}
            \lim_{k \rightarrow 0}A = \frac {\alpha_1 - \gamma_1^2}{\alpha_1}, \qquad \lim_{k \rightarrow 0} (\omega_k^{(0)})^2 = -\frac {\alpha_1 \beta_3 - \gamma_2^2}{\alpha_1 - \gamma_1^2}\,.
        \end{align}
        We also have 
        \begin{align}
           \lim_{k \rightarrow 0} (\omega_k^{(\pm 1)})^2 = -\beta_3 \,.
        \end{align}
        For $G^{(0)}$ and $\frac 12 (G^{(+1)}+G^{(-1)})$ to have the same limit as $k\to 0$, we therefore require
        \begin{align}
            \frac {\alpha_1 - \gamma_1^2}{\alpha_1}= 1, \qquad -\frac {\alpha_1 \beta_3 - \gamma_2^2}{\alpha_1 - \gamma_1^2} = -\beta_3\,.
        \end{align}
        The first condition imposes $\gamma_1 = 0$, while the second one imposes $\gamma_2 = 0$. From the general Lagrangian \eqref{spin1generalfreewithparityodd}, this case completely decouples the temporal mode $\sigma_0$ from the other helicity modes and brings us again back to the flat space version of CCM. We therefore discard this case.
        \item  $\alpha_1 = \gamma_1^2, \; \alpha_2 \neq 0$: In this case we have 
        \begin{align}
            \lim_{k \rightarrow 0} A = 1, \qquad \lim_{k \rightarrow 0}(\omega_k^{(0)})^2 = -\beta_3\,,
        \end{align}
        and therefore $G^{(0)}$ and $\frac 12 (G^{(+1)}+G^{(-1)})$ have the same limit as $k\to 0$ leading to no additional conditions. 
    \end{itemize}
\end{itemize}
To briefly conclude, assuming the dynamics are different to those of special case 1, i.e. the temporal mode is playing a non-trivial role, we require $\alpha_1 = \gamma_1^2$ and $\alpha_2 \neq 0$ by analyticity. 
\paragraph{\ref{boundchecklist} Check the bound} With $\alpha_1 = \gamma_1^2, \; \alpha_2 \neq 0$, the dispersion relation for the longitudinal mode reduces to
\begin{align}
    (\omega_k^{(0)})^2 = -\frac {\gamma_1^2 (\beta_1 + \beta_2)k^4 + (\gamma_1^2\beta_3 + \alpha_2(\beta_1 + \beta_2) - \gamma_2^2)k^2 + \alpha_2 \beta_3}{\alpha_2}\,.
\end{align}
To check that the bound condition \ref{bound} holds, we check \eqref{dispersionbound}. For large $k$, the left hand side of that inequality is dominated by the $\mathcal{O}(k^2)$, but the right hand side is $\mathcal{O}(k)$. So, imposing the bound for arbitrarily large $k$ requires $(\beta_1+ \beta_2) \gamma_1^2 = 0$. For now we are not imposing a cut-off but in Section \ref{EFTflatspace} we will work within a finite regime of validity. Therefore, we require $\gamma_1^2 = 0$ or $\beta_1 + \beta_2 = 0$. We study these two cases separately:
\begin{itemize}
    \item $\gamma_1^2 = 0$: since we already have $\alpha_1 =\gamma_1^2$ by analyticity, we have $\alpha_1 = \gamma_1^2 = 0$ and $\alpha_2 \neq 0$, meaning that the $\sigma_0$ dependence in the Lagrangian is 
    \begin{align}
2 \mathcal{L} \supset  \alpha_2  (\sigma_0)^2 + 2 \gamma_2 \sigma_0 \partial_i \sigma_i \,.
\end{align}
We can now use the field redefinition
    \begin{align}
        \sigma_0 \mapsto \sigma_0 - \frac {\gamma_2}{\alpha_2} \partial_i \sigma_i \,,
    \end{align}
    to decouple $\sigma_0$ and $\sigma_i$ thereby landing us back on case 1 of the special case corresponding to the flat space version of CCM. We therefore again ignore this possibility.
    \item $\beta_1 + \beta_2 = 0$: in this case, the dispersion relation of the longitudinal mode further reduces to 
    \begin{align}
        (\omega_k^{(0)})^2 = \frac {(\gamma_2^2 - \gamma_1^2 \beta_3)k^2}{\alpha_2}- \beta_3\,.
    \end{align}
    Recall that we didn't impose the condition of having a positive dispersion relation earlier due its complicated form but with this simple form imposing positivity and consistency with the bound \eqref{dispersionbound} is straightforward and yields $0< \frac {\gamma_2^2 -  \gamma_1^2 \beta_3}{\alpha_2} \leq 1, \; \beta_3 < 0$. In fact, from the Lagrangian \eqref{genericcaseLagrangianintegratedoutflat} after integrating out $\sigma_0$, in order for $\sigma_L$ to not be a ghost for all $k$, we require $\frac{\alpha_1}{\alpha_2}> 0$. Given that  $\alpha_1 = \gamma_1^2 \neq 0$, we therefore require $\alpha_2 >0$ so from positivity and the bound we require $0<\gamma_2^2 - \beta_3 \gamma_1^2 \leq \alpha_2, \beta_3 < 0$.
\end{itemize}
For the transverse modes, the case 1 discussion of the special cases apply and hence we also have $\beta_2 \geq -1$. 

\paragraph{\ref{temporalchecklist} Check the temporal commutators} With $\alpha_{1} = \gamma_1^{2} \neq 0$, $\alpha_{2} \neq 0$ and $\beta_1 + \beta_2 = 0$, the Green's functions arising from the temporal mode are 
\begin{align}
    G_{\text{mixed}} &= \frac{i \gamma_1 k}{\alpha_2} \cos\left(\omega_k^{(0)}(t_1 -t_2) \right) - \frac{i \gamma_2 k}{\alpha_2  } \frac{\sin\left(\omega_k^{(0)}(t_1 -t_2) \right)}{\omega_k^{(0)}}\,, \\
    G_{\text{temporal}} &= -\frac{i((\omega_k^{(0)})^2 + \beta_3)}{\alpha_2}\frac{\sin\left(\omega_k^{(0)}(t_1 -t_2) \right)}{\omega_k^{(0)}} \,.
\end{align}
As required, $G_{\text{mixed}}$ is $k$ times an analytic function of $k^2$ for all choices of parameters, while $G_{\text{temporal}}$ is an analytic function of $k^2$ also for all choices. Analyticity of the temporal commutators therefore imposes no new constraints. For the bound, we imposed $0<\gamma_2^2 -  \gamma_1^2 \beta_3 \leq \alpha_2$ and $\beta_3 < 0$, in which case \eqref{dispersionbound} is satisfied, and hence these temporal commutators also satisfy the bound \eqref{boundeqn}.

\paragraph{Summary} Finally, combining all the above, we have the following constraints for this generic case:
    \begin{keyeqn}
    \begin{equation} \label{generalisedProcacoefficients}
    \begin{split}
  \text{Generic case:} ~~~~  &\alpha_1 =\gamma_1^2 \neq 0, \quad \alpha_2 > 0, \quad \beta_1 + \beta_2 = 0, \\ 
       &0<\gamma_2^2 - \beta_3 \gamma_1^2 \leq \alpha_2, \qquad -1 \leq \beta_2 < 0, \qquad \beta_3 < 0, \qquad \beta_4^2 < 4\beta_2 \beta_3 \,.
        \end{split}
    \end{equation}
    \end{keyeqn}
    Upon rescaling $\sigma_0 \mapsto -\gamma_1^{-1} \sigma_0$, the condition $\alpha_1 =\gamma_1^2$ leads to the gauge invariant kinetic term $F_{0i}F^{0i}$, while the condition $\beta_1 + \beta_2 = 0$ leads to the gauge invariant term $F_{ij}F^{ij}$. It is very interesting that microcausality favours gauge invariant kinetic terms even when Lorentz boosts are broken. If we relabel the free theory parameters by: $\beta_2 \mapsto - c_T^2$ (speed of sound of transverse modes), $\beta_3 \mapsto -M^2$ (mass of longitudinal and transverse modes), $\frac{\alpha_2}{\gamma_1^2} \mapsto m^2$ (mass of temporal mode), $-\frac{\gamma_2}{\gamma_1} \mapsto \beta$ (1-derivative mode mixing), and $\beta_4 \mapsto \kappa$ (parity violation), we can write the final Lagrangian as
\begin{keyeqn}
    \begin{align}
    2\mathcal{L}& = \underbrace{- F_{0i}F^{0i} -\frac{c_T^2}{2} F_{ij}F^{ij}}_{\text{gauge-invariant kinetic terms}} + \underbrace{m^2(\sigma_0)^2 - M^2 (\sigma_i)^2}_{\text{mass terms}} + \underbrace{2\beta \sigma_0 \partial_i \sigma_i}_{\text{1-derivative term}} + \underbrace{2\kappa  \epsilon_{ijk} \sigma_i \partial_j \sigma_k}_{\text{parity-violating term}}\,, \label{flatspacegeneralisedProca}
\end{align}
\end{keyeqn}
which is subject to the conditions:\footnote{Notice that the massless limit $m,M\to 0$ is forbidden by our requirement that $A$ in~\eqref{genericcaseLagrangianintegratedoutflat} does not vanish or, in other words, that there are three dynamical degrees of freedom in the theory.}
\begin{keyeqn}
    \begin{align}
        0< \beta^2 + M^2 \leq m^2, \qquad 0< c_T^2 \leq 1, \qquad m^2 > 0, \qquad M^2 > 0, \qquad \kappa^2 < 4c_T^2M^2\,.
    \end{align}
\end{keyeqn}
Physically, the first two inequalities correspond to subluminal propagation of the longitudinal and transverse modes respectively, as usual the positive mass squared conditions ensure plane wave solutions rather than unstable ones that grow exponentially fast, while the last one prevents parity violation from being too big to again ensure stable plane wave solutions. In terms of these parameters, the Lorentz-invariant Proca theory corresponds to $c_T^2=1$, $m^2 = M^2$, $\beta = \kappa=0$. We note that in the limit where $m \rightarrow M$, we have $\beta \rightarrow 0$ meaning that the one-derivative term must vanish as the masses become equal.

\begin{figure} 
\centering
    \begin{tikzpicture}[node distance=2cm]
\node (constraint) [process] {Are both $\alpha_1 = \alpha_2 = 0$?};
\node (specialcases) [process, right of=constraint, xshift=2.5cm] {Is $\gamma_1 = 0$?};
\node (gamma1specialcase) [failure, below of=specialcases, yshift=-0.2cm]{no plane waves};
\node (gamma2specialcase) [process, right of=specialcases, xshift=1.5cm] {Is $\gamma_2 = 0$?};
\node (CCM) [good, right of=gamma2specialcase, xshift=1.5cm] {CCM};
\node (weirdspecialcase) [failure, below of=gamma2specialcase, yshift=-0.2cm]{violates \ref{analyticitycondition3}};
\node (genericcase) [process, below of=constraint, yshift=-2cm]{Is $\alpha_1-\gamma_1^2 = 0$?};
\node (alpha2) [process, right of=genericcase, xshift=+2.5cm]{Is $\alpha_2 = 0$?};
\node (gamma1gamma2) [process, right of=alpha2, xshift=1.5cm] {Is $\gamma_1 =\gamma_2= 0$?};
\node (CCM2) [good, right of=gamma1gamma2, xshift=+1.5cm]{CCM};
\node (alpha2violateanalyticity) [failure, below of=alpha2, yshift=-0.2cm]{violates \ref{analyticitycondition3}};
\node (gamma1gamma2violateanalyticity) [failure, below of=gamma1gamma2, yshift=-0.2cm]{violates \ref{analyticitycondition3}};
\node (alpha22) [process, below of=genericcase, yshift=-2cm]{Is $\alpha_2=0$?};
\node(stronglycoupled) [failure, below of=alpha22, yshift=-0.2cm]{strongly coupled};
\node(beta1beta2) [process, right of=alpha22, xshift=+2.5cm]{Is $\beta_1+\beta_2=0$?};
\node (gamma1)[process, right of=beta1beta2, xshift=+1.5cm]{Is $\gamma_1=0$?};
\node (finalCCM) [good,right of=gamma1, xshift=+1.5cm]{CCM};
\node (boostlessProca) [good, below of=beta1beta2,yshift=-0.2cm]{Boostless Proca};
\node (violatebound) [failure, below of=gamma1, yshift=-0.2cm]{violates \ref{bound}};

\draw [arrow] (constraint) -- node[anchor=south] {Yes} (specialcases);
\draw [arrow] (specialcases) -- node[anchor=west]{No} (gamma1specialcase);
\draw [arrow] (specialcases) -- node[anchor=south]{Yes}(gamma2specialcase);
\draw [arrow] (gamma2specialcase) -- node[anchor=south]{Yes} (CCM);
\draw [arrow] (gamma2specialcase) -- node[anchor=west]{No} (weirdspecialcase);
\draw [arrow] (constraint) -- node[anchor=east]{No} (genericcase);
\draw [arrow] (genericcase) -- node[anchor=south]{No} (alpha2);
\draw [arrow] (alpha2) -- node[anchor=south]{No} (gamma1gamma2);
\draw [arrow] (gamma1gamma2) -- node[anchor=south]{Yes} (CCM2);
\draw [arrow] (alpha2) -- node[anchor=west]{No} (alpha2violateanalyticity);
\draw [arrow] (gamma1gamma2) -- node[anchor=west]{No} (gamma1gamma2violateanalyticity);
\draw [arrow] (genericcase) -- node[anchor=east]{Yes} (alpha22);
\draw[arrow] (alpha22) -- node[anchor=east]{Yes}(stronglycoupled);
\draw [arrow] (alpha22)-- node[anchor=south]{No}(beta1beta2);
\draw [arrow] (beta1beta2) -- node[anchor=south]{No}(gamma1);
\draw [arrow] (gamma1) -- node[anchor=south] {Yes} (finalCCM);
\draw [arrow](beta1beta2) -- node[anchor=west]{Yes}(boostlessProca);
\draw [arrow](gamma1) -- node[anchor=west]{No} (violatebound);
\end{tikzpicture}
\caption{Flowchart showing the constraints on \eqref{spin1generalfreewithparityodd} in flatspace.}
\label{FlowchartFlat}
\end{figure}
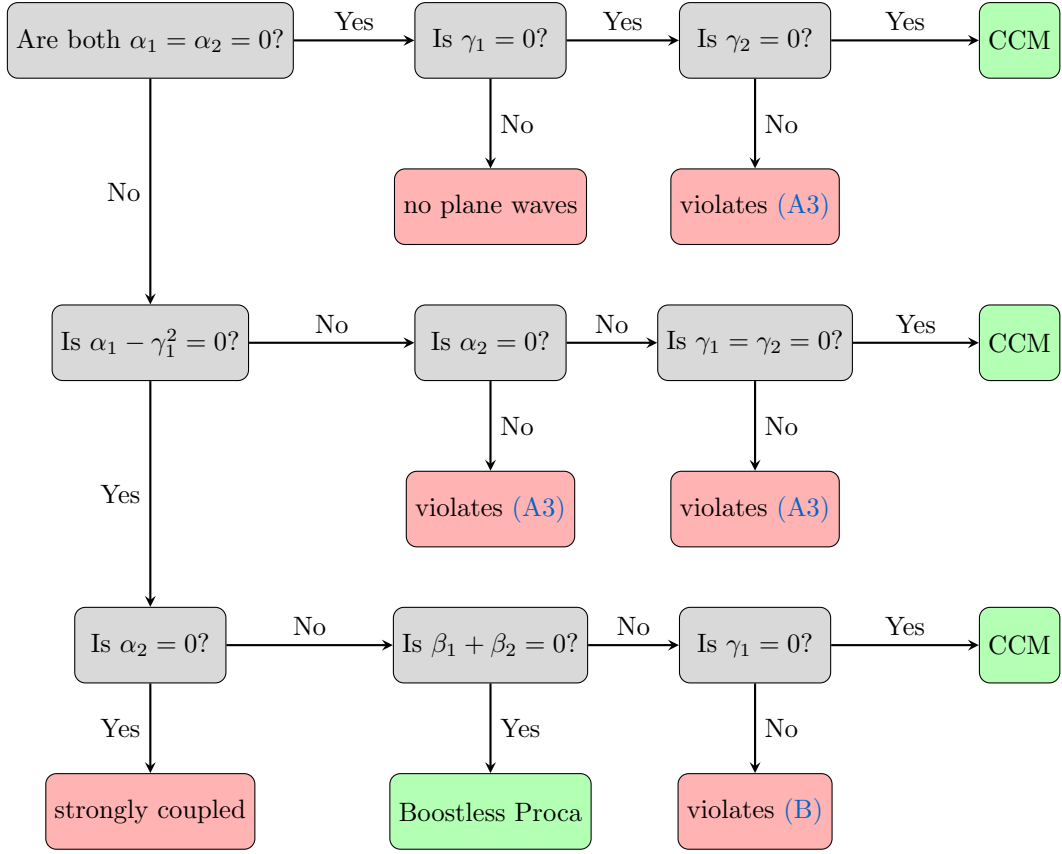

We summarise our findings in the flowchart of figure \ref{FlowchartFlat}. Note that this flowchart is valid for any choice of $\beta_4/\kappa$ so it does not appear.
\subsection{Finite regime of validity} \label{EFTflatspace}
In the previous subsections, we assumed an infinite regime of validity for the free theory in \eqref{spin1generalfreewithparityodd} and imposed the conditions of microcausality for all momenta. However, since we are really describing the fluctuations around some symmetry-breaking vacuum, we expect this description to only be valid for a finite range of momenta. We therefore now work with a restricted set of momenta with a cut-off $|\operatorname{Re} k|, |\operatorname{Im} k| \ll \Lambda$. The special cases of Section \ref{specialcases} reduced to the flat-space version of CCM physics and satisfy microcausality for all momenta (as long as the speeds of sound are subluminal). So there is not much to add when we impose a cut-off. We therefore move directly to the generic cases and ask how our final result \eqref{flatspacegeneralisedProca} can be corrected.  

\paragraph{Generic cases} To show more clearly how working within a finite regime of validity allows us to deviate from the solution we found in the generic case i.e. \eqref{flatspacegeneralisedProca}, we reparametrise the Lagrangian \eqref{spin1generalfreewithparityodd} such that it is given by \eqref{flatspacegeneralisedProca} plus additional terms that were previously imposed to vanish. We therefore have 
\begin{align} \label{FiniteRegimeTheory}
2\mathcal{L} & = - F_{0i}F^{0i} - \frac{c_T^2}{2} F_{ij}F^{ij} + m^2(\sigma_0)^2 - M^2 (\sigma_i)^2 + 2\beta \sigma_0 \partial_i \sigma_i + 2\kappa  \epsilon_{ijk} \sigma_i \partial_j \sigma_k + \widetilde{\alpha} \partial_i \sigma_0 \partial_i \sigma_0 + \widetilde{\beta} \partial_i \sigma_i \partial_j \sigma_j\,.
\end{align}
Note that we have performed the same rescaling of $\sigma_0$ as above and have therefore assumed $\gamma_1 \neq 0$. We now repeat the above procedure.

The dispersion relation for the longitudinal mode is now  \begin{align} \label{LongitudinalModeDRNew}
    (\omega_k^{(0)})^2 = -\frac {(\widetilde{\alpha}+1)\widetilde{\beta} k^4 + (-(\widetilde{\alpha}+1)M^2 + m^2 \widetilde{\beta} - \beta^2)k^2 - m^2M^2}{\widetilde{\alpha}k^2 + m^2}\,,
\end{align}
while that of the transverse modes is
\begin{align}
    (\omega_k^{(\pm 1)})^2 = c_T^2 k^2 +M^2 \mp 2 \kappa k\,. 
\end{align}
We now turn our attention back to analyticity and demand that it holds for $|\operatorname{Re} k|, |\operatorname{Im} k| \ll \Lambda$. This means we can allow for singularities that cannot be accessed within this regime of validity. The analyticity conditions \ref{analyticitycondition1} and \ref{analyticitycondition2} are again automatically satisfied so we turn to \ref{analyticitycondition3}. In terms of these new parameters, the Green's function for the longitudinal mode is
\begin{align}
    G^{(0)}(t_1,t_2,k) & = -\frac{(1 + \tilde{\alpha})k^2 + m^2}{\tilde{\alpha}k^2 + m^2}\frac {i\sin (\omega_k^{(0)} (t_2-t_1))}{\omega_k^{(0)}} \,,
\end{align}
and analyticity of $G^{(0)}$ in $k^2$ requires the poles in $(\omega_k^{(0)})^2$ at
\begin{align}
    k = \pm\frac {im}{\sqrt{\widetilde{\alpha}}}\,,
\end{align}
to be outside the regime of validity. For this, we need to impose
\begin{align}
|\widetilde{\alpha}| < \frac {m^2}{\Lambda^2} \,.\label{extraalphabound}
\end{align}
If we want the cut-off of the theory to be much greater than the mass of the temporal mode, then we require $|\tilde{\alpha}| \ll 1$.\footnote{If we take the speed of sound of the longitudinal mode to be subluminal, the requirement $|\tilde{\alpha}| \ll 1$ arises from demanding that strong coupling scale to be much greater than the mass of the longitudinal/transverse modes.} We therefore see that the $\tilde{\alpha}$ correction in \eqref{FiniteRegimeTheory} must be perturbatively small. It is then simple to check that $G^{(0)}$ and $\frac{1}{2}(G^{(+1)} + G^{(-1)})$ have the same limit as $k \rightarrow 0$ and therefore once we have imposed \eqref{extraalphabound}, analyticity is satisfied within the finite regime of validity.

In addition to analyticity, we need to impose the bound. However, as we discussed earlier, the bound is only useful in the limit of large momenta which takes us outside the regime of validity of the theory. If $\tilde{\beta} = 0$ then the bound simply tells us that the speed of sounds need to be subluminal. However, if $\tilde{\beta} \neq 0$, then it is clear that generically the bound will be violated once the $k^4$ term in \eqref{LongitudinalModeDRNew} dominates the dispersion relation which is inevitable if the theory is considered viable up to arbitrarily large momenta. We therefore learn that when $\tilde{\beta} \neq 0$, the theory \textit{must} be treated within a finite regime of validity and within that regime, the $k^4$ term cannot dominate otherwise corrections that are higher order in $k^{2n}$ will also become important and any control we have over the theory will be lost. With $|\tilde{\alpha}| \ll 1$, we can approximate the dispersion relation of the longitudinal mode by 
 \begin{align} 
    (\omega_k^{(0)})^2 = -\frac {\widetilde{\beta} k^4 + (-M^2 + m^2 \widetilde{\beta} - \beta^2)k^2 - m^2M^2}{ m^2}\,. 
\end{align}
Give that $M \neq 0$, in the absence of fine tuning that makes the $k^2$ term small, we can compare the $\tilde{\beta} k^4$ term to the $M^2 k^2$ term and we require
\begin{align}
 \frac{|\tilde{\beta}|k^4}{M^2 k^2} < 1\,,   
\end{align}
and given that the $k^4$ term is largest close to the cut-off, we require 
\begin{align}
|\tilde{\beta}| < \frac{M^2}{\Lambda^2}\,.  
\end{align}
If we again want the cut-off to be much greater than the mass of the transverse and longitudinal modes we require $|\tilde{\beta}| \ll 1$. We therefore see that the $\tilde{\beta}$ correction in \eqref{FiniteRegimeTheory} must also be perturbatively small.
\section{de Sitter space} \label{sec:dS}
We now turn our attention to dS space and repeat the analysis of the previous section. In dS space, the equations of motion coming from \eqref{spin1generalfreewithparityodd} are
\begin{align}
\frac{\delta S}{\delta \sigma_{0}} &= 0 \implies (\alpha_1 k^2 + \alpha_2 a^2(\eta)) \sigma_0 + \gamma_1 k \partial_{\eta} \sigma_L - \gamma_2 a(\eta) k \sigma_L = 0\,, \label{eomsigma0dS}\\
\frac{\delta S}{\delta \sigma_{L}} &= 0 \implies -\partial_{\eta}^2\sigma_L + [(\beta_1 + \beta_2)k^2 + \beta_3a^2(\eta)] \sigma_L- \gamma_1 k \partial_{\eta} \sigma_0 -  \gamma_2 a(\eta) k\sigma_0  = 0\,, \label{eomsigmaLdS} \\
\frac{\delta S}{\delta \sigma^{(\pm 1)}} &= 0 \implies -\partial_{\eta}^2 \sigma^{(\pm 1)} + (\beta_2 k^2 + \beta_3 a^2(\eta) \pm 2\beta_4 a(\eta) k)\sigma^{(\pm 1)}  = 0 \,.\label{eomsigmaTdS}
 \end{align}
Similar to the flat space case, apart from the special cases ($\alpha_1 = \alpha_2 = 0$) listed in Section \ref{specialcasesdS}, we solve \eqref{eomsigma0dS} to obtain $\sigma_0$ in terms of the longitudinal mode $\sigma_L$, and substitute the solution back into \eqref{eomsigmaLdS}. We first require the constraints to hold for all $k\in \mathbb{C}$ in Sections \ref{specialcasesdS}, \ref{genericcasesdS}, and \ref{parityviolationdS} (where we add parity violation), and investigate in Section \ref{EFTdS} how our derived parameter space can be relaxed if we only require the physical constraints to hold within a finite regime of validity. 

\subsection{Special cases} \label{specialcasesdS}
Similar to the flat space case, we divide the $\alpha_1 = \alpha_2=0$ case into three subcases depending on whether $\gamma_1$ and/or $\gamma_2$ also vanish. 
\subsubsection*{Case 1: $\alpha_1 = \alpha_2 = \gamma_1 = \gamma_2 = 0$} This is the CCM case \cite{Bordin:2018pca}, and we apply the procedure listed above. For now we work in the parity-even limit with $\beta_4 = 0$. Note that our field differs from that in \cite{Bordin:2018pca} by a factor of the scale factor.
\paragraph{\ref{BDchecklist} Check Bunch-Davies conditions} To check whether Bunch-Davies conditions can be imposed, note that since there is no temporal mode, $\sigma_L$ already has a canonical kinetic term. Therefore, we take the limit of the equation of motion \eqref{eomsigmaLdS} as $k \eta \to -\infty$ which reduces it to the flat-space equation with no mass term. Stable plane wave solutions are therefore realised as long as $\beta_1 + \beta_2 < 0$. For the transverse modes, in the same limit the equation of motion also corresponds to the flat-space equation with vanishing mass so stable plane wave solutions requires $\beta_2 < 0$. 
\paragraph{\ref{computeGreenchecklist} Compute the Green's functions} Subject to the Wronskian conditions \eqref{CCMWronskian} and Bunch-Davies conditions, we have the mode functions
\begin{align}
    \sigma^{(h)} &= \frac {\sqrt{\pi}}2 (-\eta)^{1/2} e^{i\pi(\nu + 1/2)/2} H_{\nu}^{(1)}(-c_s^{(h)}k\eta)\,, \label{CCMdSmodefunction}
\end{align}
where
\begin{align}
    \nu = \sqrt{\frac 14 + \frac {\beta_3}{H^2}}, \qquad c_s^{(0)} = \sqrt{-\beta_1 - \beta_2}, \qquad c_s^{(\pm 1)} = \sqrt{-\beta_2}\,. \label{eq:paramsCCM}
\end{align}
Assuming that $\nu \notin \mathbb{N}$, the helicity mode Green's functions can be computed as
\begin{align}
G^{(h)}(\eta_1,\eta_2,k) & = \frac {i \pi (\eta_1 \eta_2)^{1/2}}{2 \sin(\pi \nu)}\left [\left (\frac {\eta_1}{\eta_2}\right )^{\nu} C_{\nu} \left (-\frac {(c_s^{(h)})^2 k^2 \eta_1^2}4\right ) C_{-\nu} \left (-\frac {(c_s^{(h)})^2 k^2 \eta_2^2}4\right ) - (\eta_1 \leftrightarrow \eta_2)\right ]\,, \label{CCMdSGreen}
\end{align}
where 
\begin{align}
   C_{\nu}(z) & = \sum_{m=0}^{\infty} \frac {1}{\Gamma(m+\nu+1)} \frac {z^m}{m!}\,,
\end{align}
is the Bessel-Clifford function \cite{BesselClifford} that is related to Hankel function of the first kind by
\begin{align}
H^{(1)}_{\nu}(z) & = -\frac i{\sin (\pi \nu)} \left [\left (\frac z2 \right )^{-\nu}C_{-\nu}\left (-\frac {z^2}4\right ) - e^{-\nu \pi i}\left (\frac z2\right )^{\nu}C_{\nu}\left (-\frac {z^2}4\right )\right ]\,. \label{HankelitoBesselClifford}
\end{align}
The Bessel-Clifford function is an entire function in both $z$ and $\nu$ and given that the momentum dependence of \eqref{CCMdSGreen} enters only via the Bessel-Clifford function with argument $k^2$, these Green's functions are analytic in $k^2$. The non-analytic terms in $H_{\nu}^{(1)}(-c_s^{(h)} k \eta)$ of the form $(-c_s^{(h)}k \eta)^{\pm \nu}$ therefore cancel in the commutator. Note that \eqref{CCMdSGreen} holds for both $\nu \in \mathbb{R}\setminus \mathbb{N}$ and $\nu = i \mu$, where $\mu \in \mathbb{R}$.

In the case where $\nu \in \mathbb{N}$, the expressions \eqref{CCMdSGreen} and \eqref{HankelitoBesselClifford} should be understood in the limit $\nu \to n$. Note that
\begin{align}
    C_{-n}(z) & = \sum_{m=n}^{\infty} \frac {z^m}{m!(m-n)!} = \sum_{m=0}^{\infty} \frac {z^{m+n}}{m!(m+n)!} = z^nC_n(z)\,.
\end{align}
Combined with the $(\eta_1/\eta_2)^{\nu}$ factor, the numerator becomes symmetric under $\eta_1 \leftrightarrow \eta_2$, and vanishes as $\nu \to n$. In addition, $\sin(\pi\nu) \to 0$, so we can use the L'Hopital rule to compute the limit as
\begin{align}
    \lim_{\nu \to n} G^{(h)}(\eta_1,\eta_2,k) & = \frac {i(\eta_1\eta_2)^{1/2}}{2(-1)^n} \left .\frac {\partial}{\partial \nu} \right |_{\nu = n}\left [\left (\frac {\eta_1}{\eta_2}\right )^{\nu} C_{\nu} \left (-\frac {(c_s^{(h)})^2 k^2 \eta_1^2}4\right ) C_{-\nu} \left (-\frac {(c_s^{(h)})^2 k^2 \eta_2^2}4\right ) - (\eta_1 \leftrightarrow \eta_2)\right ]\,. \label{Glimitinteger}
\end{align}
The derivative of $C_{\nu}$ with respect to $\nu$ is
\begin{align}
\lim_{\nu \to n}\frac {\partial}{\partial \nu} C_{\nu}(z) & = -\lim_{\nu \to n}\sum_{m=0}^{\infty} \frac {\psi(m+\nu+1)}{\Gamma (m+\nu +1)}\frac {z^m}{m!} \\
& = -\sum_{m=0}^{\infty} \frac {-\gamma + \sum_{p=1}^{m+n} (1/p)}{(m+n)!}\frac {z^m}{m!}\,,
\end{align}
where $\gamma$ is the Euler-Mascheroni constant. The harmonic series in the numerator grows much slower than factorials in the denominator, and hence this series converges absolutely for all $z$. In addition, the derivative of $C_{-\nu}$ with respect to $\nu$ is
\begin{align}
    \lim_{\nu \to n} \frac {\partial}{\partial \nu}C_{-\nu}(z) & = -\lim_{\nu \to n}\sum_{m=0}^{\infty} \frac {\psi(m-\nu+1)}{\Gamma (m-\nu +1)}\frac {z^m}{m!} \\
    & = -\sum_{m=0}^{n-1} \frac {(-1)^{n-m}(n-m-1)!z^m}{m!} - \sum_{m=n}^{\infty} \frac {-\gamma + \sum_{p=1}^{m-n} (1/p)}{(m-n)!}\frac {z^m}{m!}\,,
\end{align}
which again converges absolutely for all $z$. Hence, the expression \eqref{Glimitinteger} is analytic in $k^2$ which confirms that the Green's function is analytic for all $\nu$. 
\paragraph{\ref{analyticitychecklist} Check the analyticity conditions} We now check the analyticity conditions \ref{analyticitycondition1}, \ref{analyticitycondition2}, and \ref{analyticitycondition3} in turn.
\begin{itemize}
    \item To check \ref{analyticitycondition1}, since $C_{\nu}(z)$ is entire in $z$, and all the arguments of $C_{\nu}$ in \eqref{CCMdSGreen} are analytic in $k^2$, all Green's functions are analytic in $k^2$, so $\frac 12 (G^{(+1)} + G^{(-1)})$ is analytic in $k^2$, satisfying \ref{analyticitycondition1}.
    \item To check \ref{analyticitycondition2}, since we are not considering parity violation, we have $\frac 12 (G^{(+1)} - G^{(-1)}) = 0$, and hence \ref{analyticitycondition2} is satisfied trivially. We include parity violation in Section \ref{parityviolationdS}.
    \item To check \ref{analyticitycondition3}, we have seen that all Green's functions are analytic in $k^2$, including $G^{(0)}$. We now check that as $k\to 0$, $G^{(0)}$ and $\frac 12 (G^{(+1)} + G^{(-1)})$ tend to the same limit. To this end, note that
    \begin{align}
        \lim_{z\to 0} C_{\nu}(z) = \frac 1{\Gamma(\nu+1)} \,, \label{BesselCliffordlimit}
    \end{align}
    and hence we have
    \begin{align}
        \lim_{k\to 0} G^{(h)}(\eta_1,\eta_2,k) & = \frac {i\pi (\eta_1\eta_2)^{1/2}}{2\sin(\pi \nu)\Gamma (\nu +1)\Gamma(-\nu+1)} \left [\left (\frac {\eta_1}{\eta_2}\right )^{\nu} - \left (\frac {\eta_2}{\eta_1}\right )^{\nu}\right ] \\
        & = \frac {i(\eta_1 \eta_2)^{1/2}}{2\nu} \left [\left (\frac {\eta_1}{\eta_2} \right )^{\nu} - \left (\frac {\eta_2}{\eta_1}\right )^{\nu}\right ] \,.\label{helicitymodeGreenfunctionlimitdSCCM}
    \end{align}
    This expression is independent of $c_s^{(h)}$ and is therefore the same for all helicity modes. It follows that $G^{(0)}$ and $\frac 12 (G^{(+1)} + G^{(-1)})$ tend to the same limit, satisfying \ref{analyticitycondition3}.\footnote{For $\nu =0$, we take the limit of this expression as $\nu \to 0$, and we find that $$\lim_{k,\nu \to 0} G^{(h)}(\eta_1,\eta_2,k) = i(\eta_1 \eta_2)^{1/2} \ln \left (\frac {\eta_1}{\eta_2}\right ) \neq 0.$$}
\end{itemize}
Therefore, similar to flat space CCM, the analyticity conditions do not provide new constraints on the parameters in \eqref{spin1generalfreewithparityodd}.
 \paragraph{\ref{boundchecklist} Check the bound} To check the bound it is better to write the Green's functions in terms of the Hankel functions. We have
 \begin{align}
G^{(h)}(\eta_1,\eta_2,k) & = \frac{\pi}{4} (\eta_1 \eta_2)^{1/2} [H_{\nu}^{(1)}(-c_s^{(h)}k\eta_1)H_{\nu}^{(2)}(-c_s^{(h)}k\eta_2) -  H_{\nu}^{(1)}(-c_s^{(h)}k\eta_2)H_{\nu}^{(2)}(-c_s^{(h)}k\eta_1)]\,. \label{CCMdSGreenHankel}
\end{align}
The fastest growth of Hankel functions takes place along the imaginary axis where 
\begin{align}
&H_{\nu}^{(1)}(z) \sim e^{|\operatorname{Im} z|}\,, \qquad z \rightarrow -i \infty\,, \\
&H_{\nu}^{(2)}(z) \sim e^{|\operatorname{Im} z|}\,, \qquad z \rightarrow + i \infty \,.
\end{align}
Given that both $\eta_1 < 0$ and $\eta_2 <0$, it is not possible to access these limiting behaviours for $H_{\nu}^{(1)}(-c_s^{(h)}k\eta_1)$ and $H_{\nu}^{(2)}(-c_s^{(h)}k\eta_2)$ simultaneously. We therefore have
\begin{align}
  |  G(\eta_1, \eta_2, k) | \lesssim  e^{c_s^{(h)} |\operatorname{Im} k| |\eta_1-\eta_2|}\,,
\end{align}
and the bound is satisfied as long as $c_s^{(h)} \leq 1$ for all helicities. The same conclusion was arrived at in \cite{Hui:2025aja}.
\paragraph{\ref{temporalchecklist} Check the temporal commutators} In this case $\sigma_0$ decouples and is therefore not part of the free theory of the vector field. There is therefore no need to check its commutators. 

\paragraph{Summary} In conclusion, for this case, corresponding to the parity-even CCM case of \cite{Bordin:2018pca}, stability and microcausality imposes:
\begin{keyeqn}
    \begin{equation} \label{CCMdSconstraints}
    \begin{split}
       \text{Case 1:} ~~~
        -1 \leq \beta_1+\beta_2 < 0, \qquad -1 \leq \beta_2 < 0\,.
        \end{split}
    \end{equation}
    \end{keyeqn}

\subsubsection*{Case 2: $\alpha_1 = \alpha_2 = \gamma_1 = 0$ but $\gamma_2 \neq 0$} \label{dSspecialcase2}
Similar to the flat space case, \eqref{eomsigma0dS} now imposes $\sigma_L = 0$, and hence $G^{(0)}(\eta_1,\eta_2,k) = 0$, and is obviously analytic. The transverse mode commutator remains the same as before. Now we check the analyticity condition \ref{analyticitycondition3}, where we require $\frac 12 (G^{(+1)} + G^{(-1)}) \to 0$ as $k\to 0$. From \eqref{helicitymodeGreenfunctionlimitdSCCM}, for $\eta_1 \neq \eta_2$, we have
\begin{align}
    \lim_{k\to 0} G^{(h)} (\eta_1,\eta_2, k) = \frac {i(\eta_1\eta_2)^{1/2}}{2\nu} \left [\left (\frac {\eta_1}{\eta_2}\right )^{\nu} - \left (\frac {\eta_2}{\eta_1}\right )^{\nu}\right ] \neq 0\,,
\end{align}
for all $\nu$. Therefore, this case does not satisfy \ref{analyticitycondition3}, and is ruled out.
\subsubsection*{Case 3: $\alpha_1 = \alpha_2 = 0$ but $\gamma_1 \neq 0$} Similar to the flat space case, \eqref{eomsigma0dS} becomes a first-order differential equation, which cannot satisfy Bunch-Davies conditions in the far past. Therefore, this case is ruled out.
\subsection{Generic cases} \label{genericcasesdS}
When at least one of $\alpha_1$ and $\alpha_2$ is non-zero, \eqref{eomsigma0dS} is a constraint equation for $\sigma_0$, which has the solution
\begin{align}\label{Sigma0Solutiuon}
    \sigma_0 = -\frac {k}{\alpha_1k^2 + \alpha_2 a^2(\eta)} (\gamma_1 \partial_{\eta} \sigma_L - \gamma_2 a(\eta) \sigma_L)\,.
\end{align}
This can be substituted back into \eqref{spin1generalfreewithparityodd} to obtain the Lagrangian (dropping boundary terms)
\begin{align} \label{genericcaseLagrangianintegratedout}
2\mathcal{L}& \supset C \partial_{\eta} \sigma_L \partial_{\eta} \sigma_L +  D \sigma_L^2 \,,
\end{align}
where 
\begin{align}
    C &= \frac {\alpha_2+(\alpha_1 - \gamma_1^2)H^2k^2\eta^2}{\alpha_2 + \alpha_1 H^2k^2\eta^2}\,, \\
    D &= \frac {\beta_3}{H^2\eta^2} + k^2\left (\beta_1 + \beta_2 -\frac {\gamma_2(-\alpha_2\gamma_1 H  + \alpha_1 \gamma_1 H^3k^2\eta^2+ \alpha_2 \gamma_2 + \alpha_1 \gamma_2 H^2k^2\eta^2)}{(\alpha_2 +\alpha_1H^2k^2\eta^2)^2}\right)\,.
\end{align}
Note that we want to keep $C >0$ throughout the evolution which will put constraints on the parameters of \eqref{spin1generalfreewithparityodd}. We will come back to this once we have imposed the conditions from microcausality. By varying the action, we obtain the following equation of motion for $\sigma_L$:
\begin{align}
&-\left (\frac {\alpha_2+(\alpha_1 - \gamma_1^2)H^2k^2\eta^2}{\alpha_2 + \alpha_1 H^2k^2\eta^2}\right )\partial_{\eta}^2 \sigma_L + \frac {2 \alpha_2 \gamma_1^2H^2k^2 \eta}{(\alpha_2+\alpha_1 H^2k^2\eta^2)^2}\partial_{\eta}\sigma_L \nonumber \\ +&\left[\frac {\beta_3}{H^2\eta^2} + k^2\left (\beta_1 + \beta_2 -\frac {\gamma_2(-\alpha_2\gamma_1 H  + \alpha_1 \gamma_1 H^3k^2\eta^2+ \alpha_2 \gamma_2 + \alpha_1 \gamma_2 H^2k^2\eta^2)}{(\alpha_2 +\alpha_1H^2k^2\eta^2)^2}\right) \right]\sigma_L   = 0\,. \label{eomsigmaLdSaftersub}
\end{align}
\paragraph{\ref{BDchecklist} Check Bunch-Davies conditions} To check whether Bunch-Davies conditions can be satisfied, we need to canonically normalise $\sigma_L$ to $\hat{\sigma}_L = \sqrt{C}\sigma_L$ and consider whether the equations of motion on $\hat{\sigma}_L$ reduces to a wave equation as $k \eta \to -\infty$. In terms of the canonically normalised field, we have
\begin{align}
    -\partial_{\eta}^2 \hat{\sigma}_L + E\hat{\sigma}_L = 0 \,, \label{canonicallynormalisedequation}
\end{align}
where 
\begin{align}
    E = \frac {\sum_{n=0}^5 E_n(Hk\eta)^{2n}}{H^2\eta^2(\alpha_2 + \alpha_1H^2k^2\eta^2)^2(\alpha_2 + (\alpha_1 - \gamma_1^2)H^2k^2\eta^2)^2}\,,
\end{align}
with the $\eta$-independent coefficients
\begin{align}
    E_0 & = \alpha_2^4 \beta_3 \,, \\
    E_1& = \alpha_2^3 [\alpha_2(\beta_1+\beta_2)+\beta_3(4\alpha_1-\gamma_1^2)-\gamma_1^2H^2+\gamma_1\gamma_2H-\gamma_2^2]\,, \\
    E_2 & = \alpha_2^2 [3\alpha_1\beta_3(2\alpha_1 - \gamma_1^2)+\alpha_2(4\alpha_1 -\gamma_1^2)(\beta_1+\beta_2)\nonumber \\
    & \quad +(\gamma_1H -\gamma_2)((3\alpha_1-\gamma_1^2)\gamma_2 + 2H\alpha_1\gamma_1)]\,,\\
    E_3 & = \alpha_1\alpha_2  [\alpha_1\beta_3(4\alpha_1 - 3\gamma_1^2)+3\alpha_2(2\alpha_1 - \gamma_1^2)(\beta_1+\beta_2)\nonumber \\
    &\quad+(\alpha_1-\gamma_1^2)(3H^2\gamma_1^2-2\gamma_2^2)  -\alpha_1\gamma_2(\gamma_1H+\gamma_2)]\,,\\
    E_4 & = \alpha_1^2[\alpha_2(4\alpha_1 - 3\gamma_1^2)(\beta_1+\beta_2)+(\alpha_1-\gamma_1^2)(\alpha_1\beta_3-\gamma_1\gamma_2H-\gamma_2^2)]\,,\\
    E_5& = \alpha_1^3 (\alpha_1-\gamma_1^2)(\beta_1+\beta_2)\,.
\end{align}
We now separate the discussion into several cases according to the values of $\alpha_1$, $\alpha_2$ and $\alpha_1-\gamma_1^2$: 
\begin{enumerate} [label=(\alph*)]
    \item \label{caseadS} $\alpha_1 = 0$, $\alpha_2 \neq 0$, and $\alpha_1-\gamma_1^2 =0$: the field redefinition for the non-canonically normalised field
    \begin{align} \label{dSFieldRedefinition}
        \sigma_0 \mapsto \sigma_0 - \frac {\gamma_2}{\alpha_2 a(\eta)} \partial_i \sigma_i\,,
    \end{align}
    turns off the coupling between $\sigma_0$ and $\sigma_i$, and hence we are in the CCM case which we covered in Section \ref{specialcasesdS}.
    \item \label{casebdS} $\alpha_1 = 0$, $\alpha_2 \neq 0$, and $\alpha_1 - \gamma_1^2 \neq 0$: when $|k \eta| \to \infty$ we have
  \begin{align}
    E \sim \frac{k^2}{\alpha_2^2(\alpha_1-\gamma_1^2)^2} \frac {\sum_{n=0}^5 E_n(Hk\eta)^{2n}}{H^6 k^6 \eta^6}\,,
\end{align}
and to satisfy the Bunch-Davies condition we require $E_5 = E_4 = 0$ and $E_3 \neq 0$ which is not the case since when $\alpha_1=0$ we have $E_3 = 0$. In this case we therefore cannot satisfy the Bunch-Davies condition. 
    
\item \label{casecdS} $\alpha_1 \neq 0$, $\alpha_2 = 0$, and $\alpha_1 - \gamma_1^2 = 0$: in this case, since $C \equiv 0$, the first step of canonical normalisation cannot even work. From the Lagrangian \eqref{genericcaseLagrangianintegratedout} we see that in this case the kinetic term of the longitudinal mode vanishes and hence it is infinitely strongly coupled. This case is therefore ruled out.
\item \label{weirdgenericcasedS} $\alpha_1 \neq 0$, and $\alpha_1 - \gamma_1^2 \neq 0$: regardless of whether $\alpha_2 = 0$ or not, as we take the limit $| k\eta| \to \infty$, we have 
\begin{align}
E \sim \frac{k^2}{\alpha_1^2(\alpha_1-\gamma_1^2)^2}\frac{\sum_{n=0}^5 E_n (H k \eta)^{2n}}{H^{10} k^{10} \eta^{10}} \sim \frac{E_5 k^2}{\alpha_1^2(\alpha_1-\gamma_1^2)^2}\,,
\end{align}
which can yield plane wave solutions as long as $E_5 \neq 0$. Given that $\alpha_1 \neq 0$ and $\alpha_1 - \gamma_1^2 \neq 0$ we therefore require $\beta_1+\beta_2 \neq 0$ to satisfy the Bunch-Davies condition.
\item \label{generalisedProcadS} $\alpha_1 \neq 0$, $\alpha_2 \neq 0$, and $\alpha_1-\gamma_1^2 = 0$: as $|k \eta| \to \infty$, we have
 \begin{align}
    E \sim \frac{k^2}{\alpha_1^2 \alpha_2^2} \frac {\sum_{n=0}^5 E_n(Hk\eta)^{2n}}{H^6 k^6 \eta^6}\,.
\end{align}
To satisfy the Bunch-Davies condition we therefore require $E_4 = E_5 =0 $ and $E_3 \neq 0$. $E_5 = 0$ since $\alpha_1 - \gamma_1^2 = 0$, while for $E_4$ we have 
\begin{align}
    E_4 = \alpha_1^3\alpha_2(\beta_1+\beta_2)\,.
\end{align}
Since $\alpha_1 \neq 0$ and $\alpha_2 \neq 0$, we therefore require $\beta_1+\beta_2 = 0$ to satisfy the Bunch-Davies condition. We then have 
\begin{align}
    E_3 & = \gamma_1^4\alpha_2[\gamma_1^2\beta_3 - \gamma_1\gamma_2 H - \gamma_2^2]\neq 0 \implies E \sim \frac {(\gamma_1^2\beta_3 - \gamma_1\gamma_2H - \gamma_2^2)k^2}{\alpha_2}\,,
\end{align}
as $|k \eta| \to \infty$. For stable plane wave solutions we therefore require $(\gamma_1^2\beta_3 - \gamma_1\gamma_2 H - \gamma_2^2) / \alpha_2<0$. In fact, in this case, for the non-canonically normalised field $\sigma_L$, the kinetic term has the coefficient
\begin{align}
    C & = \frac {\alpha_2}{\alpha_2 + \gamma_1^2H^2k^2\eta^2}\,,
\end{align}
which is required to be positive for all $\eta$ (or for all $k$) so that $\sigma_L$ is not a ghost. This implies $\alpha_2 > 0$ which is the Higuchi bound \cite{Higuchi:1986py} now generalised to our more general set-up.\footnote{It is interesting that the Higuchi-like bound we have here is a bound on the mass of the non-dynamical mode rather than the dynamical ones. In the Proca theory, these masses are the same by symmetry and therefore the bound applies to both.} Combining the two inequalities, this case requires $\gamma_1^2\beta_3 - \gamma_1\gamma_2 H - \gamma_2^2 < 0 < \alpha_2$ to satisfy the Bunch-Davies condition.
\end{enumerate}
Note that all other cases involve $\alpha_1 = \alpha_2 = 0$, which have been dealt with in Section \ref{specialcasesdS}. This analysis suggests that after imposing Bunch-Davies conditions, only CCM, cases \ref{weirdgenericcasedS} and \ref{generalisedProcadS} are viable and proceed to the next stages where we compute the Green's functions and consider analyticity. 
\paragraph{\ref{computeGreenchecklist} Compute the Green's functions} We consider the two cases identified above separately.
\begin{itemize} 
\item Case \ref{weirdgenericcasedS}: since the parameter space in this case remains largely unconstrained, it is difficult to solve the equation of motion \eqref{eomsigmaLdSaftersub} directly. We will postpone the discussion of this case to the next step where we consider analyticity.
\item Case \ref{generalisedProcadS}: in this case, $\alpha_1 - \gamma_1^2 = 0$, and we also deduced $\beta_1 + \beta_2 = 0$ in the previous discussion. Therefore, the equation of motion greatly simplifies, and we can compute the mode functions directly to find
\begin{align} \label{CaseeMode}
\sigma_L & = \frac {\sqrt{\pi}}{4\sqrt{\alpha_2}c_L} e^{i\pi (\nu+1/2)/2} (-\eta)^{1/2} \left [[H\gamma_1 (1-2\nu) + 2\gamma_2] H_{\nu}^{(1)}(-c_Lk\eta) - 2\gamma_1 H c_Lk\eta H_{\nu+1}^{(1)}(-c_Lk\eta)\right ]\,,
\end{align}
where
\begin{align} \label{LongModecs}
c_L & = \sqrt{\frac {\gamma_2^2 + \gamma_1 \gamma_2 H - \beta_3 \gamma_1^2}{\alpha_2}}, \qquad \qquad \nu = \sqrt{\frac 14 + \frac {\beta_3}{H^2}}\,.
\end{align}
In this case, we imposed the Wronskian condition
\begin{align}
C(\sigma_L \partial_{\eta}\sigma_L^{*} - \sigma_L^{*}\partial_{\eta}\sigma_L) = i \,,
\end{align}
and it is straightforward to check that in the limit $|k \eta| \to \infty$ the canonically normalised field's mode functions are exactly those of flat space with $\omega = c_L k$. Note that unlike the CCM mode function \eqref{CCMdSmodefunction}, which is given by a single Hankel function, this mode function is a sum of Hankel functions of different orders which is familiar from the Proca theory in dS space \cite{Lee:2016vti} (which is a subset of this case \ref{generalisedProcadS}). 

The longitudinal mode Green's function is then
\begin{align}
    G^{(0)}(\eta_1,\eta_2,k) & = \frac {i\pi H^2 (\eta_1 \eta_2)^{1/2}}{2\alpha_2c_L^2 \sin(\pi \nu)} \left (\frac {\eta_1}{\eta_2}\right )^{\nu} \times\nonumber \\
    &\left [\left (\frac {\gamma_1}2 - \gamma_1 \nu + \frac {\gamma_2}H \right ) C_{\nu} \left (-\frac {c_L^2k^2\eta_1^2}4\right ) + \frac{1}{2}\gamma_1 c_L^2 k^2 \eta_1^2 C_{\nu + 1} \left (-\frac {c_L^2 k^2 \eta_1^2}4\right )\right ] \times \nonumber \\
    &\left [\left (\frac {\gamma_1}2 - \gamma_1 \nu + \frac {\gamma_2}H \right ) C_{-\nu} \left (-\frac {c_L^2k^2\eta_2^2}4\right ) - 2\gamma_1 C_{-\nu - 1} \left (-\frac {c_L^2 k^2 \eta_2^2}4\right )\right ] - (\eta_1 \leftrightarrow \eta_2)\,, \label{dSgeneralisedProcaGreen}
\end{align}
where $C_{\nu}(z)$ is again the Bessel-Clifford function. Using the recurrence relation 
\begin{align}
z C_{\nu+1}(z) + \nu C_{\nu}(z) = C_{\nu-1}(z)\,,
\end{align}
we can write 
\begin{align}
    G^{(0)}(\eta_1,\eta_2,k) & = \frac {i\pi H^2 (\eta_1 \eta_2)^{1/2}}{2\alpha_2c_L^2 \sin(\pi \nu)} \left (\frac {\eta_1}{\eta_2}\right )^{\nu} \times\nonumber \\
    &\left [\left (\frac {\gamma_1}2 + \gamma_1 \nu + \frac {\gamma_2}H \right ) C_{\nu} \left (-\frac {c_L^2k^2\eta_1^2}4\right ) - 2\gamma_1 C_{\nu - 1} \left (-\frac {c_L^2 k^2 \eta_1^2}4\right )\right ] \times \nonumber \\
    &\left [\left (\frac {\gamma_1}2 - \gamma_1 \nu + \frac {\gamma_2}H \right ) C_{-\nu} \left (-\frac {c_L^2k^2\eta_2^2}4\right ) - 2\gamma_1 C_{-\nu - 1} \left (-\frac {c_L^2 k^2 \eta_2^2}4\right )\right ] - (\eta_1 \leftrightarrow \eta_2)\,. \label{dSgeneralisedProcaGreen2}
\end{align}
Since the transverse modes are not coupled to the non-dynamical $\sigma_0$, their Green's functions are the same as \eqref{CCMdSGreen} for $h=\pm 1$.
\end{itemize}
\paragraph{\ref{analyticitychecklist} Check the analyticity conditions} We now check the analyticity conditions in each case. Since the transverse modes Green's function are unchanged, the conditions \ref{analyticitycondition1} and \ref{analyticitycondition2} are automatically satisfied and so we concentrate on \ref{analyticitycondition3}. 
\begin{itemize}
    \item Case \ref{weirdgenericcasedS}: we can try a series solution for $\sigma_L$. A reasonable assumption is that the mode function of $\sigma_L$ is regular in the vicinity of some $\eta_c<0$. Therefore, we seek a series solution of the form
    \begin{align}
        \sigma_L & = \sum_{n=0}^{\infty} a_n (\eta - \eta_c)^n\,, \label{seriesansatz}
    \end{align}
    where $a_n$ are complex coefficients. Since this solution is subject to the Wronskian condition
    \begin{align}
    \sigma_L \partial_{\eta}\sigma^{*}_L - \partial_{\eta}\sigma_L \sigma^*_L & = \frac {i(\alpha_2 + \alpha_1 H^2k^2\eta^2)}{\alpha_2 + (\alpha_1 - \gamma_1^2) H^2k^2\eta^2}\,,
    \end{align}
    we can expand both sides in a series of $\eta - \eta_c$. Comparing the zeroth-order terms yields
    \begin{align}
        a_0 a_1^* - a_1 a_0^* & = \frac {i(\alpha_2 + \alpha_1 H^2k^2\eta_c^2)}{\alpha_2 + (\alpha_1 - \gamma_1^2) H^2k^2\eta_c^2}\,. \label{Wronskiancoefficients}
    \end{align}
    Since \ref{analyticitycondition3} requires the longitudinal mode Green's function to be analytic in $k^2$ for all $\eta_1$ and $\eta_2$, we take $\eta_1$ and $\eta_2$ to be within the radius of convergence of \eqref{seriesansatz}, specifically $\eta_1 - \eta_c = \eta_c - \eta_2 = \epsilon$, and we expand the Green's function in powers of $\epsilon$:
    \begin{align}
        G^{(0)}(\eta_1, \eta_2,k) & = \sigma_L (\eta_1,k) \sigma^*_L (\eta_2,k) - \sigma^*_L(\eta_1,k) \sigma_L(\eta_2,k) \\
        & = (a_0 + a_1 \epsilon) (a^*_0 - a^*_1\epsilon) - (a^*_0 + a^*_1 \epsilon) (a_0 - a_1\epsilon) + \mathcal{O}(\epsilon^2) \\
        & = 2(a^*_0 a_1 - a^*_1 a_0)\epsilon + \mathcal{O}(\epsilon^2)\\
& =  -\frac {2i(\alpha_2 + \alpha_1 H^2k^2\eta_c^2)}{\alpha_2+(\alpha_1 - \gamma_1^2)H^2k^2\eta_c^2}\epsilon + \mathcal{O}(\epsilon^2)\,, \label{dSweirdgenericcasefirstorderGreen}
    \end{align}
    where we used the Wronskian condition \eqref{Wronskiancoefficients} in the last line. Applying the same argument to the transverse mode Green's functions, using the Wronskian condition \eqref{CCMWronskian}, yields
    \begin{align}
        G^{(\pm 1)} (\eta_1, \eta_2,k) = -2i \epsilon + \mathcal{O}(\epsilon^2)\,.
    \end{align}
    Since the full Green's function is analytic in $k^2$ for every $\epsilon$ within the radius of convergence, the leading coefficient in $\epsilon$ has to be analytic in $k^2$. In particular, \ref{analyticitycondition3} requires
    \begin{align}
        G^{(0)} - \frac 12 (G^{(+1)} + G^{(-1)}) = -\frac {2i\gamma_1^2 H^2k^2\eta_c^2}{\alpha_2 + (\alpha_1 - \gamma_1^2) H^2 k^2 \eta_c^2} \epsilon + \mathcal{O}(\epsilon^2) \,,\label{checkA3limitdSweirdcase}
    \end{align}
    to be analytic in $k^2$, and tend to 0 as $k\to 0$. If $\alpha_2=0$ then the limit requires $\gamma_1=0$, while if $\alpha_2 \neq 0$, analyticity requires $\gamma_1 = 0$ since in this case we have $\alpha_1 - \gamma_1^2 \neq 0$. So for any value of $\alpha_2$ we require $\gamma_1 = 0$. 
    
    This allows us to simplify the equation of motion \eqref{eomsigmaLdSaftersub} to
    \begin{align}
        \left [-\partial_{\eta}^2 + \frac {\beta_3}{H^2\eta^2} + k^2 \left (\beta_1 + \beta_2 -\frac {\gamma_2^2}{\alpha_2 + \alpha_1 H^2 k^2 \eta^2}\right )\right ]\sigma_L & = 0\,,
    \end{align}
    where the first-order term vanishes. We use the same series ansatz \eqref{seriesansatz}, but this time we expand up to the third order, and use the equation of motion instead of just the Wronskian. The zeroth-order term gives
    \begin{align}
        2a_2 & = \left [\frac {\beta_3}{H^2\eta_c^2} + k^2 \left (\beta_1 + \beta_2 -\frac {\gamma_2^2}{\alpha_2 + \alpha_1 H^2 k^2 \eta_c^2} \right ) \right ]a_0\,,
    \end{align}
    while the linear term gives
    \begin{align}
        6a_3 & = Fa_0 + \left [\frac {\beta_3}{H^2\eta_c^2} + k^2 \left (\beta_1 + \beta_2 -\frac {\gamma_2^2}{\alpha_2 + \alpha_1 H^2 k^2 \eta_c^2} \right ) \right ]a_1\,,
    \end{align}
    where $F$ is some function of $k$ and $\eta_c$, but the exact details are irrelevant, since we can compute the Green's function to the third order without it:
    \begin{align}
    G^{(0)}(\eta_1,\eta_2,k) & = \sigma_L (\eta_1,k) \sigma^*_L (\eta_2,k) - \sigma^*_L(\eta_1,k) \sigma_L(\eta_2,k) \\
        & =  2(a^*_0a_1 - a^*_1 a_0) \epsilon + 2(a_3a^*_0 - a_0a^*_3 + a_1a_2^* - a_2a_1^*)\epsilon^3 + \mathcal{O}(\epsilon^5)\\
        & = 2(a^*_0a_1 - a^*_1 a_0) \left (\epsilon + \frac 23 \left [\frac {\beta_3}{H^2\eta_c^2} + k^2 \left (\beta_1 + \beta_2 -\frac {\gamma_2^2}{\alpha_2 + \alpha_1 H^2 k^2 \eta_c^2} \right )\right ]\epsilon^3\right ) + \mathcal{O}(\epsilon^5)\\
        & = -2i \left (\epsilon + \frac 23 \left [\frac {\beta_3}{H^2\eta_c^2} + k^2 \left (\beta_1 + \beta_2 -\frac {\gamma_2^2}{\alpha_2 + \alpha_1 H^2 k^2 \eta_c^2} \right )\right ]\epsilon^3\right ) + \mathcal{O}(\epsilon^5)\,, \label{Step2analyticity}
    \end{align}
    where we used the Wronskian condition \eqref{Wronskiancoefficients} with $\gamma_1=0$. Since this case has $\alpha_1 \neq 0$ by the Bunch-Davies condition, in order for the $\epsilon^3$ coefficient to be analytic in $k^2$, we require either $\gamma_2=0$ or $\alpha_2 = 0$. If $\gamma_2 = 0$, then since we already have $\gamma_1 = 0$, $\sigma_0$ and $\sigma_i$ completely decouple in \eqref{spin1generalfreewithparityodd} and we are back to the CCM case covered in Section \ref{specialcasesdS}. If instead $\alpha_2 = 0$, the equation of motion simplifies even further to
    \begin{align}
        \left [-\partial_{\eta}^2 + (\beta_1 + \beta_2)k^2 + \left (\beta_3 - \frac {\gamma_2^2}{\alpha_1}\right ) \frac 1{H^2\eta^2}\right ] \sigma_L & = 0\,,
    \end{align}
    which is the same equation of motion as the transverse modes \eqref{eomsigmaTdS} with different parameters, and hence the mode function is
    \begin{align}
        \sigma_L & = \frac {\sqrt{\pi}}2 (-\eta)^{1/2} e^{i\pi(\nu^{(0)} + 1/2)/2} H_{\nu^{(0)}}^{(1)}(-c_s^{(0)}k\eta)\,,
    \end{align}
    where 
    \begin{align}
        \nu^{(0)} = \sqrt{\frac 14 + \frac {\beta_3 - \gamma_2^2/\alpha_1}{H^2}}, \qquad \qquad c_s^{(0)} = \sqrt{-\beta_1 - \beta_2}\,.
    \end{align}
    The longitudinal mode Green's function is then
    \begin{align}
        G^{(0)}(\eta_1,\eta_2,k) & = \frac {i \pi (\eta_1 \eta_2)^{1/2}}{2 \sin(\pi \nu^{(0)})}\left [\left (\frac {\eta_1}{\eta_2}\right )^{\nu^{(0)}} C_{\nu^{(0)}} \left (-\frac {(c_s^{(0)})^2 k^2 \eta_1^2}4\right ) C_{-\nu^{(0)}} \left (-\frac {(c_s^{(0)})^2 k^2 \eta_2^2}4\right ) - (\eta_1 \leftrightarrow \eta_2)\right ]\,.
    \end{align}
    The analyticity condition \ref{analyticitycondition3} requires $G^{(0)}$ to be analytic in $k^2$, which it is, and for it to tend to the same limit as $\frac 12 (G^{(+1)} + G^{(-1)})$ as $k\to 0$. The limit for the transverse mode Green's functions have been computed in \eqref{helicitymodeGreenfunctionlimitdSCCM}, and that for the longitudinal mode is similar, with $\nu$ replaced by $\nu^{(0)}$. These two limits are equivalent for all $\eta_1$ and $\eta_2$ only if $\nu = \nu^{(0)}$, which requires $\gamma_2=0$. As discussed, $\gamma_2 = 0$ then implies decoupling between $\sigma_0$ and $\sigma_i$, and we are back in the CCM case. This case can therefore be discarded as it leads to nothing new.
\item Case \ref{generalisedProcadS}: from \eqref{dSgeneralisedProcaGreen2}, since $C_{\nu}(z)$ is an entire function in $z$, $G^{(0)}$ is analytic in $k^2$ and so the first part of \ref{analyticitycondition3} is satisfied. Now we need to check whether the $k\to 0$ limits of $G^{(0)}$ and $\frac 12 (G^{(+1)} + G^{(-1)})$ are the same. To this end, we use \eqref{BesselCliffordlimit} to compute the limit of the longitudinal mode Green's function:
\begin{align}
    \lim_{k\to 0} G^{(0)} (\eta_1,\eta_2,k) & = \frac {i\pi H^2 (\eta_1\eta_2)^{1/2}}{2\alpha_2 c_L^2 \sin (\pi \nu)} \left (\frac {\eta_1}{\eta_2}\right )^{\nu} \frac {\left (\frac {\gamma_1}2 - \gamma_1 \nu + \frac {\gamma_2}H\right ) \left (\frac {\gamma_1}2 + \gamma_1 \nu + \frac {\gamma_2}H \right )}{\Gamma(1+\nu) \Gamma(1-\nu)} - (\eta_1 \leftrightarrow\eta_2 ) \\
    & = \frac {i H^2 (\eta_1 \eta_2)^{1/2}}{2 \nu \alpha_2 c_L^2 } \left (\frac {\eta_1}{\eta_2}\right )^{\nu} \left [\frac {\gamma_1^2}4 + \frac {\gamma_1 \gamma_2}H + \frac {\gamma_2^2}{H^2} - \gamma_1^2\nu^2\right ] - (\eta_1 \leftrightarrow \eta_2 ) \\
    & = \frac {i(\eta_1\eta_2)^{1/2}}{2\nu} \left [\left (\frac {\eta_1}{\eta_2}\right )^{\nu} - \left (\frac {\eta_2}{\eta_1}\right )^{\nu}\right ] \,,
\end{align}
which is exactly the same as that for the transverse modes given in \eqref{helicitymodeGreenfunctionlimitdSCCM}. Hence $G^{(0)}$ and $\frac 12 (G^{(+1)} + G^{(-1)})$ tend to the same limit as $k\to 0$, and \ref{analyticitycondition3} is satisfied with no additional conditions on the Lagrangian parameters. 
\end{itemize}
In summary, we can focus on case \ref{generalisedProcadS} with any other cases ruled out or reduced to the CCM case of Section \ref{specialcasesdS}.

\paragraph{\ref{boundchecklist} Check the bound} Given that the transverse mode functions are the same as in Section \ref{specialcasesdS}, the bound for the transverse modes is unchanged and requires $-1 \leq \beta_2 <0$. For the longitudinal mode, we again need to consider the behaviour of the mode function \eqref{CaseeMode} at $k \eta \to \pm i \infty$ where the mode function grows/decays exponentially in a manner that is independent of the Hankel function's order parameter. To satisfy the bound we therefore need $c_L \leq 1$ where $c_L$ is given in \eqref{LongModecs}. Combining this condition with that of having stable plane wave solutions in the far past, yields $0 < \gamma_2^2 + H \gamma_1 \gamma_2 - \beta_3 \gamma_1^2 \leq \alpha_2$.

\paragraph{\ref{temporalchecklist} Check the temporal commutators} Using \eqref{CaseeMode} and \eqref{Sigma0Solutiuon}, we have
\begin{align} \label{eq:s0modecasee}
    \sigma_0 = \frac{\sqrt{\pi} H c_L k}{2 \sqrt{\alpha_2}} e^{i \pi(\nu +1/2)/2}(- \eta)^{3/2}H_{\nu}^{(1)}(- c_L k \eta)\,,
\end{align}   
and so for the same reasons as above, the bound is guaranteed to be satisfied for both $G_{\text{mixed}}$ and $G_{\text{temporal}}$ with $c_L \leq 1$. To check analyticity we need explicit expressions and have 
\begin{align}
G_{\text{temporal}}(\eta_1,\eta_2,k) & = \frac {i \pi H^2 c_L^2 k^2 (\eta_1 \eta_2)^{3/2}}{2 \alpha_2 \sin(\pi \nu)}\left [\left (\frac {\eta_1}{\eta_2}\right )^{\nu} C_{\nu} \left (-\frac {c_L^2 k^2 \eta_1^2}4\right ) C_{-\nu} \left (-\frac {c_L^2 k^2 \eta_2^2}4\right ) - (\eta_1 \leftrightarrow \eta_2)\right ]\,, 
\end{align}
which is analytic in $k^2$ as required, and 
\begin{align}
G_{\text{mixed}}(\eta_1,\eta_2,k) & = -\frac {i\pi H^2k}{2\alpha_2 \sin(\pi\nu)} \left [\eta_1^{3/2 + \nu}\eta_2^{1/2-\nu} C_{\nu} \left (-\frac {c_L^2k^2\eta_1^2}4\right ) \times \right .\nonumber \\
& \qquad \left .\left [\left (\frac {\gamma_1}2 - \frac {\gamma_1 \nu}2 + \frac {\gamma_2}H\right ) C_{-\nu} \left (-\frac {c_L^2k^2\eta_2^2}4\right ) - 2\gamma_1 C_{-\nu-1} \left (-\frac {c_L^2k^2\eta_2^2}4\right )\right ] - (\nu \leftrightarrow -\nu)\right ]\,,
\end{align}
which is $k$ times an analytic function in $k^2$, as required. As above, for $\nu \rightarrow 0$ we take the limit.  
\paragraph{Summary} In summary, we are left with case \ref{generalisedProcadS} where the conditions on the parameters of \eqref{spin1generalfreewithparityodd} are:
\begin{keyeqn}
    \begin{equation} \label{generaliseddSProcacoefficients}
    \begin{split}
        \alpha_1 =\gamma_1^2 \neq 0 , \qquad \beta_1 + \beta_2 = 0, \qquad 
        0 < \gamma_2^2 + \gamma_1 \gamma_2 H - \beta_3 \gamma_1^2 \leq \alpha_2, \qquad -1 \leq \beta_2 < 0\,.
        \end{split}
    \end{equation}
    \end{keyeqn}
    As we did in the flat-space case, we can now rescale $\sigma_0 \mapsto -\gamma_1^{-1} \sigma_0$ and perform the following relabeling of parameters: $\beta_2 \mapsto - c_T^2$ (speed of sound of transverse modes), $\beta_3 \mapsto -M^2$ (mass of longitudinal and transverse modes), $\frac{\alpha_2}{\gamma_1^2} \mapsto m^2$ (mass of temporal mode), $-\frac{\gamma_2}{\gamma_1} \mapsto \beta$ (1-derivative mode mixing). We can write the final parity-even Lagrangian as\footnote{Here we have extracted all scale factors and are therefore raising and lowering indices with the flat metric $\eta_{\mu\nu}$.}
\begin{keyeqn}
    \begin{align}
    2\mathcal{L}& = \underbrace{- F_{0i}F^{0i} -\frac{c_T^2}{2} F_{ij}F^{ij}}_{\text{gauge-invariant kinetic terms}} + \underbrace{m^2 a^2(\eta)(\sigma_0)^2 - M^2 a^2(\eta) (\sigma_i)^2}_{\text{mass terms}} + \underbrace{2\beta a(\eta) \sigma_0 \partial_i \sigma_i}_{\text{1-derivative term}} \,.\label{dSspacegeneralisedProca}
\end{align}
\end{keyeqn}
This is subject to the constraints:
\begin{keyeqn}
    \begin{equation} \label{generaliseddSProcaconstraints}
    \begin{split} 
  0 <  c_T^2 \leq 1, \qquad     0 < \beta^2 - H \beta + M^2 \leq m^2 \,.
        \end{split}
    \end{equation}
    \end{keyeqn}
As in the flat-space case, we therefore see that in the presence of the temporal mode,  gauge invariant kinetic terms are required by microcausality. Note that we require $m^2 > 0$ such that the longitudinal mode is not a ghost but there is no such condition on $M^2$. In the dS-invariant limit we have $m^2 = M^2$ in which case $m^2 > 0$ corresponds to the Higuchi bound \cite{Higuchi:1986py}. This Lagrangian is the same as the one briefly discussed in the superluminality section of \cite{Lee:2016vti} with the addition of the one-derivative term. 
\subsection{Adding parity violating terms} \label{parityviolationdS}
In the previous subsection, we worked with the parity-even theory with $\beta_4 = 0$. Here we include the parity-odd term and since only the transverse modes are affected, we only need to slightly modify the previous discussion. With $\beta_4 \neq 0$, the transverse mode functions are now modified to
\begin{align}
     \sigma^{(\pm 1)} & = \frac {e^{\pi \kappa_{\pm}/2}}{\sqrt{2c_Tk}}W_{-i\kappa_{\pm},\nu}(2ic_Tk\eta)\,, \qquad \beta_2 = -c_T^2,\qquad \kappa_{\pm} = \pm \frac {\beta_4}{c_T H}, \qquad \nu = \sqrt{\frac 14 + \frac {\beta_3}{H^2}}\,,
\end{align}
where $W_{a,b}(z)$ is the Whittaker $W$-function. It is simple to check that this mode function satisfies the Wronskian condition \eqref{CCMWronskian}. Note that by the Bunch-Davies condition, we require $\beta_2 <0$ so we defined $\beta_2 = -c_T^2$, with $c_T \in \mathbb{R}$. Assuming $2\nu \notin \mathbb{N}$, the corresponding Green's function is then 
\begin{align}
G^{(\pm 1)} & (\eta_1,\eta_2,k ) = -\frac {i\pi e^{-ic_Tk (\eta_1+\eta_2)} (\eta_1\eta_2)^{1/2}}{\sin(2\pi \nu)} \left (\frac {\eta_2}{\eta_1}\right )^{\nu} \times \nonumber \\
    & \; {\bf M}\left(\frac 12 - \nu + i\kappa_{\pm}, 1-2\nu, 2ic_T k \eta_1 \right) {\bf M}\left(\frac 12 + \nu + i\kappa_{\pm}, 1+ 2\nu, 2ic_Tk\eta_2 \right) - (\eta_1 \leftrightarrow \eta_2)\,, \label{nuintegerWhittaker}
\end{align}
where ${\bf M}(a,b,z) = M(a,b,z)/\Gamma(b)$ is the confluent hypergeometric function divided by $\Gamma(b)$. ${\bf M}(a,b,z)$ which is entire in $z$ for all values of $a$ and $b$. When $b = 1+2\nu \in \mathbb{Z}^+$, we have
\begin{align}
    {\bf M}(a,b,z) & = \frac {z^{1-b} {\bf M}(a-b+1,2-b,z)}{(a-b+1)\cdots (a-1)}\,.
\end{align}
Therefore, 
\begin{align}
    & \left (\frac {\eta_2}{\eta_1}\right )^{\nu} {\bf M}\left (\frac 12 - \nu + i\kappa_{\pm},1-2\nu,2ic_Tk\eta_1\right ){\bf M}\left (\frac 12 + \nu + i\kappa_{\pm}, 1+2\nu, 2ic_Tk\eta_2\right ) \\
    =\; &\frac {e^{i\pi \nu}(4c_T^2k^2\eta_1\eta_2)^{-\nu}}{\left (\frac 12 - \nu + i\kappa_{\pm}\right )_{2\nu}} {\bf M}\left (\frac 12 - \nu + i\kappa_{\pm}, 1-2\nu, 2ic_Tk\eta_1\right ){\bf M}\left (\frac 12 - \nu + i\kappa_{\pm}, 1-2\nu, 2ic_Tk\eta_2\right )\,,
\end{align}
which is symmetric under $\eta_1 \leftrightarrow \eta_2$. Therefore, the numerator in \eqref{nuintegerWhittaker} vanishes as $1+2\nu$ tends to an integer. In addition, $\sin(2 \pi \nu) \to 0$, so we can use the L'Hopital rule to compute the limit of \eqref{nuintegerWhittaker} as $1+2\nu \to n \in \mathbb{Z}^+$. We then only need to show that the derivative is well-defined in this limit. To this end, we compute
\begin{align}
    \left .\frac {\partial}{\partial u}\right |_{u\to n} {\bf M}\left (\frac u2 + i\kappa_{\pm},u,z\right ) & =  \left .\frac {\partial}{\partial u}\right |_{u\to n}\sum_{m=0}^{\infty} \frac {(u/2 + i\kappa_{\pm})_m}{\Gamma(u+m)} \frac {z^m}{m!} \\
    & = \sum_{m=0}^{\infty} \underbrace{\frac {(n/2 + i\kappa_{\pm})_m}{(m+n-1)!}}_{\mathcal{O}(1), \text{ as }m\to \infty}\frac {z^m}{m!} \underbrace{\left [\sum_{p=0}^{m-1}\left (\frac 1{n+2 p + 2i\kappa_{\pm}}-\frac 1{n+p}\right )- \psi(n)\right ]}_{\sim \log m, \text{ as }m \to \infty}\,,
\end{align}
where $\psi$ is the digamma function. The derivative still converges absolutely for any $z$ since the logarithm in the numerator is dominated by factorial in the denominator, and is therefore still analytic in $z$. This means for all $\nu$, the Green's function \eqref{nuintegerWhittaker} is analytic in $k$. We now turn to the three analyticity conditions:
\begin{itemize}
    \item \ref{analyticitycondition1}: $G^{(\pm 1)}(\eta_1,\eta_2,k)$ is analytic in $k$, so for $G^{(+1)} + G^{(-1)}$ to be analytic in $k^2$ we need to show that it is a symmetric combination of $k$. Using the identity
    \begin{align}
        {\bf M}(a,b,z) = e^z  {\bf M}(b-a,b,-z)\,, \label{hypergeometricidentity}
    \end{align}
    and $\kappa_{+} = - \kappa_{-}$, we have 
     \begin{align}
        G^{(+1)}(\eta_1, \eta_2, -k) + G^{(-1)}(\eta_1,\eta_2,-k) = G^{(+1)}(\eta_1, \eta_2, k) + G^{(-1)}(\eta_1,\eta_2,k)\,.
    \end{align}
    The condition \ref{analyticitycondition1} is therefore satisfied.  
    \item \ref{analyticitycondition2}: For this condition we need to show that $G^{(+1)} - G^{(-1)}$ is $k$ times an analytic function of $k^2$. We therefore need to show that it is an odd combination of $k$. Using \eqref{hypergeometricidentity} and $\kappa_{+} = - \kappa_{-}$ again, we have 
    \begin{align}
        G^{(+1)}(\eta_1, \eta_2, -k) - G^{(-1)}(\eta_1,\eta_2,-k) = G^{(-1)}(\eta_1, \eta_2, k) - G^{(+1)}(\eta_1,\eta_2,k)\,,
    \end{align}
    so \ref{analyticitycondition2} is satisfied.  
    \item \ref{analyticitycondition3}: $G^{(0)}$ is analytic from our previous discussion so we just need to verify that $G^{(0)}$ and $\frac 12 (G^{(+1)} + G^{(-1)})$ tend to the same limit as $k\to 0$. To this end, note that
    \begin{align}
        {\bf M}(a,b,z) = \frac 1{\Gamma(b)} + \mathcal{O}(z)\,,
    \end{align}
    for small $z$, and hence
    \begin{align}
        \lim_{k\to 0} G^{(\pm 1)}(\eta_1,\eta_2,k) & = \frac {i(\eta_1\eta_2)^{1/2}}{2\nu} \left [\left (\frac {\eta_1}{\eta_2}\right )^{\nu} - \left (\frac {\eta_2}{\eta_1}\right )^{\nu}\right ] \,,  
    \end{align}
    which is the same limit as in the parity-even case c.f. \eqref{helicitymodeGreenfunctionlimitdSCCM}. The condition \ref{analyticitycondition3} therefore remains satisfied when $\beta_4 \neq 0$. 
\end{itemize}
Therefore, adding the parity-violating term does not violate any of the analyticity conditions. 

For the bound we write the Green's function as 
\begin{align}
G^{(\pm 1)}(\eta_1, \eta_2, k) = \frac{e^{\pi \kappa_{\pm}}}{2 c_T k}W_{-i \kappa_{\pm}, \nu}(2 i c_T k \eta_1)W_{i \kappa_{\pm}, \nu}(-2 i c_T k \eta_2)- (\eta_1 \leftrightarrow \eta_2)\,,
\end{align}
and note that the Whittaker function $W_{a,b}(z)$ grows fastest and exponentially when $z \rightarrow - \infty$: $W_{a,b}(2z) \sim e^{|z|}$ as $z \rightarrow -\infty$. It is not possible to simultaneously probe the exponential growth of both Whittaker functions in $G^{(\pm 1)}(\eta_1, \eta_2, k)$ given that $\eta_1 < 0$ and $\eta_2 <0$ and therefore the bound is satisfied as long as $c_T \leq 1$, as before. In summary, we can add the $\beta_4$ term, $2\beta_4 a(\eta) \epsilon_{ijk} \sigma_i \partial_j \sigma_k$, to the Lagrangian \eqref{dSspacegeneralisedProca} without spoiling microcausality. 

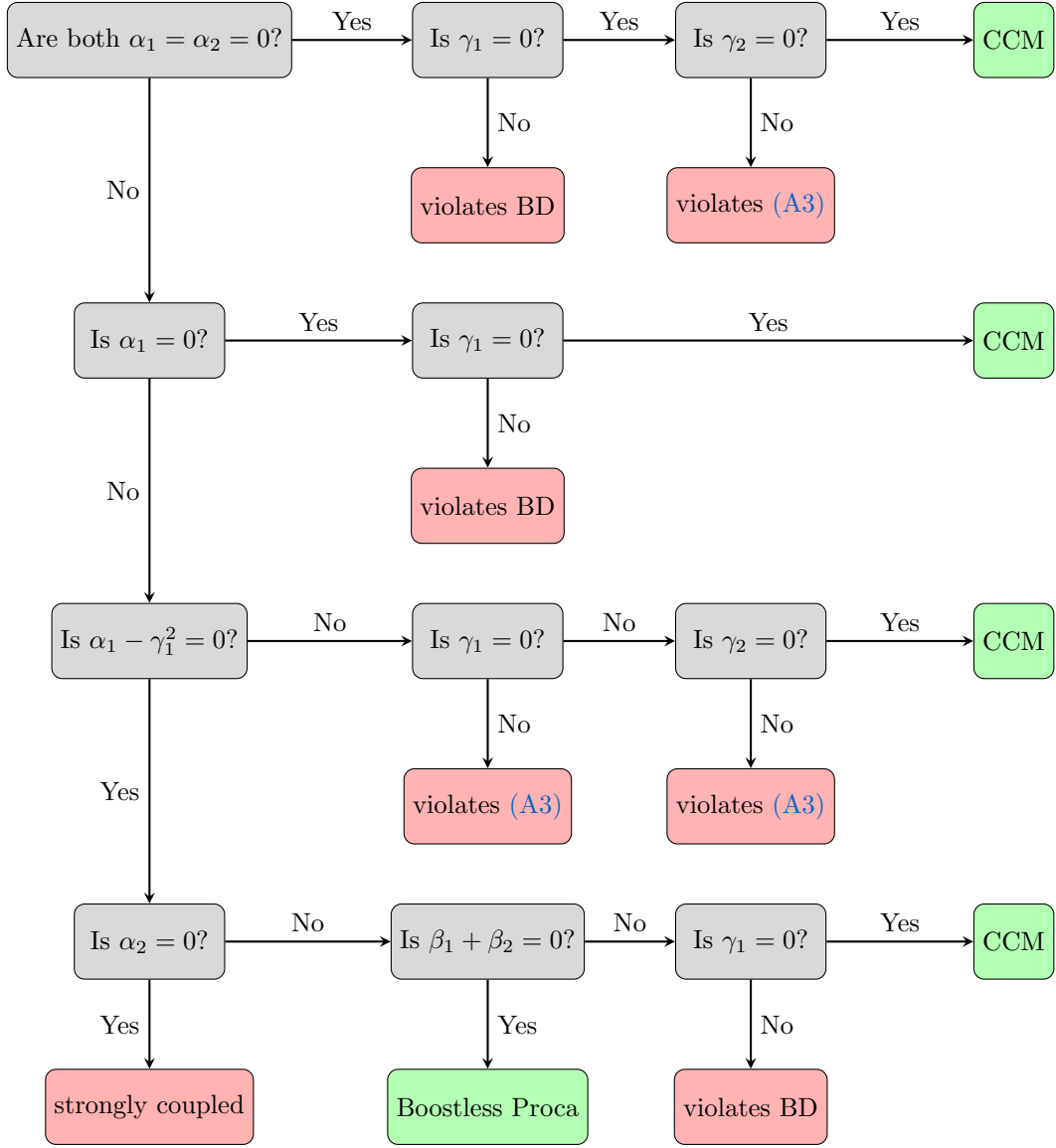
\begin{figure}
\centering
    \begin{tikzpicture}[node distance=2cm]
\node (constraint) [process] {Are both $\alpha_1 = \alpha_2 = 0$?};
\node (specialcases) [process, right of=constraint, xshift=2.5cm] {Is $\gamma_1 = 0$?};
\node (gamma1specialcase) [failure, below of=specialcases, yshift=-0.2cm]{violates BD};
\node (gamma2specialcase) [process, right of=specialcases, xshift=1.5cm] {Is $\gamma_2 = 0$?};
\node (CCM) [good, right of=gamma2specialcase, xshift=1.5cm] {CCM};
\node (weirdspecialcase) [failure, below of=gamma2specialcase, yshift=-0.2cm]{violates \ref{analyticitycondition3}};
\node(firstalpha1)[process, below of=constraint, yshift = -2cm]{Is $\alpha_1=0$?};
\node(firstgamma1)[process, right of=firstalpha1, xshift = +2.5cm]{Is $\gamma_1=0$?};
\node (gamma1violateBD) [failure, below of=firstgamma1, yshift=-0.2cm]{violates BD};
\node (secondCCM) [good, right of=firstgamma1, xshift=+5cm]{CCM};
\node (genericcase) [process, below of=firstalpha1, yshift=-2cm]{Is $\alpha_1-\gamma_1^2 = 0$?};
\node (gamma12) [process, right of=genericcase, xshift=+2.5cm]{Is $\gamma_1 = 0$?};
\node (gamma1gamma2) [process, right of=gamma12, xshift=1.5cm] {Is $\gamma_2= 0$?};
\node (CCM2) [good, right of=gamma1gamma2, xshift=+1.5cm]{CCM};
\node (alpha2violateanalyticity) [failure, below of=gamma12, yshift=-0.2cm]{violates \ref{analyticitycondition3}};
\node (gamma1gamma2violateanalyticity) [failure, below of=gamma1gamma2, yshift=-0.2cm]{violates \ref{analyticitycondition3}};
\node (alpha22) [process, below of=genericcase, yshift=-2cm]{Is $\alpha_2=0$?};
\node(stronglycoupled) [failure, below of=alpha22, yshift=-0.2cm]{strongly coupled};
\node(beta1beta2) [process, right of=alpha22, xshift=+2.5cm]{Is $\beta_1+\beta_2=0$?};
\node (gamma1)[process, right of=beta1beta2, xshift=+1.5cm]{Is $\gamma_1=0$?};
\node (finalCCM) [good,right of=gamma1, xshift=+1.5cm]{CCM};
\node (boostlessProca) [good, below of=beta1beta2,yshift=-0.2cm]{Boostless Proca};
\node (violatebound) [failure, below of=gamma1, yshift=-0.2cm]{violates BD};

\draw [arrow] (constraint) -- node[anchor=south] {Yes} (specialcases);
\draw [arrow] (specialcases) -- node[anchor=west]{No} (gamma1specialcase);
\draw [arrow] (specialcases) -- node[anchor=south]{Yes}(gamma2specialcase);
\draw [arrow] (gamma2specialcase) -- node[anchor=south]{Yes} (CCM);
\draw [arrow] (gamma2specialcase) -- node[anchor=west]{No} (weirdspecialcase);
\draw [arrow] (constraint) -- node[anchor=east]{No} (firstalpha1);
\draw [arrow](firstalpha1) -- node[anchor=south]{Yes}(firstgamma1);
\draw [arrow](firstgamma1) -- node[anchor=west]{No}(gamma1violateBD);
\draw [arrow](firstgamma1) -- node[anchor=south] {Yes}(secondCCM);
\draw[arrow] (firstalpha1) -- node[anchor=east]{No}(genericcase);
\draw [arrow] (genericcase) -- node[anchor=south]{No} (gamma12);
\draw [arrow] (gamma12) -- node[anchor=south]{No} (gamma1gamma2);
\draw [arrow] (gamma1gamma2) -- node[anchor=south]{Yes} (CCM2);
\draw [arrow] (gamma12) -- node[anchor=west]{No} (alpha2violateanalyticity);
\draw [arrow] (gamma1gamma2) -- node[anchor=west]{No} (gamma1gamma2violateanalyticity);
\draw [arrow] (genericcase) -- node[anchor=east]{Yes} (alpha22);
\draw[arrow] (alpha22) -- node[anchor=east]{Yes}(stronglycoupled);
\draw [arrow] (alpha22)-- node[anchor=south]{No}(beta1beta2);
\draw [arrow] (beta1beta2) -- node[anchor=south]{No}(gamma1);
\draw [arrow] (gamma1) -- node[anchor=south] {Yes} (finalCCM);
\draw [arrow](beta1beta2) -- node[anchor=west]{Yes}(boostlessProca);
\draw [arrow](gamma1) -- node[anchor=west]{No} (violatebound);
\end{tikzpicture}
\caption{Flowchart showing the constraints on \eqref{spin1generalfreewithparityodd} in de Sitter space.}
\label{FlowchartdS}
\end{figure}

We summarise our results in the flowchart in figure \ref{FlowchartdS}. The flowchart is valid for any choice of $\beta_4/\kappa$ so it does not appear.
\subsection{Finite regime of validity} \label{EFTdS}
As we did in the flat-space case, we now only impose the conditions of microcausality within a finite regime of validity defined by $|k\eta| < \Lambda / H$ i.e. where we impose a cut-off on the physical momentum $k_{\text{physical}} = k / a(\eta)$. 

Similar to the flat-space case, the $\alpha_1 = \alpha_2 = 0$ cases other than CCM are eliminated by analyticity when $k\to 0$, or by the Bunch-Davies condition. Therefore, working with a finite regime of validity does not alter the analysis for these cases. The CCM theory satisfies analyticity for arbitrary parameters so there is nothing to add when we work with a finite regime of validity. 
\paragraph{Generic cases} We can go through cases \ref{caseadS} to \ref{generalisedProcadS} and modify the discussion accordingly.
\begin{itemize}
    \item Case \ref{caseadS}: as above, this case is equivalent to CCM after the field redefinition \eqref{dSFieldRedefinition} and therefore working with a finite regime of validity does not alter this case.  
    \item Case \ref{casebdS}: as above, in this case we cannot impose the Bunch-Davies boundary condition on our solutions and this is not altered by working with a finite regime of validity. 
    \item Case \ref{casecdS}: the argument that ruled this case out was that the longitudinal mode becomes infinitely strongly coupled. This is an IR issue and cannot  be solved by working with a finite regime of validity. 
    \item Case \ref{weirdgenericcasedS}: in this case our first conditions arose from demanding
    \begin{align}
        G^{(0)} - \frac 12 (G^{(+1)} + G^{(-1)}) = -\frac {2i\gamma_1^2 H^2k^2\eta_c^2}{\alpha_2 + (\alpha_1 - \gamma_1^2) H^2 k^2 \eta_c^2} \epsilon + \mathcal{O}(\epsilon^2) \,,
    \end{align}
    to be analytic in $k^2$, and tend to 0 as $k\to 0$. If $\alpha_2 = 0$, then since $\alpha_1 -\gamma_1^2 \neq 0$ in this case, we require $\gamma_1 = 0$. With $\alpha_2 = \gamma_1 = 0$, we can reuse the same argument above, where analyticity as $k\to 0$ demands $\gamma_2 = 0$. This discussion is unaffected by working within a finite regime of validity, and hence we are left with $\gamma_1 = \gamma_2 = 0$, which takes us back to the CCM case. If instead $\alpha_2 \neq 0$ and $\gamma_1 = 0$, then from \eqref{Step2analyticity} we want the position of the poles (in $k\eta$) to be outside the regime of validity of the theory. This imposes
\begin{align}
    |\alpha_1|\Lambda^2 \ll \alpha_2 \,,
\end{align}
and if we want the cut-off of the theory to be greater than $\alpha_2$, this imposes $|\alpha_1| \ll 1$ meaning that we can treat the $\alpha_1$ term as a perturbative correction to the free theory. Such perturbative corrections are treated in the same way as interactions in Feynman diagrams. We therefore discard them in the free theory and without $\gamma_1$ and $\alpha_1$, we can perform the field redefinition \eqref{dSFieldRedefinition} to take us perturbatively close to the CCM case. If instead $\alpha_2 \neq 0$ and $\gamma_1 \neq 0$ we go back to \eqref{checkA3limitdSweirdcase} and demand that the poles are outside of the regime of validity which requires 
\begin{align}
    |\alpha_1 - \gamma_1^2|\Lambda^2 \ll \alpha_2 \,,
\end{align}
and again if we want the cut-off of the theory to be greater than $\alpha_2$, we require $|\alpha_1 - \gamma_1^2| \ll 1$. This takes us perturbatively close to case \ref{generalisedProcadS}.

\item Case \ref{generalisedProcadS}: the only condition we imposed on the parameters was $\beta_1 + \beta_2 = 0$ which came from imposing Bunch-Davies boundary conditions. We can instead take $\beta_1 + \beta_2  \ll 1$ such that the corresponding combination of operators is treated in the same way as interactions and therefore discarded in the free theory. We therefore arrive at the same conclusion as in flat-space where the $\tilde{\alpha}$ and $\tilde{\beta}$ operators in \eqref{FiniteRegimeTheory} can be added to \eqref{dSspacegeneralisedProca} as long as they are treated as perturbative corrections with their fate ultimately decided by the UV completion.    
\end{itemize}

\section{Phenomenology}\label{sec:phenomenology}
In this section we examine the signals that the two different classes of theories found in Section~\ref{sec:dS}, CCM and boostless Proca, could leave in the inflationary non-Gaussianities, namely $n$-point inflaton correlators\footnote{Our focus is on scalar correlators only but it would be interesting to also consider those involving the graviton.}
\begin{equation}
    \langle \pi({\mathbf{k}_1})\ldots\pi({\bfk_n}) \rangle = \delta^{(3)} \left( \bfk_1+\ldots +\bfk_n \right) B_n (\mathbf{k}_1,\ldots,\mathbf{k}_n) \, .
\end{equation} 
In Subsection~\ref{sec:bispectrum} we study the bispectrum $B_3$, and show that the signals from the two set-ups are degenerate at tree level, up to contact interactions, meaning there are no new distinct signatures from boostless Proca on this observable. Given that we have fully classified the possible ways a massive vector can propagate on an inflationary background, our results show that the bispectrum cosmological collider signals coming from exchanging a CCM field are the most general. In Subsection~\ref{sec:trispectrum} we focus on the trispectrum $B_4$---which, in line with the observations of~\cite{Cheung:2025dmc}, can distinguish between the two set-ups---and we compute the tree-level exchange contribution from the boostless Proca field.

\paragraph{Mode functions}
It will be convenient to first gather the mode functions found in Section~\ref{sec:dS} for the different massive spin-1 field theories. The CCM mode functions are, cf.~\eqref{CCMdSmodefunction},
\begin{equation}
    \sigma^{(h)}_{\rm CCM} = \frac {\sqrt{\pi}}2 e^{i\pi(\nu + 1/2)/2} (-\eta)^{1/2} H_{\nu}^{(1)}(-c_s^{(h)}k\eta) \, , \label{CCMdSmodefunction2}
\end{equation}
with $\nu$ and $c_s^{(h)}$ given in~\eqref{eq:paramsCCM}. The structure is the same for each helicity with the only difference in the speed of sound. The boostless Proca mode functions are (we use the notation of \eqref{dSspacegeneralisedProca} for the parameters of the theory)
\begin{align}
    \sigma_0 & = \frac {Hkc_L\sqrt{\pi}}{2m} e^{i\pi (\nu + 1/2)/2} (-\eta)^{3/2} H_{\nu}^{(1)}(-c_Lk\eta) \,, \label{eq:GProca_s0_def} \\
	\sigma_L & = \frac {\sqrt{\pi}}{4 c_L m} e^{i\pi (\nu+1/2)/2} (-\eta)^{1/2} \left [[-H(1-2\nu) + 2\beta] H_{\nu}^{(1)}(-c_Lk\eta) +2 H c_Lk\eta H_{\nu+1}^{(1)}(-c_Lk\eta)\right ]\,, \label{eq:GProca_sL_def} \\
	\sigma^{(\pm 1)} & = \frac {\sqrt{\pi}}2 e^{i\pi (\nu+1/2)/2} (-\eta)^{1/2} H_{\nu}^{(1)}(-c_T k\eta)\,, \label{eq:GProca_sT_def}
\end{align}
with
\begin{equation} \label{OrderandSpeed}
    \nu = \sqrt{\frac 14 - \frac {M^2}{H^2}}  \qquad\text{and}\qquad c_L = \sqrt{\frac {\beta^2-H \beta+M^2}{m^2}} \,.
\end{equation}
Here different helicities have different structures which is what qualitatively distinguishes this theory from CCM. It will also be useful to write the mode function of a massive scalar with generic sound speed $c$,
\begin{equation} \label{eq:scalar_mf}
	\sigma (ck,\eta) = \frac{H\sqrt{\pi}}{2} e^{i\pi (\nu + 1/2)/2} (-\eta)^{3/2} H^{(1)}_{\nu} (-ck\eta) \, ,
\end{equation}
and the corresponding scalar bulk-to-bulk propagators
\begin{equation} \label{eq:Gscalar}
\begin{aligned}
	G^{\rm sc}_{+-} (\eta,\eta',ck) & = \big( G^{\rm sc}_{-+} (\eta,\eta',ck) \big)^* = \big( \sigma (ck,\eta) \big)^* \sigma (ck,\eta') \, , \\[3pt]
	G^{\rm sc}_{++} (\eta,\eta',ck) & = \big( G^{\rm sc}_{--} (\eta,\eta',ck) \big)^* = \sigma (ck,\eta) \big( \sigma (ck,\eta') \big)^* \theta(\eta-\eta') + \big( \sigma (ck,\eta) \big)^* \sigma (ck,\eta') \, \theta(\eta'-\eta) \, .
\end{aligned}
\end{equation}
Note that in the scalar mode function we are taking the order of the Hankel function to be that of \eqref{OrderandSpeed}. This is different to the usual expression for a massive scalar field and just means we are restricting the mass to values $m_{\rm sc}^2 \geq 2H^2$ (assuming the kinetic term is $a^2(\eta) \phi'^2$). This is fine for our purposes where we will only use the scalar mode function as a seed to compute spin-$1$ exchange diagrams.  
\subsection{Degeneracy in bispectra} \label{sec:bispectrum}

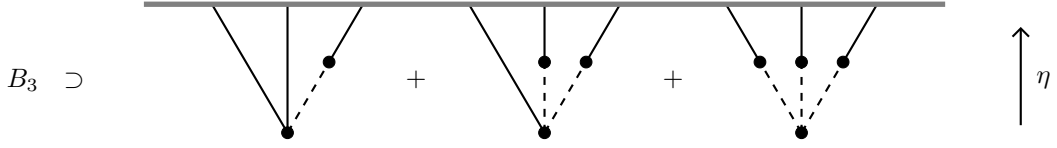
\begin{figure}[t] 
	\centering
	\begin{tikzpicture}
	
	\node[](B) at (-4.2,-1){$B_3\,\,\,\,\supset$};
	
	\node[](t) at (9,-1){$\eta$};
	\draw[-{Straight Barb[]},thick] (8.7,-1.6) -- (8.7,-0.3);
	
	\coordinate(b1) at (-1,-1.7);
	\coordinate(b2) at (2.4,-1.7);
	\coordinate(b3) at (5.8,-1.7);

    \node[](p1) at (0.7,-1){$+$};
    \node[](p2) at (4.1,-1){$+$};
	
	\filldraw[black] (b1) circle (2.2pt);
	\filldraw[black] ($(b1)+(0.55,0.935)$) circle (2.2pt);
	
	\draw[thick] (b1) -- +(-1,1.7);
	\draw[thick] (b1) -- +(0,1.7);
	\draw[dashed,thick] (b1) -- +(0.55,0.935);
	\draw[thick] ($(b1)+(0.55,0.935)$) -- ($(b1)+(1,1.7)$);

	\filldraw[black] (b2) circle (2.2pt);
	\filldraw[black] ($(b2)+(0.55,0.935)$) circle (2.2pt);
	\filldraw[black] ($(b2)+(0,0.935)$) circle (2.2pt);
	
	\draw[dashed,thick] (b2) -- +(0.55,0.935);
	\draw[thick] ($(b2)+(0.55,0.935)$) -- ($(b2)+(1,1.7)$);
	\draw[dashed,thick] (b2) -- +(0,0.935);
	\draw[thick] ($(b2)+(0,0.935)$) -- ($(b2)+(0,1.7)$);
	\draw[thick] (b2) -- +(-1,1.7);

	\filldraw[black] (b3) circle (2.2pt);
	\filldraw[black] ($(b3)+(0.55,0.935)$) circle (2.2pt);
	\filldraw[black] ($(b3)+(0,0.935)$) circle (2.2pt);
	\filldraw[black] ($(b3)+(-0.55,0.935)$) circle (2.2pt);
	
	\draw[dashed,thick] (b3) -- +(0,0.935);
	\draw[thick] ($(b3)+(0,0.935)$) -- ($(b3)+(0,1.7)$);
	\draw[dashed,thick] (b3) -- +(0.55,0.935);
	\draw[thick] ($(b3)+(0.55,0.935)$) -- ($(b3)+(1,1.7)$);
	\draw[dashed,thick] (b3) -- +(-0.55,0.935);
	\draw[thick] ($(b3)+(-0.55,0.935)$) -- ($(b3)+(-1,1.7)$);
	
	\draw[line width=2pt,gray] ($(b1)+(-1.9,1.7)$) -- ($(b3)+(1.9,1.7)$);
	
	\end{tikzpicture}
	\caption{Contributions to the bispectrum $B_3$ of the inflaton $\pi$ (solid line) from tree-level exchange of the massive spin-1 field $\sigma$ (dashed line). The gray line represents the late-time boundary $\eta=0$. Angular momentum conservation at the quadratic vertices implies that only $h=0$ modes of $\sigma$ propagate.}
    \label{fig:bispectrum}
\end{figure}

Let us first consider the contributions to the scalar bispectrum $B_3$ mediated by the tree-level exchange of a massive spin-1 field $\sigma$. As argued in~\cite{Cheung:2025dmc}, $\sigma$ always appears linearly mixed with $\pi$ in these contributions, so only its helicity-0 mode propagates (see Figure~\ref{fig:bispectrum}). The linear mixing and cubic vertices in Figure~\ref{fig:bispectrum} can contain copies of $\sigma_{0}$ and $\sigma_L$ so the relevant in-in propagators are (following the notation of~\cite{Lee:2016vti,Cheung:2025dmc})
\begin{align}
G^{\sigma}_{ab}(\eta, \eta', \bfk) &= (G^0_{0,1\to 0,1})_{ab,{\rm eff}}(\eta,\eta',k) \,, \\
	G^{\sigma}_{ab,i} (\eta,\eta',\bfk) & = \left( G^0_{1,1\to 0,1} \right)_{ab,{\rm eff}} (\eta,\eta',k) \, e^{(0)}_i (\bfk) \, , \\
	\left. G^{\sigma}_{ab,ij} (\eta,\eta',\bfk) \right|_{h=0} & = \left( G^0_{1,1\to 1,1} \right)_{ab,{\rm eff}} (\eta,\eta',k) \, e^{(0)}_i (\bfk) \, e^{(0)}_j (-\bfk) \, ,
\end{align}
where
\begin{equation}
\begin{aligned}
	\left( G^0_{0,1\to 0,1} \right)_{+-,{\rm eff}} (\eta,\eta',k) & = \Big( \left( G^0_{0,1\to 0,1} \right)_{-+,{\rm eff}} (\eta,\eta',k) \Big)^* = \big( \sigma_{0} (\eta,k) \big)^* \sigma_0 (\eta',k) \, , \\[3pt]
	\left( G^0_{0,1\to 0,1} \right)_{++,{\rm eff}} (\eta,\eta',k) & = \Big( \left( G^0_{0,1\to 0,1} \right)_{--,{\rm eff}} (\eta,\eta',k) \Big)^* \\
    & = \sigma_0 (\eta,k) \big( \sigma_0 (\eta',k) \big)^* \theta(\eta-\eta') + \big( \sigma_0 (\eta,k) \big)^* \sigma_0 (\eta',k) \theta(\eta'-\eta) \, ,
\end{aligned}
\end{equation}
and
\begin{equation}
\begin{aligned}
	\left( G^0_{1,1\to 0,1} \right)_{+-,{\rm eff}} (\eta,\eta',k) & = \Big( \left( G^0_{1,1\to 0,1} \right)_{-+,{\rm eff}} (\eta,\eta',k) \Big)^* = \big( \sigma_L (\eta,k) \big)^* \sigma_0 (\eta',k) \, , \\[3pt]
	\left( G^0_{1,1\to 0,1} \right)_{++,{\rm eff}} (\eta,\eta',k) & = \Big( \left( G^0_{1,1\to 0,1} \right)_{--,{\rm eff}} (\eta,\eta',k) \Big)^* \\
    & = \sigma_L (\eta,k) \big( \sigma_0 (\eta',k) \big)^* \theta(\eta-\eta') + \big( \sigma_L (\eta,k) \big)^* \sigma_0 (\eta',k) \theta(\eta'-\eta) \, ,
\end{aligned}
\end{equation}
and
\begin{equation} \label{eq:G0_11to11}
\begin{aligned}
	\left( G^0_{1,1\to 1,1} \right)_{+-,{\rm eff}} (\eta,\eta',k) & = \Big( \left( G^0_{1,1\to 1,1} \right)_{-+,{\rm eff}} (\eta,\eta',k) \Big)^* = \big( \sigma_L (\eta,k) \big)^* \sigma_L (\eta',k) \, , \\[3pt]
	\left( G^0_{1,1\to 1,1} \right)_{++,{\rm eff}} (\eta,\eta',k) & = \Big( \left( G^0_{1,1\to 1,1} \right)_{--,{\rm eff}} (\eta,\eta',k) \Big)^* \\ 
    & = \sigma_L (\eta,k) \big( \sigma_L (\eta',k) \big)^* \theta(\eta-\eta') + \big( \sigma_L (\eta,k) \big)^* \sigma_L (\eta',k) \, \theta(\eta'-\eta) \, .
\end{aligned}
\end{equation}
The superscript indicates the helicity of the exchanged modes which in this case is always $h=0$. The subscript $n,S \rightarrow m,S$ indicates a spin-$S$ propagator with modes coming from field components with $n$ and $m$ spatial indices. As shown in \cite{Cheung:2025dmc}, the time-ordered propagators $\left( G^0_{0,1\to 0,1} \right)_{\pm \pm,{\rm eff}} (\eta,\eta',k)$ also come with a delta-function term in addition to the time-ordered terms which arises from properly considering $\sigma_0$ as a constrained field that is integrated out. Since this delta function term is degenerate with contact diagrams, or more generally diagrams with fewer time-orderings, we will not keep track of it here. 

It can be seen from~\eqref{CCMdSmodefunction2}, \eqref{eq:GProca_s0_def}, and~\eqref{eq:GProca_sL_def} that the helicity-0 mode functions of the boostless Proca theory---$\sigma_0$ and $\sigma_L$ for the time and spatial components respectively---are related to the CCM longitudinal mode function $\sigma^{(0)}_{\rm CCM}$ by
\begin{align} 
	\sigma_0 (k,\eta ) & = \frac {c_Lk}{m \, a(\eta)} \, \sigma^{(0)}_{\rm CCM} (k,\eta)  \, , \label{eq:s0_rel_scalar} \\
	\sigma_L (k,\eta ) & = \frac{-1}{c_L \, m \, a(\eta)} \left[ \partial_{\eta} + \left( -1+\frac{\beta}{H} \right) \frac{1}{\eta} \right]  \sigma^{(0)}_{\rm CCM} (k,\eta) \, , \label{eq:sL_rel_scalar}
\end{align}
where the mode functions on both sides are evaluated for the same mass $M^2$ and sound speed $c_L$. The CCM in-in propagators for the $h=0$ mode are 
\begin{align}
     (G_{\rm CCM})_{ab,ij} (\eta,\eta',\bfk)|_{h=0} & = \left( G^0_{\rm CCM} \right)_{ab} (\eta,\eta',k) \, e^{(0)}_i (\bfk) \, e^{(0)}_j (-\bfk)\,,
\end{align}
where the CCM propagators $\left( G^0_{\rm CCM} \right)_{ab}$ are given by~\eqref{eq:Gscalar} with $\sigma \to \sigma_{\rm CCM}^{(0)}$. The $h=0$ bulk-to-bulk propagators in the boostless Proca and CCM theories are then related by
\begin{equation} \label{TemporalCorrespondence}
	\left( G^0_{0,1\to 0,1} \right)_{ab,{\rm eff}} (\eta,\eta',k) = \frac{c_L^2 k^2}{m^2 \, a(\eta) a(\eta')}  \left( G^0_{\rm CCM} \right)_{ab} (\eta,\eta',k) \,,
\end{equation}
and
\begin{equation} \label{MixedCorrespondence}
	\left( G^0_{1,1\to 0,1} \right)_{ab,{\rm eff}} (\eta,\eta',k) = - \frac{k}{m^2 \, a(\eta) a(\eta')} \left[ \partial_{\eta} + \left( -1+\frac{\beta}{H} \right) \frac{1}{\eta} \right] \left( G^0_{\rm CCM} \right)_{ab} (\eta,\eta',k) \, ,
\end{equation}
and
\begin{align}
	\left( G^0_{1,1\to 1,1} \right)_{\pm\pm,{\rm eff}} (\eta,\eta',k) = \,\, & \frac{1}{c_L^2m^2 a(\eta) a(\eta')} \left[ \partial_{\eta} + \left( -1+\frac{\beta}{H} \right) \frac{1}{\eta} \right] \left[ \partial_{\eta'} + \left( -1+\frac{\beta}{H} \right) \frac{1}{\eta'} \right] \left( G^0_{\rm CCM} \right)_{\pm\pm} (\eta,\eta',k) \nonumber \\
	& \mp \frac{i}{c_L^2 m^2 a^2(\eta) a^2(\eta')} \, \delta (\eta-\eta') \, , \label{eq:gP_CCM_rel_11} \\
	\left( G^0_{1,1\to 1,1} \right)_{\pm\mp,{\rm eff}} (\eta,\eta',k) = \,\, & \frac{1}{c_L^2m^2 a(\eta) a(\eta')} \left[ \partial_{\eta} + \left( -1+\frac{\beta}{H} \right) \frac{1}{\eta} \right] \left[ \partial_{\eta'} + \left( -1+\frac{\beta}{H} \right) \frac{1}{\eta'} \right] \left( G^0_{\rm CCM} \right)_{\pm\mp} (\eta,\eta',k) \,. \nonumber
\end{align}
Note that delta functions only arise in the correspondence between the time-ordered ($++$ and $--$) propagators and when there are at least two time derivatives acting on the CCM propagator \cite{Cheung:2025dmc}.

These relations were already found in~\cite{Cheung:2025dmc} for the Proca theory i.e. for the subcase $M^2=m^2$, $c_T=1$, and $\beta=0$ (and hence $c_L=1$).\footnote{Notice the different normalisation of the CCM field in~\cite{Cheung:2025dmc}, namely
\begin{equation}
    \sigma_i^{\rm (here)} (\eta,\bfk) = a(\eta) \, \sigma_i^{\rm (there)} (\eta,\bfk) \, .
\end{equation}} There it was argued that one can always use these relations in the bulk integrals and integrate by parts the time derivatives to rewrite any Proca exchange diagram as a linear combination of: 
\begin{enumerate}
	\item CCM exchanges coming from different interactions between the inflaton and the exchanged field. Different interactions arise when the time derivatives hit the $\pi$ propagators. As shown in \cite{Cheung:2025dmc}, if the interactions on one side are constructed within the EFToI, they remain so on the other side. The factor of $k^2$ in \eqref{TemporalCorrespondence} and factor of $k$ in \eqref{MixedCorrespondence} ensures that any inverse powers of $k$ coming from the polarisation vectors are canceled so that local interactions on the Proca side correspond to local interactions on the CCM side. 
	\item Contact terms from self-interactions of the inflaton as a consequence of the delta function in~\eqref{eq:gP_CCM_rel_11}.
\end{enumerate}

It is easy to see that this argument goes through for the boostless Proca theory with generic 
parameters $\{ M^2,m^2,c_T,\beta\}$, with only two changes with respect to \cite{Cheung:2025dmc}. The first one is that there is an extra dependence of the putative CCM and contact couplings on $M^2$ (via $c_L^2m^2$) and $\beta$, which does not affect the conclusions in any way. The second is that in~\cite{Cheung:2025dmc} the correspondence was made with the CCM field restricted to have $c_s^{(0)}=1$, since the Proca theory field is luminal. Now the boostless Proca field has a generic sound speed $0<c_L\leq 1$, just like the CCM field, and so the degeneracy is worse because it holds for any choice of parameters of the theory.

We therefore conclude that any boostless Proca exchange contribution to the scalar bispectrum at tree level is degenerate with a combination of CCM exchanges and contact contributions. This observable does therefore not allow us to distinguish between the boostless Proca and CCM set-ups. We note that if we invert the above relations to express the CCM propagators in terms of the boostless Proca ones, the relations would be non-local as we see in Appendix \ref{appBswitchingpicture}. If we wish to retain local interactions we should therefore take the CCM theory as being the general one with the CCM bispectra capturing all of the boostless Proca ones, as illustrated in Figure \ref{Venndiagramhierarchy}. 

\subsection{New observables in trispectra} \label{sec:trispectrum}

\begin{figure}[t] 
	\centering
	\begin{tikzpicture}
	
	\node[](B) at (-2.7,-1){$B_4\,\,\,\,\supset$};
	
	\node[](t) at (9.7,-1){$\eta$};
	\draw[-{Straight Barb[]},thick] (9.4,-1.6) -- (9.4,-0.3);
	
	\node[](p) at (3.55,-1){$+$};
	
	\coordinate(tr12) at (0.25,-1.7);
	\coordinate(tr22) at (2.3,-1.7);
	
	\coordinate(tr1) at (4.8,-1.7);
	\coordinate(tr2) at (7,-1.7);
	
	\filldraw[black] (tr12) circle (2.2pt);
	\filldraw[black] (tr22) circle (2.2pt);
		
	\draw[thick] (tr12) -- +(-0.75,1.7);
	\draw[thick] (tr12) -- +(0.75,1.7);
	\draw[thick] (tr12) -- +(0,1.7);
	\draw[thick] (tr22) -- +(0.5,1.7);
	\draw[dashed,thick] (tr12) -- (tr22);
	
	\filldraw[black] (tr1) circle (2.2pt);
	\filldraw[black] (tr2) circle (2.2pt);
		
	\draw[thick] (tr1) -- +(-0.5,1.7);
	\draw[thick] (tr1) -- +(0.5,1.7);
	\draw[thick] (tr2) -- +(-0.5,1.7);
	\draw[thick] (tr2) -- +(0.5,1.7);
	\draw[dashed,thick] (tr1) -- (tr2);
	
	\draw[line width=2pt,gray] ($(tr12)+(-1.65,1.7)$) -- ($(tr2)+(1.4,1.7)$);
	
	\end{tikzpicture}
	\caption{Tree-level contributions to the trispectrum $B_4$ of the inflaton $\pi$ (solid line) from single exchange of a massive spin-1 field $\sigma$ (dashed line). The gray line represents the late-time boundary $\eta=0$. Angular momentum conservation at the vertices implies that only the longitudinal mode of $\sigma$ propagates in the left diagram, while both longitudinal and transverse modes propagate in the right diagram.}
    \label{fig:trispectrum}
\end{figure}
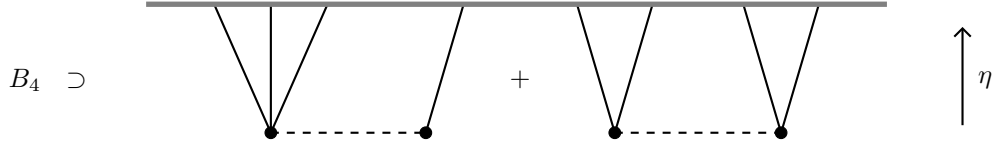

The degeneracy of the scalar bispectrum discussed in the previous subsection stems from the fact that only the helicity-0 mode of the spinning field contributes at tree level. This is no longer the case for the scalar trispectrum. The boost-breaking terms in the free theory \eqref{dSspacegeneralisedProca} necessitate certain interactions that are quadratic in $\sigma$ and contain various powers of $\pi$. They are forced upon us by the non-linearly realised symmetries of inflation. There can then also be interactions that are linear in $\sigma$. In this subsection our primary goal is to illustrate that the boostless Proca theory can be distinguished from the CCM theory and to do so we consider such interactions linear in $\sigma$. The two leading contributions involving massive spin-1 exchange correspond to the diagrams shown in Figure~\ref{fig:trispectrum}, of which the right-hand one involves all of the $h=0,\pm 1$ modes. In the following, we compute this diagram when the exchanged field is a boostless Proca field, and discuss how it can be used to distinguish this theory from CCM and from Proca.

For concreteness we will consider the parity-even cubic interaction\footnote{This interaction arises from the EFT of inflation operator $\delta g^{00} \sigma^0$, which after performing the Stückelberg trick becomes $(-2\dot{\pi}+\ldots)(\delta_{\mu}^0+\partial_{\mu}\pi) \sigma^{\mu} \supset -2a^{-2}\dot{\pi} \partial_i\pi\sigma_i$. Notice that we have used cosmic instead of conformal time coordinates here. \label{fn:EFToI_int}}
\begin{equation} \label{eq:intsEFToI}
	S_{\rm int} = \lambda \int d^3x \, d\eta \, a(\eta) \, \pi' \partial_i\pi \, \sigma_i \, ,
\end{equation}
with the linear dependence on the scale factor fixed by scale invariance. Setting $\lambda=1$ for simplicity, the right-hand diagram of Figure~\ref{fig:trispectrum} is given by 
\begin{equation} \label{eq:trisp_initial}
\begin{aligned}
	B_4
	= & \,\, i^2 (i k_2^i) (i k_4^j) \int_{-\infty}^0 d\eta \, a(\eta) \int_{-\infty}^0 d\eta' \, a(\eta') \\
	& \times \sum_{a,b=\pm 1} a b \, \partial_{\eta} G^{\pi}_a (\eta,k_1) \, G^{\pi}_a (\eta,k_2) \, G^{\sigma}_{ab,ij} (\eta,\eta', \mathbf{s}) \, \partial_{\eta'} G^{\pi}_b (\eta',k_3) \, G^{\pi}_b (\eta',k_4) + \text{perms.} \, ,
\end{aligned}
\end{equation}
with $\mathbf{s}\equiv \bfk_1+\bfk_2$. We refer the reader to \cite{Chen:2017ryl} if they are unfamiliar with the Feynman rules that compute exchange trispectra. The bulk-to-boundary propagators of $\pi$ are
\begin{equation}
    G^{\pi}_{\pm} (\eta,k) = \frac{H^2}{2c_{\pi}^3k^3} \left( 1\mp ic_{\pi} k\eta \right) e^{\pm ic_{\pi} k\eta} \, ,
\end{equation}
while the bulk-to-bulk propagators of the spin-1 field are (again following the notation of~\cite{Lee:2016vti,Cheung:2025dmc})
\begin{equation} \label{eq:btb_props}
	G^{\sigma}_{ab,ij} (\eta,\eta',\bfk) = \sum_{h=-1}^1 \left( G^h_{1,1\to 1,1} \right)_{ab,{\rm eff}} (\eta,\eta',k) \, e^{(h)}_i (\bfk) \, e^{(h)}_j (-\bfk) \, ,
\end{equation}
with $\left( G^0_{1,1\to 1,1} \right)_{ab,{\rm eff}}$ given by Eq.~\eqref{eq:G0_11to11} and 
\begin{equation}
\begin{aligned}
	\left( G^{\pm 1}_{1,1\to 1,1} \right)_{+-,{\rm eff}} & \, (\eta,\eta',k) 
	= \Big( \left( G^{\pm 1}_{1,1\to 1,1} \right)_{-+,{\rm eff}} (\eta,\eta',k) \Big)^* = \big( \sigma^{(\pm 1)} (\eta,k) \big)^* \sigma^{(\pm 1)} (\eta',k) \, , \label{eq:G1+-} \\[3pt]
	\left( G^{\pm 1}_{1,1\to 1,1} \right)_{++,{\rm eff}} & \, (\eta,\eta',k)
	= \Big( \left( G^{\pm 1}_{1,1\to 1,1} \right)_{--,{\rm eff}} (\eta,\eta',k) \Big)^* \\
    & = \sigma^{(\pm 1)} (\eta,k) \big( \sigma^{(\pm 1)} (\eta',k) \big)^* \theta(\eta-\eta') + \big( \sigma^{(\pm 1)} (\eta,k) \big)^* \sigma^{(\pm 1)} (\eta',k) \, \theta(\eta'-\eta) \, .
\end{aligned}
\end{equation}

\paragraph{Propagator relations} Our strategy to compute the trispectrum~\eqref{eq:trisp_initial} is to reduce it to a set of differential operators in the kinematic variables acting on a simpler, well-known scalar ``seed.'' The first step to achieve this is to relate the bulk-to-bulk propagators~\eqref{eq:btb_props} to the scalar ones defined in~\eqref{eq:Gscalar}. We start with the helicity-0 propagators~\eqref{eq:G0_11to11}; from~\eqref{eq:GProca_sL_def} and~\eqref{eq:scalar_mf} one can see that the longitudinal and scalar mode functions are related by\footnote{The operator $U_{\eta}$ is almost identical to the operator relating $\sigma_L$ and $\sigma_{\rm CCM}^{(0)}$ in~\eqref{eq:sL_rel_scalar}. The only difference comes from the relative normalisation of the scalar and CCM mode functions, $\sigma_{\rm CCM}^{(h)}(k,\eta)=a(\eta)\,\sigma (k,\eta)$, which results in a different numerical factor for the $1/\eta$ term.}
\begin{equation}
	\sigma_L = -U_{\eta} \, \sigma (c_Lk,\eta ) \qquad\text{with}\qquad U_{\eta} \equiv \frac{1}{c_Lm} \left[ \partial_{\eta} + \left( -2+\frac{\beta}{H} \right) \frac{1}{\eta} \right] \, ,
\end{equation} 
which implies the relation
\begin{equation}
	\left( G^0_{1,1\to 1,1} \right)_{ab,{\rm eff}} (\eta,\eta',k) = U_{\eta} \, U_{\eta'} \, G^{\rm sc}_{ab} (\eta,\eta',c_Lk) + \ldots \delta (\eta-\eta') \, .
\end{equation}
The terms with $\delta (\eta-\eta')$ only appear for $ab=\pm\pm$ and produce contact interactions that are degenerate with contributions from inflaton self-interactions, so they will be ignored. From~\eqref{eq:GProca_sT_def} and~\eqref{eq:scalar_mf} we see that $\sigma^{(\pm 1)} = a(\eta) \, \sigma (c_Tk,\eta )$, so for the helicity-1 propagators~\eqref{eq:G1+-} we similarly have
\begin{equation} \label{eq:G1Gsc_rel}
	\left( G^{\pm 1}_{1,1\to 1,1} \right)_{ab,{\rm eff}} (\eta,\eta',k) = a(\eta) \, a(\eta') \, G^{\rm sc}_{ab} (\eta,\eta',c_Tk) \, .
\end{equation}
We can then write the full bulk-to-bulk propagators in~\eqref{eq:btb_props} as
\begin{equation} \label{eq:reduced_btb_prop}
\begin{aligned}
	G^{\sigma}_{ab,ij} (\eta,\eta',\bfk) = & \,\, \Pi_{ij}^{(0)} (\hat{\bfk}) \, \Big[ U_{\eta} \, U_{\eta'} \, G^{\rm sc}_{ab} (\eta,\eta',c_Lk) + \ldots \delta (\eta-\eta') \Big] \\
	& \,\, + \Pi_{ij}^{(1)} (\hat{\bfk}) \, a(\eta) \, a(\eta') \, G^{\rm sc}_{ab} (\eta,\eta',c_Tk) \, ,
\end{aligned}
\end{equation}
where we have introduced the compact notation for the polarisation sums:
\begin{equation}
	\Pi_{ij}^{(0)} (\hat{\bfk}) \equiv e^{(0)}_i (\bfk) \, e^{(0)}_j (-\bfk) \, , \qquad\qquad
	\Pi_{ij}^{(1)} (\hat{\bfk}) \equiv \sum_{h=\pm 1} e^{(h)}_i (\bfk) \, e^{(h)}_j (-\bfk) \, .
\end{equation}
The second step is to relate the massless scalar bulk-to-boundary propagators $G^{\pi}_{\pm}$ to those of a conformally-coupled scalar (which are just plane waves). To do this we use the relation
\begin{equation} \label{eq:ws_rel}
	\partial_{\eta} G^{\pi}_a (\eta,k_A) \, G^{\pi}_a (\eta,k_B) = \frac{H^4}{4c_{\pi}^6k_A^3k_B^3} (a ic_{\pi} k_A) \, \mathcal{W}_{AB} \, e^{a i c_{\pi} k_{AB}\eta} \, ,
\end{equation}
where $a\in \{+,-\}$, $A,B\in \{1,2,3,4\}$, and we have defined the weight-shifting operator
\begin{equation} \label{eq:WSop}
	\mathcal{W}_{AB} \equiv -k_A\partial_{k_A}(1-k_B\partial_{k_B}) \, .
\end{equation}
For clarity of presentation, we split the trispectrum~\eqref{eq:trisp_initial} into the contributions from $h=0$ and $h=\pm 1$ exchanges, $B_4 = B_4^{(0)} + B_4^{(1)}$. Then, using the relations~\eqref{eq:reduced_btb_prop} and~\eqref{eq:ws_rel} we can express them as
\begin{equation}
\begin{aligned} \label{eq:tr_h0_aux1}
	B_4^{(0)} = & \, - k_2^i \, k_4^j \, \Pi_{ij}^{(0)} (\hat{\mathbf{s}}) \, \frac{H^8(c_{\pi} k_1)(c_{\pi} k_3)}{16 \, c_{\pi}^{12} (k_1k_2k_3k_4)^3} \, \mathcal{W}_{12} \, \mathcal{W}_{34} \\
	\times \int_{-\infty}^0 d\eta \, a(\eta) \int_{-\infty}^0 d\eta' \, a(\eta') & \sum_{a,b=\pm 1} e^{a i c_{\pi} k_{12}\eta} \, U_{\eta} \, U_{\eta'} \Big( G^{\rm sc}_{ab} (\eta,\eta', c_L s) \Big) e^{b i c_{\pi} k_{34}\eta'} + \text{perms.} + \text{contact} \, ,
\end{aligned}
\end{equation}
where the unimportant contact terms come from the delta function in~\eqref{eq:reduced_btb_prop} and we have defined $s\equiv |\mathbf{s}|$, and
\begin{equation}
\begin{aligned} \label{eq:tr_h1_aux1}
	B_4^{(1)} = & \, - k_2^i \, k_4^j \, \Pi_{ij}^{(1)} (\hat{\mathbf{s}}) \, \frac{H^8(c_{\pi} k_1)(c_{\pi} k_3)}{16 \, c_{\pi}^{12} (k_1k_2k_3k_4)^3} \, \mathcal{W}_{12} \, \mathcal{W}_{34} \\
	\times \int_{-\infty}^0 d\eta \, a(\eta)^2 & \int_{-\infty}^0 d\eta' \, a(\eta')^2 \sum_{a,b=\pm 1} e^{a i c_{\pi} k_{12}\eta} \, G^{\rm sc}_{ab} (\eta,\eta', c_T s) \, e^{b i c_{\pi} k_{34}\eta'} + \text{perms.} \, .
\end{aligned}
\end{equation}

\paragraph{Scalar seed} As discussed earlier, we want to write the trispectrum as a series of differential operators acting on the scalar seed correlator
\begin{equation} \label{eq:Fseed_def}
	\tilde{F} \equiv \sum_{a,b=\pm 1} ab F_{ab} = F_{++} + F_{--} - F_{+-} - F_{-+} \, ,
\end{equation}
where the $F_{ab}$ are the in-in contributions defined in~\cite{CosmoBootstrap1,Jazayeri:2022kjy},\footnote{Our seed~\eqref{eq:Fseed_def} differs from the one used in~\cite{CosmoBootstrap1,Jazayeri:2022kjy}, $F= F_{++} + F_{--} + F_{+-} + F_{-+}$. The reason is that the interaction~\eqref{eq:intsEFToI} we are using contains one time derivative which pulls down a sign $a=\pm$ from the propagator $G_a^{\pi}$ (see~\eqref{eq:ws_rel}). Consequently, the $F_{\pm\mp}$ contributions pick up a relative minus sign with respect to the $F_{\pm\pm}$ ones. In contrast, Refs.~\cite{CosmoBootstrap1,Jazayeri:2022kjy} work with a scalar $\sigma$ and use the interactions $\pi^2\sigma$, $(\partial_{\mu}\pi)^2\sigma$, $(\pi')^2\sigma$ and $(\partial_i\pi)^2\sigma$, which pull down an even number of $a=\pm$ from the propagators and hence do not yield a relative sign. See Appendix~\ref{app:Seed} for more details on the relation between $F$ and $\tilde{F}$.}
\begin{equation}
	F_{ab} (k_{12},k_{34},s) \equiv \frac{(a i)(b i)}{2H^2} \int_{-\infty}^0 \frac{d\eta}{\eta^2} \int_{-\infty}^0 \frac{d\eta'}{\eta'^2} \, e^{a i k_{12}\eta} \, G^{\rm sc}_{ab} (\eta,\eta',s) \, e^{b i k_{34}\eta'} \qquad \text{for } a,b\in \pm 1 \, .
\end{equation}
Appendix~\ref{app:Seed} contains more details about these seeds and the derivation of the explicit solution for $\tilde{F}$.

For the $h=0$ exchange term~\eqref{eq:tr_h0_aux1}, we start by integrating the differential operators $U_{\eta}$ by parts and then trading derivatives of $\eta$ by derivatives of $k_{12}$. Focusing only on the $\eta$-dependent part of the time integrals, and up to a vanishing boundary term, this gives
\begin{align}
	& \int_{-\infty}^0 \frac{d\eta}{\eta} \, e^{a i c_{\pi} k_{12}\eta} \, U_{\eta} \Big(G^{\rm sc}_{ab} (\eta,\eta', c_L s) \Big) = \frac{1}{c_Lm} \bigg( \frac{\beta}{H} -1 - k_{12} \partial_{k_{12}} \bigg) \int_{-\infty}^0 \frac{d\eta}{\eta^2}  \, e^{a i c_{\pi} k_{12}\eta} \, G^{\rm sc}_{ab} (\eta,\eta', c_L s) \, ,
\end{align}
and there is a similar relation for the $\eta'$ part. Using this and the definition~\eqref{eq:Fseed_def} we get
\begin{equation}
	B_4^{(0)} 
	= \,\, k_2^i \, k_4^j \, \Pi_{ij}^{(0)} (\hat{\mathbf{s}}) \, \frac{H^6(c_{\pi} k_1)(c_{\pi} k_3)}{8 \, c_{\pi}^{12} (k_1k_2k_3k_4)^3} \, \mathcal{W}_{12} \, \mathcal{W}_{34} \, \mathcal{D} \, \tilde{F} (c_{\pi} k_{12},c_{\pi} k_{34},c_L s) + \text{perms.} + \text{contact} \, ,
\end{equation}
where we have defined the differential operator
\begin{equation} \label{eq:Dop}
	\mathcal{D} \equiv \frac{H^2}{c_L^2m^2} \bigg( \frac{\beta}{H} - 1 - k_{12}\partial_{k_{12}} \bigg) \bigg( \frac{\beta}{H} - 1 - k_{34}\partial_{k_{34}} \bigg)  \, .
\end{equation}
For the helicity $h=\pm 1$ exchange term~\eqref{eq:tr_h1_aux1} we can use~\eqref{eq:Fseed_def} straightforwardly, getting
\begin{equation}
	B_4^{(1)} = k_2^i \, k_4^j \, \Pi_{ij}^{(1)} (\hat{\mathbf{s}}) \, \frac{H^6(c_{\pi} k_1)(c_{\pi} k_3)}{8 \, c_{\pi}^{12} (k_1k_2k_3k_4)^3} \, \mathcal{W}_{12} \, \mathcal{W}_{34} \, \tilde{F} (c_{\pi} k_{12},c_{\pi} k_{34},c_T s) + \text{perms.}\,. 
\end{equation}
Adding up the two contributions and restoring the coupling $\lambda$ we have the full analytic expression for the trispectrum,
\begin{keyeqn}
\begin{equation} \label{eq:trispectrum}
\begin{split}
	B_4 = & \,\, \frac{\lambda^2 H^6(c_{\pi} k_1)(c_{\pi} k_3)}{8 \, c_{\pi}^{12} (k_1k_2k_3k_4)^3} \, k_2^i \, k_4^j \Bigg[ \Pi_{ij}^{(0)} (\hat{\mathbf{s}}) \, \mathcal{W}_{12} \, \mathcal{W}_{34} \, \mathcal{D} \, \tilde{F} (c_{\pi} k_{12},c_{\pi} k_{34},c_L s) \\
	& \,\, + \Pi_{ij}^{(1)} (\hat{\mathbf{s}}) \, \mathcal{W}_{12} \, \mathcal{W}_{34} \, \tilde{F} (c_{\pi} k_{12},c_{\pi} k_{34},c_T s) \Bigg] + \text{perms.} + \text{contact} \, ,
\end{split}
\end{equation}
\end{keyeqn}
where $\tilde{F}$ is given by~\eqref{eq:sol_Ftilde} (when at least one of the variables $u$ and $v$ defined in~\eqref{eq:uv_defs} is $<1$) and~\eqref{eq:sol_Ftilde_uvlarge} (when both $u,v>1$).

Had we computed this trispectrum for the CCM theory, we would have obtained the same result without the $\mathcal{D}$ operator in the helicity-0 contribution.\footnote{This statement (and the rest of the discussion) holds whenever we compare CCM and boostless Proca trispectra coming from identical vertices. Notice however that these two theories couple differently to the inflation and, in fact, the EFT operator described in Footnote~\ref{fn:EFToI_int} does not exist for a CCM field. This provides another potential way of distinguishing the theories observationally, but here we assume identical interactions and focus on the mode function difference.} The reason is that the helicity-0 and helicity-1 mode functions~\eqref{CCMdSmodefunction2} of CCM have the same functional form and so they are both related to the scalar mode function~\eqref{eq:scalar_mf} by just a factor of $a(\eta)$; this, in turn, implies that the relation between CCM and scalar propagators has the functional form~\eqref{eq:G1Gsc_rel} for both helicities, so the differential operator $U_{\eta}$ and its kinematic version $\mathcal{D}$ never show up in the helicity-0 contribution. This confirms the expectation that the trispectra from CCM and boostless Proca are distinguishable, since the relative values of their helicity-0 and helicity-1 contributions are different. We illustrate this fact in the right-hand side of Figure~\ref{Venndiagramhierarchy}. While possible in principle, distinguishing the two spin-1 theories might be very hard in practice since the trispectrum signal is highly degenerate with contributions from other particles or from inflaton self-interactions. There is a specific kinematic regime---the collapsed limit---in which the signal is clean, but we will see in the next pages that the spin-1 exchange trispectrum has a relative suppression in that limit.

\paragraph{Action of differential operators} Before proceeding, let us briefly look at the action of the differential operators $\mathcal{D}$ and $\mathcal{W}_{AB}$ on the seed $\tilde{F}$. Notice that 
\begin{equation} \label{eq:uv_defs}
	\tilde{F} (c_{\pi} k_{12},c_{\pi} k_{34},c s) = (c s)^{-1} \, \hat{\tilde{F}} (u,v)\qquad\text{with} \quad u= \frac{c s}{c_{\pi} k_{12}} \, , \quad v= \frac{c s}{c_{\pi} k_{34}} \, .
\end{equation}
The $(cs)^{-1}$ prefactor can be pulled out and the operators act on a function of $u$ and $v$ only. When acting on such a function, the operator $\mathcal{D}$ in~\eqref{eq:Dop} can be written as
\begin{equation} \label{eq:DF_uv}
	\mathcal{D} \quad\rightarrow\quad \frac{H^2}{c_L^2m^2} \bigg( \frac{\beta}{H} - 1 + u\partial_u \bigg) \bigg( \frac{\beta}{H} - 1 + v\partial_v \bigg) \, .
\end{equation}
Similarly, the action of $\mathcal{W}_{12} \, \mathcal{W}_{34}$ on a generic function of $u$ and $v$ is
\begin{equation} \label{eq:WWf_uv}
	\mathcal{W}_{12} \, \mathcal{W}_{34} \quad\rightarrow\quad \left[ \frac{k_1}{k_{12}} \left( 1 + \frac{2k_2}{k_{12}} \right) u\partial_u + \frac{k_1k_2}{k_{12}^2} u^2\partial_u^2 \right] \left[ \frac{k_3}{k_{34}} \left( 1 + \frac{2k_4}{k_{34}} \right) v\partial_v + \frac{k_3k_4}{k_{34}^2} v^2\partial_v^2 \right] \, .
\end{equation}

\paragraph{Collapsed limit} We are interested in studying the trispectrum in the collapsed limit $s\to 0$, in which the correlator displays a characteristic scaling $s^{\nu}$ associated to the propagation of the exchanged field~\cite{Arkani-Hamed:2015bza}. For fields with $M^2>H^2/4$ we have
\begin{equation}
    \nu = i\mu \quad\text{with } \mu\in\mathbb{R} \, ,
\end{equation}
and the correlator oscillates in the collapsed limit. For simplicity, we will focus on the part of the solution that contains these characteristic oscillations, which constitute a clean probe of massive exchange; other terms, which only affect the background signal, will not be studied in detail although they are contained in the full solution \eqref{eq:trispectrum}. 

As one can see in~\eqref{eq:sol_Ftilde} and~\eqref{eq:sol_Ftilde_uvlarge}, the solution for the seed consists of a series part analytic in $u$ and $v$, and a non-analytic part. The differential operators~\eqref{eq:DF_uv} and~\eqref{eq:WWf_uv} only depend on $u$ via $u\partial_u$ and $u^2\partial_u^2$, and similarly for $v$, so it follows that their action on the series part of the solution will not produce any oscillations $s^{i\mu}$ in the $s\to 0$ limit, regardless of the channel we look at. We can therefore ignore the series part of the solution.

Focusing on the non-analytic part of the seed~\eqref{eq:sol_Ftilde}, notice that its building blocks are the functions $\hat{f}_{\pm}$. It is easy to see that the $t$- and $u$-channel versions of these functions do not give rise to the desired $s^{i\mu}$ behaviour in the $s\to 0$ limit. The reason is that the $t$- and $u$-channel energy ratios,
\begin{equation}
    u_t = \frac{c|\bfk_1+\bfk_4|}{c_{\pi}k_{14}} \, , \qquad v_t = \frac{c|\bfk_1+\bfk_4|}{c_{\pi}k_{23}} \, , \qquad u_u = \frac{c|\bfk_1+\bfk_3|}{c_{\pi}k_{13}} \, , \qquad v_u = \frac{c|\bfk_1+\bfk_3|}{c_{\pi}k_{24}} \, ,
\end{equation}
do not vanish as $s\to 0$ (except for very specific, degenerate choices of the momenta). When $c_{\pi}<1$ we can have $u_t,v_t>1$ and/or $u_u,v_u>1$, in which case we need to use the seed~\eqref{eq:sol_Ftilde_uvlarge} instead. The conclusion is nevertheless unchanged, because the basis functions $\hat{h}_{\pm}$ are linear combinations of $\hat{f}_{\pm}$ and the new logarithmic dependence doesn't give oscillations as $s\to 0$ either. We conclude that we can ignore the $t$- and $u$-channel permutations of the trispectrum for our current interests. We therefore only consider the $s$-channel contributions coming from the non-analytic part of the seed, which we will denote by $B_{4,{\rm osc}}$. Since $u,v\to 0$ in the collapsed limit, we use the solution~\eqref{eq:sol_Ftilde}. For $s\to 0$ we get 
\begin{equation} \label{eq:collapsed_lim}
\begin{aligned}
	\lim_{s\to 0} B_{4,{\rm osc}} = & - {\rm Re} \Bigg\{ \frac{\lambda^2 H^6 \big( 1- i\sinh (\pi \mu) \big)}{16\pi \, c_{\pi}^{11} (k_1k_2k_3k_4)^2 ( k_{12}k_{34} )^{\frac{5}{2}}} \, \Gamma \left( \frac{3}{2} + i\mu \right)^2 \Gamma \left( - i\mu \right)^2 \times \\ 
	& \times \Bigg[ k_1k_3 \left( k_{12} + \left( \frac{3}{2} +i\mu \right) k_2 \right) \left( k_{34} + \left( \frac{3}{2} +i\mu \right) k_4 \right) \hat{k}_2^i \, \hat{k}_4^j + 
	\left( \parbox{4.5em}{\centering
	$s$-channel \\[-3pt] perms.
	} \right)
	\Bigg] \times \\
	& \times \Bigg[ \Pi_{ij}^{(0)} (\hat{\mathbf{s}}) \, \frac{H^2}{c_L^2m^2} \left( \frac{1}{2}-\frac{\beta}{H}-i\mu \right)^2 \left( \frac{c_L^2s^2}{4c_{\pi}^2 k_{12}k_{34}} \right)^{i\mu} + \Pi_{ij}^{(1)} (\hat{\mathbf{s}}) \, \left( \frac{c_T^2s^2}{4c_{\pi}^2 k_{12}k_{34}} \right)^{i\mu} \Bigg] \Bigg\} \, ,
\end{aligned}
\end{equation}
where we have assumed $u\leq v$ without loss of generality and the permutations in the second line are only the three remaining $s$-channel ones. The powers $s^{i2\mu}$ yield the characteristic oscillations while the well-known Boltzmann suppression comes from the $\mu$-dependent factors in the first line,
\begin{equation}
    -\big( 1-i \sinh (\pi\mu) \big) \, \Gamma \left( \frac{3}{2} + i\mu \right)^2 \Gamma \left( - i\mu \right)^2 \, \xrightarrow[\, \mu\gg1 \,]{} \, 2\pi^2 \mu e^{-\pi\mu} \, .
\end{equation}
We have verified this result making use of analyticity of the retarded propagator: the Feynman propagator used to compute the $++$ and $--$ diagrams can be written as a sum of the retarded propagator and a factorised term with no time-ordering. Since the integrals that compute the correlator are convergent, analyticity of the retarded Green's functions means that the corresponding contributions to the trispectrum cannot contribute to the non-analytic oscillations. Any cosmological collider signals therefore come from factorised time integrals (the factorised parts of the $++$ and $--$ diagrams and the $+-$ and $-+$ diagrams) and can therefore be computed directly. We performed this bulk calculation and recovered \eqref{eq:collapsed_lim}.  

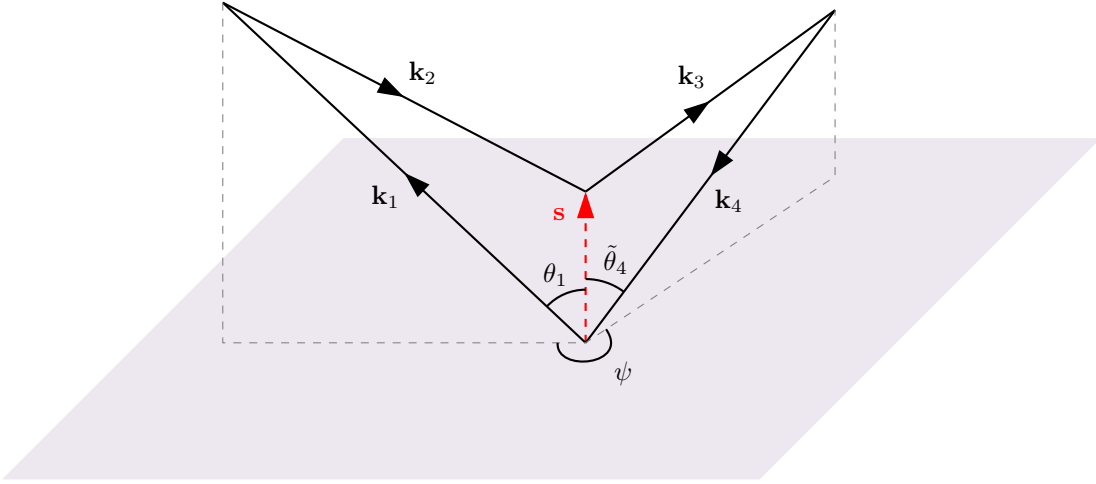
\begin{figure} \centering
\begin{tikzpicture}
	[decoration={ markings,
	mark=at position 0.5 with {\arrow{Stealth[inset=0pt, length=10pt, angle'=35]}}}] 
		
	\filldraw[lightgray!30] (2.3,-1.8) -- ++(4.5,4.5) -- ++(-10,0) -- ++(-4.5,-4.5) -- ++(10,0);

	\coordinate (A) at (0,0);
	\coordinate (B) at (-4.8,4.5);
	\coordinate (C) at (0,2);
	\coordinate (D) at (3.3,4.4);
	\coordinate (aux1) at (-4.8,0);
	\coordinate (aux2) at (3.3,2.2);
	
	\draw[arrows={-Stealth[inset=0pt, length=10pt, angle'=35]}, red, dashed, thick] (A) -- (C) node [midway,xshift=-10,yshift=20]{$\mathbf{s}$};
	\draw[postaction={decorate}, thick] (A) -- (B) node [midway,xshift=-7,yshift=-9]{$\bfk_1$};
	\draw[postaction={decorate}, thick] (B) -- (C) node [midway,xshift=7,yshift=9]{$\bfk_2$};
	\draw[postaction={decorate}, thick] (C) -- (D) node [midway,xshift=-7,yshift=9]{$\bfk_3$};
	\draw[postaction={decorate}, thick] (D) -- (A) node [midway,xshift=7,yshift=-9]{$\bfk_4$};
	
	\draw[dashed, gray] (B) -- (aux1) -- (A) -- (aux2) -- (D);
	 
	\pic [draw, thick, angle radius=20] {angle = C--A--B};
	\node at ([shift={(-0.4,0.9)}]A) {$\theta_1$};
	
	\pic [draw, thick, angle radius=24] {angle = D--A--C};
	\node at ([shift={(0.4,1.1)}]A) {$\tilde{\theta}_4$};
	
	\draw[thick] (-0.37,0) .. controls (-0.4,-0.4) and (0.6,-0.3) .. (0.27,0.18);
	\node at ([shift={(0.5,-0.4)}]A) {$\psi$};
	
\end{tikzpicture}
\caption{The momentum configuration of the trispectrum. The four momentum vectors $\bfk_{1,\ldots,4}$ close into a quadrilateral due to momentum conservation. The angles between each vector $\bfk_a$ and $\mathbf{s}$ are denoted $\theta_a$, with $\tilde{\theta}_4 \equiv \pi - \theta_4$, while $\psi$ denotes the angle between the planes $\text{span}(\bfk_1,\bfk_2)$ and $\text{span}(\bfk_3,\bfk_4)$. In~\eqref{eq:collapsed_lim_final} we take the collapsed limit $s\to 0$ keeping $k_1$, $k_4$, $\theta_1$, $\tilde{\theta}_4$, and $\psi$ fixed.} \label{fig:collapsed}
\end{figure}

The momentum magnitudes and polarisation contractions vary with $s$ too, but in order to get the explicit dependence we need to specify how we take the collapsed limit. Here we choose to do it by keeping the vectors $\bfk_1$ and $\bfk_4$ fixed and shortening the vector $\mathbf{s}= \bfk_1+\bfk_2$ while keeping its direction fixed, cf. Figure~\ref{fig:collapsed}. In practice this means writing all kinematics in terms of the variable set $\{k_1,k_4,\theta_1,\theta_4,\psi,s\}$ (angles are defined in Figure~\ref{fig:collapsed}) and keeping all fixed except $s\to 0$. The necessary substitutions are
\begin{equation}
    k_2=\sqrt{k_1^2+s^2-2sk_1\cos\theta_1} \, , \qquad\quad k_3=\sqrt{k_4^2+s^2+2sk_4\cos\theta_4} \, ,
\end{equation}
and\footnote{The $(-1)^{a+b}$ comes from the fact that the vectors $\bfk_2$ and $\bfk_4$ have a $\pi$ radians shift in their azimuthal angle with respect to $\bfk_1$ and $\bfk_3$.}
\begin{equation}
    \hat{k}_a^i \, \hat{k}_b^j \, \Pi_{ij}^{(0)} (\hat{\mathbf{s}}) = \cos \theta_a \cos \theta_b \, , \qquad\quad
	\hat{k}_a^i \, \hat{k}_b^j \, \Pi_{ij}^{(1)} (\hat{\mathbf{s}}) = (-1)^{a+b} \sin \theta_a \sin \theta_b \cos \psi \,,
\end{equation}
where we can remove the dependence on the angles $\theta_2$ and $\theta_3$ using the relations
\begin{equation}
\begin{aligned}
	k_1 \sin\theta_1 = k_2 \sin\theta_2 \, , & \qquad k_3 \sin\theta_3 = k_4 \sin\theta_4 \, , \\
	k_1 \cos\theta_1 + k_2 \cos\theta_2 = s \, , & \qquad k_3 \cos\theta_3 + k_4 \cos\theta_4 = s \, .
\end{aligned}
\end{equation}
Using all of this in~\eqref{eq:collapsed_lim}, the leading order result in the collapsed limit is
\begin{keyeqn}
\begin{align}
	& \lim_{s\to 0} B_{4,{\rm osc}} = - {\rm Re} \Bigg\{ \frac{\lambda^2 H^6 s^2}{4\pi (4c_{\pi}^2 k_{1}k_{4} )^{\frac{11}{2}}} \big( 1- i\sinh (\pi\mu) \big) \, \Gamma \left( \frac{3}{2} + i\mu \right)^2 \Gamma \left( - i\mu \right)^2 \times \nonumber \\ 
	\times \Bigg[ & \Big( 3 +i2\mu - (11 +i2\mu) \cos (2\theta_1) \Big) \Big( 25 + i6\mu +(11 +i2\mu) \cos (2\theta_4) \Big) \frac{H^2}{c_L^2m^2} \left( \frac{1}{2}-\frac{\beta}{H}-i\mu \right)^2 \left( \frac{c_L^2s^2}{16c_{\pi}^2 k_1k_4} \right)^{i\mu} \nonumber \\
    & + \sin (2\theta_1) \sin (2\theta_4) \cos \psi \, (11+i2\mu)^2 \left( \frac{c_T^2s^2}{16c_{\pi}^2 k_1k_4} \right)^{i\mu} \Bigg] \Bigg\} \, . \label{eq:collapsed_lim_final}
\end{align}
\end{keyeqn}

\paragraph{Scaling} Before discussing the oscillations, we notice that the collapsed limit signal above has an overall suppression $\propto s^2$. One can see that this suppression is not explicit in~\eqref{eq:collapsed_lim}, i.e. it does not come from the time integrals evaluated at the collapsed limit. Instead it comes from the kinematic (angle) structure---more precisely from the cancellation of the leading ($\propto s^0$) contribution when summing over $s$-channel permutations of the polarization contractions. This was already pointed out for the Proca field exchange contribution to the trispectrum in~\cite{Qin:2022lva}, and an analogous suppression is present in the odd-spin exchange contributions to the bispectrum~\cite{Lee:2016vti}. It carries over to our boostless Proca theory because it is a consequence of the spin-1 kinematics. 

This scaling implies that the spin-1 exchange oscillations are suppressed with respect to inflaton contact self-interactions and to scalar exchange, both of which scale like $\propto s^0$ in the collapsed limit~\cite{Creminelli:2012ed,Qin:2022lva}. Also, the $t$- and $u$-channel contributions to~\eqref{eq:trispectrum} scale like $\propto s^0$ when $s\to 0$, so the oscillations are subdominant even in the absence of other fields and interactions. Although this makes it very hard in practice to find the signal~\eqref{eq:collapsed_lim_final} in the data, in the following we discuss how one can in principle read off the parameters of the theory since we find this spin-$1$ example to be an excellent proof of principle that:

\begin{enumerate}
    \item Microcausality leads to new spinning propagators that yield new signals in inflationary observables.
    \item The trispectrum is powerful enough to distinguish between different theories of massive spinning fields.
\end{enumerate} 
This motivates us to extend this analysis to higher-spins where we also expect to see new signals without the suppression from summing over permutations.

\paragraph{Obtaining the parameters} The frequency of the oscillations in the collapsed limit $s\to 0$ is uniquely determined by $\mu$, irrespective of the other parameters of the theory and the specific kinematic configuration in which the limit is approached. To read off the remaining parameters, it is necessary to isolate and compare the contributions from the different helicities.

\begin{figure} \centering
    \includegraphics{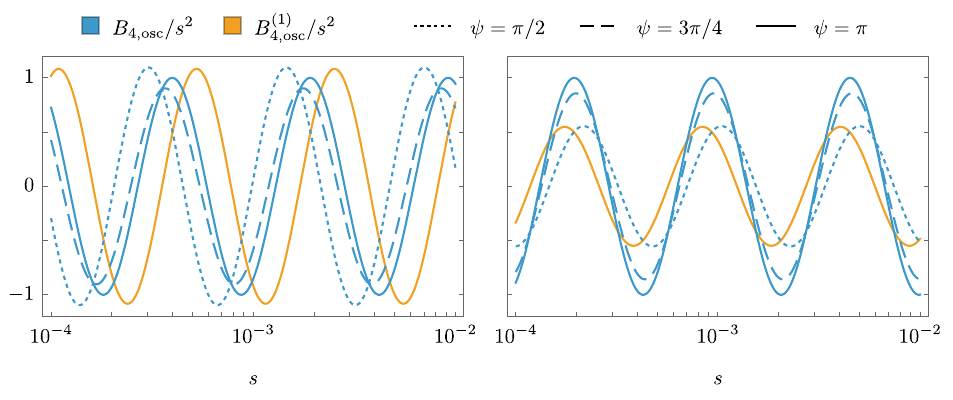}
    \caption{Oscillatory part of the trispectrum in the collapsed limit (Eq.~\eqref{eq:collapsed_lim_final}) for different values of the angle $\psi$. The orange line shows only the $h=1$ contribution. The fixed kinematic variables are $k_1=2k_4=2$ and $\theta_1=\theta_4/3=\pi/4$. We set $\mu=2$, $\beta=H/2$, $c_{\pi}=1$, and $\lambda$ is chosen such that the amplitude of the oscillations is 1 when  $\psi=\pi$. The spin-1 sound speeds are $c_L=0.5<c_T=0.8$ (left) and $c_L=0.7>c_T=0.5$ (right). } \label{fig:plotpsi}
\end{figure}

The $h=0$ and $h=1$ contributions inside the curly brackets of~\eqref{eq:collapsed_lim_final} can be written as complex numbers in polar coordinates $A_h\exp (i\varphi_h)$, such that
\begin{equation}
    \lim_{s\to 0} B_{4,{\rm osc}} = {\rm Re} \left\{ A_0 e^{i\varphi_0} + A_1 e^{i\varphi_1} \right\} = A_0 \cos (\varphi_0) + A_1 \cos (\varphi_1) \, .
\end{equation}
The amplitudes of the two different contributions are
\begin{align}
    A_0 & = \frac{\pi \lambda^2 H^6 s^2}{(4c_{\pi}^2 k_{1}k_{4} )^{\frac{11}{2}}} \frac{4\mu^2+1}{16\mu} \csch (\pi\mu) \bigg| \Big( 3 +i2\mu - (11 +i2\mu) \cos (2\theta_1) \Big) \Big( 25 + i6\mu +(11 +i2\mu) \cos (2\theta_4) \Big) \bigg| \, , \nonumber \\[4pt]
    A_1 & = \frac{\pi \lambda^2 H^6 s^2}{(4c_{\pi}^2 k_{1}k_{4} )^{\frac{11}{2}}} \frac{4\mu^2+1}{16\mu} \csch (\pi\mu) \big| (11+i2\mu)^2 \big| \sin (2\theta_1) \sin (2\theta_4) \cos \psi \, , \label{eq:amps}
\end{align}
while the phases are given by
\begin{equation} \label{eq:phases}
\begin{aligned}
	\varphi_0 = & \,\, {\rm Arg} \left[ \Big( 3 +i2\mu - (11 +i2\mu) \cos (2\theta_1) \Big) \Big( 25 + i6\mu +(11 +i2\mu) \cos (2\theta_4) \Big) \left( \frac{1}{2}-\frac{\beta}{H}-i\mu \right)^2 \right] \\
    & + {\rm Arg} \left[ \big( i\sinh (\pi \mu) -1 \big) \, \Gamma \left( \frac{3}{2} + i\mu \right)^2 \Gamma \left( - i\mu \right)^2 \right] +\mu \log \left( \frac{c_L^2s^2}{16c_{\pi}^2k_1k_4} \right) \, , \\[4pt]
    \varphi_1 = & \,\, {\rm Arg} \left[ \big( i\sinh (\pi \mu) -1 \big) \, \Gamma \left( \frac{3}{2} + i\mu \right)^2 \Gamma \left( - i\mu \right)^2 \, (11 +i2\mu)^2 \right] +\mu \log \left( \frac{c_T^2s^2}{16c_{\pi}^2k_1k_4} \right) \, .
\end{aligned}
\end{equation}
As one can see in~\eqref{eq:amps}, only the helicity-1 contribution depends on $\psi$, which allows us to tune it independently of the helicity-0 one by varying this angle. This is shown in Figure~\ref{fig:plotpsi} for two theories with different sound speed choices. In this way, a measure of the amplitude and phase of the oscillations in a configuration with $\cos\psi=0$ would directly give $A_0$ and $\varphi_0$ (blue dotted line in the figure). The helicity-1 quantities $A_1$ and $\varphi_1$ (orange line) can then be extracted from the amplitude and phase shift of the signal as the angle $\psi$ is varied away from $\cos\psi=0$ (blue dashed and solid lines).

The amplitudes $A_0$ and $A_1$ do not depend on $c_L$, $c_T$ or $\beta$ so they do not allow us to distinguish between the dS-invariant Proca theory and the generalised version found in this work. Having measured $\mu$ from the frequency of oscillations, they only fix the quantity $\lambda^2H^6/c_{\pi}^{11}$. The phases, however, have a richer dependence: given $\mu$, $\varphi_1$ fixes the ratio $c_T/c_{\pi}$ while $\varphi_0$ is sensitive to $c_L/c_{\pi}$ and $\beta$. The degeneracy with $c_{\pi}$ can be removed by fitting the phase difference
\begin{equation}
\begin{aligned}
	\varphi_0 - \varphi_1 = {\rm Arg} \left[ \left( \frac{3 +i2\mu}{11+i2\mu} - \cos (2\theta_1) \right) \left( \frac{25 + i6\mu}{11+i2\mu} + \cos (2\theta_4) \right) \left( \frac{1}{2}-\frac{\beta}{H}-i\mu \right)^2 \right] +2\mu \log \left( \frac{c_L}{c_T} \right) \, ,
\end{aligned}
\end{equation}
instead of the individual phases. This quantity allows for a clean detection of deviations from the dS-invariant case $\beta=0$, $c_L=c_T$. That said, the oscillatory signal in the collapsed limit of the trispectrum is not enough to separately determine $\beta$, $c_L$ and $c_T$. To break the degeneracy between these parameters we need to consider the analytic terms in $s^2$ in the collapsed limit which we have checked can indeed be used to e.g. measure $\beta$, or work away from the collapsed limit where again $\beta$ can be determined.

The situation is similar if we aim to distinguish CCM from boostless Proca. As discussed below~\eqref{eq:trispectrum}, given identical interaction vertices the only difference at the trispectrum level is the presence of the operator $\mathcal{D}$. This operator doesn't have any effect on the amplitude $A_0$, and its effect on the phase $\varphi_0$ is degenerate with the sound speed ratio $c_L/c_{\pi}$. An unambiguous distinction between CCM and boostless Proca therefore requires considering the series part of the trispectrum or fitting the signal away from the collapsed limit.

\section{Conclusion and outlook} \label{Conclusion}
In recent years, the cosmological collider program has been very active from both a theory and a phenomenological point of view; providing a research avenue that ties together QFT calculations in curved spacetime with cosmological observations. Our primary motivation for this work has been to understand what the most general signals of new spinning particles are in the end-of-inflation correlation functions that ultimately form a statistical description of the universe at the start of the hot big bang. How we couple such new states to inflaton/graviton perturbations is straightforward when working within the EFToI \cite{Cheung:2007st} and an immense effort has furnished us with numerous innovative computational techniques enabling us to efficiently compute such signals. The final piece of the puzzle is a description of how these new states propagate on an inflationary background, and in this work we set out to derive the most general Lagrangians to describe this evolution. Although we believe our approach to be spin-independent, we have focused on a spin-$1$ state and found a number of interesting results. By imposing the symmetries of scale invariance and rotations, and further demanding consistency of our theory with Bunch-Davies vacuum conditions and propagation that obeys microcausality, we found that spin-$1$ theories are of the CCM type of \cite{Bordin:2018pca} or within a class we dubbed \textit{boostless Proca} which is characterised by 5 different Lagrangian parameters. 

Microcausality, thanks to the Paley-Weiner theorem, requires the retarded propagator to be analytic in the full complex momentum plane and to satisfy an exponentially boundedness condition. In a theory with non-dynamical modes, which is characteristic of massive spinning particles on any background, these conditions are very non-trivial to satisfy since upon integrating out the non-dynamical modes, non-localities for the dynamical modes can arises. For spin-$1$, we find that microcausality necessitates gauge invariant kinetic terms, allows for two mass terms (one for the temporal mode and one for the spatial modes), a one-derivative term, and a single parity-odd term.

We also investigated the phenomenology of this boostless Proca theory by coupling the field to the inflaton and computing cosmological correlators. We focused on the bispectrum and trispectrum. At the level of bispectra, our new boostless Proca theories do not produce distinct signal from those of CCM theories meaning that the CCM signals are very general and make no assumptions on the underlying theory beyond symmetries, Bunch-Davies conditions and microcausality. However, this degeneracy is broken when it comes to trispectra, where signals from boostless Proca theories cannot be mimicked by CCM physics and new shapes arise. In the collapsed limit of the trispectrum, we found that all 5 parameters of the boostless Proca theory can be read off with some extracted from the non-analytic cosmological collider oscillations and others from the analytic parts.

There are a few directions for future research:
\begin{itemize}
\item {\bf Higher spins:} our results for spin-$1$ provide us with motivation to explore higher spins where we again expect to find new theories and new signals in cosmological observables. For higher spins, our expectation is that new signals will arise in the bispectrum and given the heightened activity in searching for such signals, we believe that this endeavour is an important one especially given that even spin signals are not suppressed in the squeezed/collapsed limit. 
For instance, the scalar bispectrum from a new type of massive spin-$2$ particle may lead to new collider shapes and thus can be of interest for the on-going observational search on primordial non-Gaussianity. In addition, while higher-spin dS-invariant theories are constrained by the Higuchi bound \cite{Higuchi:1986py}, there is no such bound on the boost-breaking CCM theories. These new types of theories, being an ``intermediate'' theory between the two, might be constrained by a weaker Higuchi-like bound. Meanwhile, higher-spin exchanges can contribute to tensor non-Gaussianities, which offer an additional channel for probing these particle species.
We leave the possible new phenomenologies from the general higher-spin theories to future work. 
During our analysis, we eliminated the most generic case \ref{weirdgenericcasedS} using only the Wronskian and the equations of motion i.e. we didn't need to find explicit solutions for the mode functions. Therefore, in the higher-spin cases, we anticipate to be able to carry out the same analysis without solving the equations of motion directly. The main challenge when going to higher spin is the proliferation in the number of terms we can write down in the free theory. 
\item {\bf Interacting theory:} we focused on the free theory in this paper, but we might be able to use microcausality to constrain interactions too. Indeed, thanks to Jacobi identity, we anticipate the nested commutators $[\hat{\sigma}_{\mu}(\eta_1,\bfx_1), [\hat{\sigma}_{\nu}(\eta_2,\bfx_2), \hat{\sigma}_{\rho}(\eta_3,\bfx_3)]]$ to vanish whenever both pairs $(\eta_1,\bfx_1), (\eta_2,\bfx_2)$ and $(\eta_1,\bfx_1), (\eta_3,\bfx_3)$ are spacelike separated\cite{Hui:2025aja}. Extracting constraints from such commutators would be very interesting for our spin-$1$ field but perhaps also for the curvature perturbation where non-localities arise upon integrating out the non-dynamical parts of the metric \cite{Maldacena:2002vr}. It might well be the case that such constraints first arise at loop level. 
\item {\bf Adding time dependence:} we assumed scale invariance in this work, so that we have control over the time dependence of the parameters. However, the most general EFT in the unitary gauge should allow for any time dependence (albeit possibly with slow-roll suppression). It would be interesting to repeat our analysis allowing for a more general time dependence where analyticity must still hold. The main obstacle here would be in solving for the mode functions but perhaps as mentioned above this might not be necessary and analyticity of the Wronksian might prove sufficient. Our perspective in Appendix \ref{appBswitchingpicture} where we find a local basis for the equations of motion might also be helpful in this regard. 
\end{itemize}
\paragraph*{Acknowledgements} We thank Katsuki Aoki, Giovanni Cabass, Ed Copeland, Zongzhe Du, Lam Hui, Austin Joyce, Leonardo Juliano, Hayden Lee, Shinji Mukohyama, Antonio Padilla, Enrico Pajer, Guilherme Pimentel, Zhehan Qin, Nicole Righi, Paul Saffin, Xi Tong, Yi Wang, Zhong-Zhi Xianyu, Masahide Yamaguchi, and Yuhang Zhu for helpful discussions. We would like to especially thank Sadra Jazayeri for his early involvement in the project and for many useful discussions throughout. T.C. is supported by an STFC studentship [grant number ST/Y509437/1] and a scholarship from the University of Nottingham. The work of C.D.P. is supported by the Scuola Normale Superiore. D.S. is supported by a UKRI Stephen Hawking Fellowship [grant number EP/W005441/1] and a Nottingham Research Fellowship from the University of Nottingham. DGW is supported by GRF projects 16306425 and 16308926 from the Research Grants Council of Hong Kong, and the Stephen Hawking Centre for Theoretical Cosmology in Cambridge. For the purpose of open access, the authors have applied a CC BY public copyright licence to any Author Accepted Manuscript version arising.
\appendix
\appendixpage
\addappheadtotoc

\section{Analyticity, exponential boundedness and contour integration} \label{sec:contourintegration}
Here we give a perhaps more direct perspective on the role of the analyticity and exponential boundedness conditions \ref{analyticity} and \ref{bound} that ensure the commutator vanishes outside of the lightcone by performing the Fourier transform from $G(x^0,y^0, \bfk)$ to $G(x^0,y^0,\bfx - {\bf y})$.\footnote{We thank Xi Tong for discussions on this topic.} Since the Green's function is really a distribution, in checking that it vanishes outside of the lightcone we should integrate it against a test function. The set up would be to consider a test function that vanishes \textit{inside} the lightcone but is non-zero outside the lightcone, then integrate the Green's function acting on this test function over all space at fixed times. Inside the lightcone the integral vanishes by construction so if the full integrated result vanishes, the Green's function must vanish outside of the lightcone. Instead, we regulate the momentum integral by deforming the contour into the complex plane.  

\paragraph{$1+1$ dimension} Let's start with the example of a scalar Green's function in $1+1$-dimensional space. The position-space Green's function is given by
\begin{align}
    G(x^0,y^0,x-y) = \int_{-\infty}^{\infty}\frac {dk}{2\pi} G(x^0,y^0,k) e^{ik(x-y)}\,.
\end{align}
We can compute this by contour integration, but depending on the sign of $x-y$, we close the contour in different directions, as shown below.\footnote{Here, we regulated the contours for convergence, so that the argument of the arc goes from $\pm \epsilon$ to $\pm (\pi - \epsilon)$.}
\begin{figure}[H]
\begin{subfigure}[h]{0.45\linewidth}
\includegraphics[width=\linewidth]{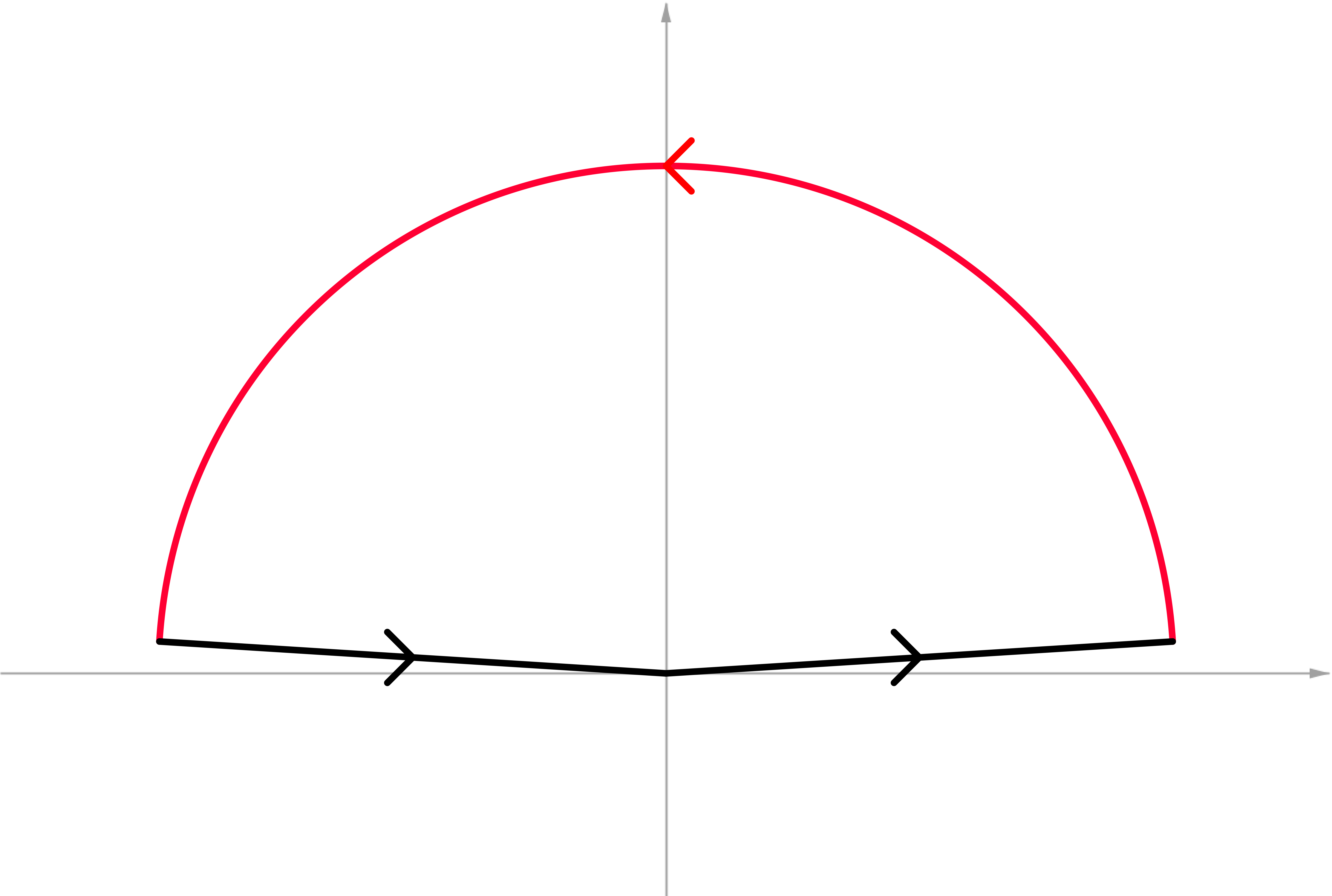}
\caption{\centering When $x-y>0$, we close the contour in upper half complex plane}
\label{contourup}
\end{subfigure}
\hfill
\begin{subfigure}[h]{0.45\linewidth}
\includegraphics[width=\linewidth]{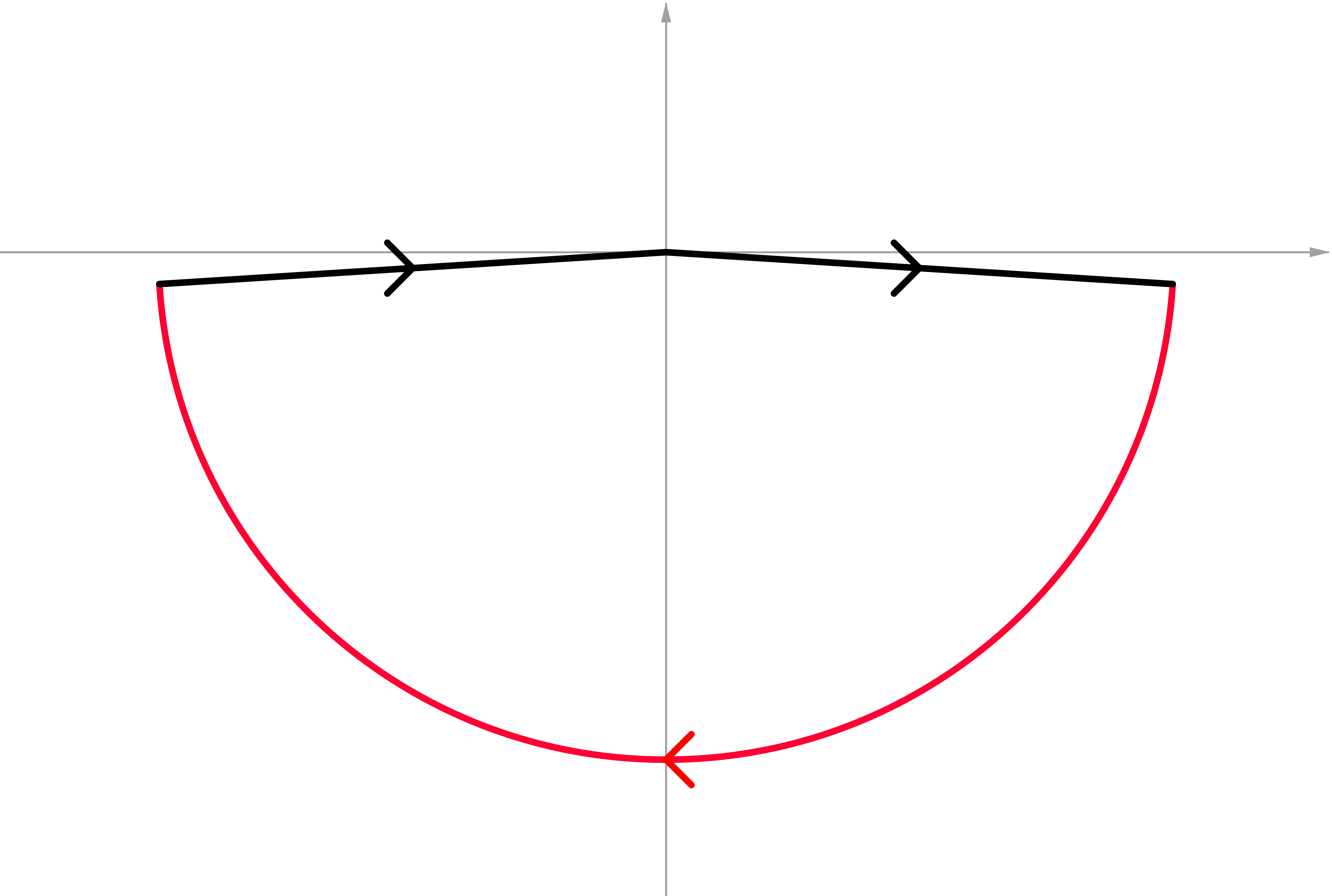}
\caption{\centering When $x-y<0$, we close the contour in lower half complex plane}
\label{contourdown}
\end{subfigure}%
\caption{\centering Closing contours differently depending on sign of $x-y$ to ensure $(\operatorname{Im}k)(x-y)\geq0$.}
\label{closingcontour}
\end{figure}
The analyticity condition \ref{analyticity} ensures that whichever direction we close the contour, the full contour integral vanishes by Cauchy's theorem. The bound \ref{bound} then ensures that for spacelike separation $|x^0-y^0| < |x-y|$, the contribution from the arc vanishes when $|k|\to \infty$. To see this, first denote \begin{align}
    \Delta = |x-y| - |x^0-y^0| > 0\,,
\end{align} 
which is a constant given $x$, $y$, $x^0$, and $y^0$. Assume $x-y>0$ and we close the contour from above, then along the arc of radius $R$, we have
\begin{align}
    \left |\int_{\text{arc}}dk \; G(x^0,y^0,k) e^{ik(x-y)}\right | & \leq  \pi R \max_{k=Re^{i\theta}} \left |G(x^0,y^0,k) \right | \left |e^{ik(x-y)}\right | \tag{estimation lemma}\\
    & \leq \pi R C(D+R)^N \max_{k=Re^{i\theta}}\exp (\operatorname{Im}k |x^0-y^0| - \operatorname{Im}k |x-y|) \tag{assuming \ref{bound}}\\
    & = \pi R C(D+R)^N \exp(-R \Delta \sin \epsilon)\,,
\end{align}
on the deformed contour. For any given $\epsilon > 0$ and $\Delta >0$, this tends to $0$ as $R\to \infty$. Therefore, the position-space Green's function vanishes whenever $\Delta > 0$, i.e. outside the lightcone. To check that the Green's function vanishes for all spacelike separations we also need to consider the case where $x-y < 0$ which follows in an almost identical manner once we close the contour from below. Analyticity is required in the full complex plane so that causality is respected for both $x-y>0$ and $x-y < 0$. Here we directly see why the behaviour of the Green's function as $|k| \rightarrow \infty$ is important. In concrete examples we often have $N=-1$ in which case we can employ Jordan's lemma to show that the contribution from the arc vanishes. 

Now we consider the momentum-space Green's function for our vector field $G_{ij}(x^0,y^0,\bfk)$ and Fourier transform to position space. We separate the momentum-space Green's function into different tensor structures using \eqref{ijcommutator}, and compute the Fourier transform of each individually.
\paragraph{The $\delta_{ij}$ coefficient} Denoting
\begin{align}
    G^{+}(x^0,y^0,k) & = \frac 12(G^{(+1)}(x^0,y^0,k) + G^{(-1)}(x^0,y^0,k))\,,
\end{align}
for convenience, we compute the Fourier transform
\begin{align}
\int_{\bfk} G^{+}(x^0,y^0,k) e^{i\bfk \cdot (\bfx - {\bf y})} & = \int^{2\pi}_0 d\phi \int^{\infty}_0 \frac {k^2 dk}{(2\pi)^3}  \int^{\pi}_0 \sin \theta d\theta \; G^{+}(x^0,y^0,k) e^{ik|\bfx - {\bf y}| \cos \theta} \\
& = \frac 1{(2\pi)^2} \int^{\infty}_0 k^2 dk \frac {G^{+}(x^0,y^0,k)}{ik|\bfx - {\bf y}|}\left (e^{ik|\bfx - {\bf y}|} - e^{-ik|\bfx - {\bf y}|}\right ) \\
& = -\frac {i}{(2\pi)^2|\bfx - {\bf y}|} \left (\int^{\infty}_0 dk \; G^{+}(x^0,y^0,k) (k e^{ik|\bfx - {\bf y}|}) \right . \nonumber \\
& \qquad \qquad \qquad \left . + \int^0_{-\infty} dk \; G^{+}(x^0,y^0,-k)(ke^{ik|\bfx - {\bf y}|}) \right )\,.
\end{align}
If $G(x^0,y^0,k)$ is symmetric under $k \leftrightarrow -k$, then we can combine the two integrals together to 
\begin{align}
\int_{\bfk} G^{+}(x^0,y^0,k) e^{i\bfk \cdot (\bfx - {\bf y})} & = -\frac i{(2\pi)^2|\bfx - {\bf y}|} \int^{\infty}_{-\infty} dk \; G^{+}(x^0,y^0,k)(ke^{ik|\bfx - {\bf y}|})\,.
\end{align}
If in addition, $G(x^0,y^0,k)$ is analytic in the upper half plane, we can close the contour as in Figure \ref{contourup}. Then as in the $1+1$ dimensional case, the bound \ref{bound} ensures that the integral along the large arc at infinity vanishes and then by Cauchy's theorem the position-space Green's function vanishes outside the lightcone. Note that since the exponential factor is $e^{ik|\bfx - {\bf y}|}$ we were forced to close the contour in the upper-half complex plane and therefore we only need analyticity in the upper-half complex plane. However, to combine the integrals into one integral along the full real line we required $G(x^0,y^0,k)$ to be symmetric under $k\leftrightarrow -k$, and therefore analyticity in the upper-half complex plane implies analyticity in the entire complex plane. This is precisely the condition \ref{analyticitycondition1}.
\paragraph{The $\hat{k}_k$ coefficient} This time, we denote
\begin{align}
    G^{-}(x^0,y^0,k) & = \frac 12 (G^{(+1)}(x^0,y^0,k) - G^{(-1)}(x^0,y^0,k))\,,
\end{align}
and the Fourier transform works only slightly differently with an extra $\hat{k}_k$ in the integrand. We have
\begin{align}
\int_{\bfk}G^{-}(x^0,y^0,k) \hat{k}_k e^{i\bfk \cdot (\bfx - {\bf y})} & = \partial_{(x-y)_k} \int_{\bfk} \frac {G^{-}(x^0,y^0,k)}{ik} e^{i\bfk \cdot (\bfx- {\bf y})}\,.
\end{align}
From the previous discussion, we require $G^{-}(x^0,y^0,k)/k$ to be symmetric under $k\leftrightarrow -k$ and analytic in $k^2$, which is exactly the condition \ref{analyticitycondition2}, for this Green's function to vanish outside the lightcone.
\paragraph{The $\hat{k}_i\hat{k}_j$ coefficient} Lastly, we denote
\begin{align}
    G^L(x^0,y^0,k) = G^{(0)}(x^0,y^0,k) - G^{+}(x^0,y^0,k)\,.
\end{align}
The Fourier transform now has an extra $\hat{k}_i\hat{k}_j$ in the integrand, so we write
\begin{align}
    \int_{\bfk}G^{L}(x^0,y^0,k) \hat{k}_i \hat{k}_j e^{i\bfk \cdot (\bfx - {\bf y})} & = \partial_{(x-y)_i}\partial_{(x-y)_j} \int_{\bfk} \frac {G^{L}(x^0,y^0,k)}{-k^2} e^{i\bfk \cdot (\bfx- {\bf y})}\,.
\end{align}
We now require $G^L(x^0,y^0,k)/k^2$ to be symmetric under $k\leftrightarrow -k$ and analytic in $k^2$, which is exactly the condition \ref{analyticitycondition3}.
\section{Switching picture: local equations of motion} \label{appBswitchingpicture}
For the Proca theory in flat space, the temporal and longitudinal equations of motion are 
\begin{align}
\frac{\delta S}{\delta \sigma_{0}} &= 0 \implies ( k^2 + m^2)\sigma_0 - k \partial_t \sigma_L  = 0 \,, \label{ProcaEoM1} \\
\frac{\delta S}{\delta \sigma_{L}} &= 0 \implies -\partial_t^2 \sigma_L - m^2 \sigma_L  + k \partial_t \sigma_0  = 0 \,. \label{ProcaEoM2}
\end{align}
In the bulk of the paper we have taken the approach of solving \eqref{ProcaEoM1} for $\sigma_0$ and plugging the solution into \eqref{ProcaEoM2} to give us a dynamical equation for $\sigma_L$. This yields the system  
\begin{align}
\partial_t^2 \sigma_L  + (k^2+m^2) \sigma_L = 0\,, \qquad    \sigma_0 & = \frac {k \partial_t \sigma_L}{ k^2 + m^2} \,.
\end{align}
We therefore have a local equation of motion for $\sigma_L$ and an apparently non-local constraint that then fixes $\sigma_0$. However, we can \textit{switch picture} such that $\sigma_0$ solves a dynamical second-order differential equation, while $\sigma_L$ satisfies a constraint equation:
\begin{align}
\partial_t^2 \sigma_0  + (k^2+m^2) \sigma_0 = 0\,, \qquad    \sigma_L & = -\frac { 1}{k}\partial_t \sigma_0 \,,
\end{align}
where the constraint equation is the familiar $\partial_{\mu}A^{\mu} = 0$ equation. In this case both equations are local (up to a factor of $1/k$ coming from the normalisation of the polarisation vector) and therefore the non-locality in the original picture can be eliminated. 

Turning to our more general description without any restrictions on the Lagrangian parameters, the original picture equations of motion are
\begin{align}
   & \partial_t^2 \sigma_L- \left(\frac {\alpha_1 (\beta_1 + \beta_2)k^4 + (\alpha_1\beta_3 + \alpha_2(\beta_1 + \beta_2) - \gamma_2^2)k^2 + \alpha_2 \beta_3}{(\alpha_1 - \gamma_1^2)k^2 + \alpha_2} \right) \sigma_L = 0\,, \\ 
  &  \sigma_0 = -\frac {k}{\alpha_1 k^2 + \alpha_2} (\gamma_1 \partial_t - \gamma_2) \sigma_L\,,
\end{align}
which are both in general non-local. We can now again switch picture to find
\begin{align}
   & \partial_t^2 \sigma_0- \left(\frac {( \alpha_1 k^2 + \alpha_2)(k^2(\beta_1+\beta_2) +\beta_3) - \gamma_2^2 k^2}{(\alpha_1 - \gamma_1^2)k^2 + \alpha_2} \right) \sigma_0 = 0\,, \\
   & \sigma_L = -\frac{(\alpha_1 - \gamma_1^2)k^2+ \alpha_2}{k(k^2(\beta_1 + \beta_2)\gamma_1^2 + \beta_3 \gamma_1^2 -\gamma_2^2)} (\gamma_1 \partial_t \sigma_0 + \gamma_2 \sigma_0 )\,,
\end{align}
where again both equations are generically non-local. The conditions we found from microcausality are $\alpha_1 = \gamma_1^2$ and $\beta_1 + \beta_2  =0$ which yields 
\begin{align}
   & \partial_t^2 \sigma_0- \frac{1}{\alpha_2}((\gamma_1^2 \beta_3 - \gamma_2^2) k^2 + \alpha_2 \beta_3 ) \sigma_0 = 0\,, \\
   & \sigma_L = -\frac{\alpha_2}{k( \beta_3 \gamma_1^2 -\gamma_2^2)} (\gamma_1 \partial_t \sigma_0 + \gamma_2 \sigma_0 )\,,
\end{align}
which therefore makes both equations local as in the Proca case. So, the microcausality conditions ensure that the equations of motion can be expressed in a way that makes locality manifest.

In the dS case, the original picture equations of motion read
\begin{align}
     &\partial_{\eta}^2 \sigma_L - \frac {2\alpha_2\gamma_1^2H^2k^2\eta}{F_1F_2} \partial_{\eta}\sigma_L + \nonumber \\
     &\left (\frac {-F_1^2[\beta_3 + (\beta_1+\beta_2)H^2k^2\eta^2]+(\gamma_2^2+\gamma_1\gamma_2H)H^2k^2\eta^2F_1 -2\alpha_2\gamma_1\gamma_2H^3k^2\eta^2}{F_1F_2H^2\eta^2}\right )\sigma_L = 0
     \end{align}
     and
     \begin{align}
\sigma_0 &= -\frac {H^2k\eta^2}{F_1}\left (\gamma_1 \partial_{\eta} + \frac {\gamma_2}{H\eta}\right )\sigma_L\,,
\end{align}
where
\begin{align}
    F_1 & = \alpha_2 + \alpha_1 H^2k^2\eta^2 \\
    F_2 & = \alpha_2 + (\alpha_1-\gamma_1^2)H^2k^2\eta^2\,.
\end{align}
Note that both the constraint and the equation of motion are non-local. However, we can switch the picture to obtain
\begin{align}
\sigma_L & = -\frac {1}{Gk}\left (\gamma_1 F_2 \partial_{\eta} -\frac {\gamma_2F_2 +2\alpha_2\gamma_1H}{H\eta}\right ) \sigma_0 \,,
\end{align}
and 
\begin{align}
    \partial_{\eta}^2 \sigma_0 + \left (\frac {2}{\eta} - \frac {4\alpha_2}{F_2\eta} - \frac {2(\beta_1+\beta_2)\gamma_1^2H^2k^2\eta}{G}\right ) \partial_{\eta} \sigma_0 + \frac {\sum_{n=0}^3G_n(Hk\eta)^{2n}}{F_2GH^2\eta^2}\sigma_0 = 0 \,,
\end{align}
where
\begin{align}
    G & = \gamma_1^2[\beta_3 + (\beta_1+\beta_2)H^2k^2\eta^2]- \gamma_1\gamma_2H -\gamma_2^2 \,,\\
    G_0 & = -\alpha_2 (\beta_3 - 2H^2) (\beta_3\gamma_1^2-\gamma_1\gamma_2H - \gamma_2^2) \,, \\
    G_1 & = \alpha_2(\beta_1 + \beta_2) (-2\beta_3 \gamma_1^2 + \gamma_2^2 + 3\gamma_1\gamma_2 H + 6\gamma_1^2H^2)  \nonumber \\
    & \quad - (\beta_3\gamma_1^2- \gamma_1\gamma_2H - \gamma_2^2)(\alpha_1\beta_3 - \gamma_1\gamma_2H -\gamma_2^2) \,, \\
    G_2 & = -(\beta_1 + \beta_2)[\alpha_2\gamma_1^2(\beta_1+ \beta_2) - (\alpha_1-\gamma_1^2)(\gamma_1\gamma_2H+\gamma_2^2)\nonumber \\
    & \quad +2\gamma_1(\alpha_1\beta_3 \gamma_1 -\alpha_1\gamma_2H - \gamma_1\gamma_2^2)] \,, \\
    G_3 & = -\alpha_1\gamma_1^2(\beta_1+ \beta_2)^2\,.
\end{align}
The conditions from microcausality, i.e. $\alpha_1 = \gamma_1^2$ and $\beta_1+\beta_2 = 0$ reduces these equations to
\begin{align} \label{eq:cons_eq_dS}
    \sigma_L & = -\frac {\alpha_2}{k(\beta_3\gamma_1^2 -\gamma_1\gamma_2 H - \gamma_2^2)} \left (\gamma_1 \partial_{\eta} -\frac {2\gamma_1H+\gamma_2}{H\eta}\right )\sigma_0 \,,
\end{align}
and
\begin{align}
    \partial_{\eta}^2 \sigma_0 - \frac 2{\eta} \partial_{\eta} \sigma_0 + \left (\frac {2}{\eta^2} - \frac {\beta_3}{H^2\eta^2} - \frac {(\beta_3\gamma_1^2-\gamma_1\gamma_2H - \gamma_2^2)k^2}{\alpha_2}\right ) \sigma_0= 0\,,
\end{align}
which are both manifestly local (again apart from the factor of $1/k$ for the $\sigma_L$ equation). We believe this switching picture perspective will be useful when generalising our results to higher spins.

\section{Trispectrum scalar seed} \label{app:Seed}

In this Appendix we obtain the explicit expression for the scalar exchange correlator~\eqref{eq:Fseed_def} used in Subsection~\ref{sec:trispectrum} as a seed to derive the spin-1 exchange trispectrum. Our starting point will be the closely-related scalar seed $F$ derived in~\cite{CosmoBootstrap1}. There they use the fact that the massive scalar propagators of~\eqref{eq:Gscalar} satisfy 
\begin{equation}
\begin{aligned}
	\left( \eta^2\partial_{\eta}^2 -2\eta\partial_{\eta} +k^2\eta^2 - \nu^2 + \frac{9}{4} \right) G^{\rm sc}_{\pm\pm} (\eta,\eta',k) & = \mp i H^2 \eta^2\eta'^2 \delta (\eta-\eta') \, , \\
	\left( \eta^2\partial_{\eta}^2 -2\eta\partial_{\eta} +k^2\eta^2 - \nu^2 + \frac{9}{4} \right) G^{\rm sc}_{\pm\mp} (\eta,\eta',k) & = 0 \, , \label{eq:difopGpm}
\end{aligned}
\end{equation}
to derive the relations
\begin{equation} \label{eq:diffeqsFab}
\begin{aligned}
	\left( (k_{12}^2-s^2) \partial_{k_{12}}^2 + 2k_{12} \partial_{k_{12}} - \nu^2 + \frac{1}{4} \right) F_{\pm\pm} & = \frac{1}{2k_T} \, , \\
	\left( (k_{12}^2-s^2) \partial_{k_{12}}^2 + 2k_{12} \partial_{k_{12}} - \nu^2 + \frac{1}{4} \right) F_{\pm\mp} & = 0 \, .
\end{aligned}
\end{equation}
Then they define the sum
\begin{equation}
	F \equiv \sum_{a,b=\pm 1} F_{ab} = F_{++} + F_{--} + F_{+-} + F_{-+} \, ,
\end{equation}
which by~\eqref{eq:diffeqsFab} obeys the equation
\begin{equation}
	\left( (k_{12}^2-s^2) \partial_{k_{12}}^2 + 2k_{12} \partial_{k_{12}} - \nu^2 + \frac{1}{4} \right) F = \frac{1}{k_T} \, ,
\end{equation}
and solve it to find $F$. To do this it is convenient to work with different variables, namely the energy ratios
\begin{equation}
	u\equiv \frac{s}{k_{12}} \, , \qquad v\equiv \frac{s}{k_{34}} \, ,
\end{equation}
which for physical kinematics satisfy
\begin{equation}
	0\leq u,v \leq 1 \, ,
\end{equation}
but we will be interested in studying the singular limits of the correlators, which requires to analytically continue away from this region. The advantage of these variables is that we can define
\begin{equation}
	\hat{F}_{ab} \equiv s F_{ab} \, , \qquad \hat{F} \equiv s F \, ,
\end{equation}
and by scale symmetry $\hat{F}_{ab}$ and $\hat{F}$ are functions of only $u$ and $v$. The differential equation becomes 
\begin{equation}
	\left( \Delta_u - \nu^2 + \frac{1}{4} \right) \hat{F} = \frac{u v}{u+v} \, ,
\end{equation}
where
\begin{equation}
	\Delta_u \equiv u^2(1-u^2) \partial_u^2 - 2u^3 \partial_u \, .
\end{equation}
 
In our case we are finding a different scalar seed, cf.~\eqref{eq:Fseed_def}. It is obvious from~\eqref{eq:diffeqsFab} that this object still solves the same differential equation as $F$, i.e. for $\hat{\tilde{F}} \equiv s \tilde{F}$ we have
\begin{equation} \label{eq:DiffEqF}
	\left( \Delta_u - \nu^2 + \frac{1}{4} \right) \hat{\tilde{F}} = \frac{u v}{u+v} \, ,
\end{equation}
but its boundary conditions are slightly different. In particular, $\hat{F}$ was fixed in~\cite{CosmoBootstrap1} by requiring $u\leftrightarrow v$ symmetry, regularity at $u\to 1$ and correct normalisation of the $u\to -1$ singularity; the two first conditions hold the same for $\hat{\tilde{F}}$, but the normalisation at $u\to -1$ changes. In any case, we will not attempt to solve~\eqref{eq:DiffEqF}---our strategy for finding $\hat{\tilde{F}}$ will be to solve just for the mixed components $\hat{F}_{\pm\mp}$, which are simpler, and then use
\begin{equation} \label{eq:FFtild_rel}
	\hat{\tilde{F}} = \hat{F} - 2 \big( \hat{F}_{+-} + \hat{F}_{-+} \big) \, .
\end{equation}

\paragraph{Three-point function} We define\footnote{This object is proportional to the tree-level wavefunction coefficient $\psi_{\varphi\varphi\sigma}$ of two conformally-coupled scalars and one massive scalar coming from the interaction $g\varphi^2\sigma$, namely
\begin{equation}
    \psi_{\varphi\varphi\sigma} (k_1,k_2,s) = - \frac{g}{H^3\eta_0^2 \big( \sigma (s,\eta_0) \big)^*} \, f(k_{12},s) \, .
\end{equation}
}
\begin{equation} \label{eq:f_def}
	f (k_{12},s) = \frac{1}{\sqrt{s}} \hat{f} (u) \equiv \frac{i}{H} \int_{-\infty}^0 \frac{d\eta}{\eta^2} \, e^{i k_{12}\eta} \big( \sigma (s,\eta) \big)^* \, , 
\end{equation}
so that
\begin{equation} \label{eq:Fpm_f_rel}
	\hat{F}_{+-} = \frac{1}{2} \, \hat{f} (u) \, \hat{f} (v)^* \, , \qquad \hat{F}_{-+} = \frac{1}{2} \, \hat{f} (u)^* \hat{f} (v) \, .
\end{equation}
The time integrals are well defined only for certain values of the kinematics. Since they ultimately depend on the energy ratios $u$ and $v$, we can explore the domains of analyticity without loss of generality by fixing $s \in \mathbb{R}^+$ and analytically continuing in the external energies $k_a$. The integral~\eqref{eq:f_def} is then well defined for
\begin{equation} \label{eq:analyticity_dom_f3}
	f (k_{12},s) : \{ (k_{12},s) | \, {\rm Im} (k_{12}) < 0 , s \in \mathbb{R}^+ \} \, , \qquad\Rightarrow\qquad \hat{f} (u) : \{ u | \, {\rm Im} (u) > 0 \} \,  .
\end{equation}
For $k_{12}+s\to 0$ ($u\to -1$) the integral~\eqref{eq:f_def} diverges. In that limit, its value is dominated by the far-past contribution, so we can substitute
\begin{equation}
	\lim_{\eta\to -\infty} \big( \sigma (k,\eta ) \big)^* = \frac{-H}{\sqrt{2k}} \, \eta \, e^{ik\eta} \, ,
\end{equation}
to get~\cite{Jazayeri:2022kjy} (we erase the upper integration limit to emphasise that we are only picking up the far-past contribution)
\begin{equation} \label{eq:lim-1f3}
	\lim_{u\to -1+i\epsilon} \hat{f} (u) \simeq \frac{1}{\sqrt{2}} \int_{-\infty} \frac{d\eta}{\eta} \, e^{i (k_{12}+s) \eta} = \frac{1}{\sqrt{2}} \int_{-\infty} \frac{d\xi}{\xi} \, e^{i (1+1/u) \xi} = \frac{-i}{\sqrt{2}} \log (1+u) \, .
\end{equation}
Notice that in~\eqref{eq:difopGpm} the differential operator is just the equation of motion for $\sigma$. Therefore, $\hat{f}$ must also satisfy the homogeneous equation~\eqref{eq:DiffEqF},
\begin{equation} \label{eq:DiffEqf}
	\left( \Delta_u - \nu^2 + \frac{1}{4} \right) \hat{f} = 0 \, .
\end{equation}
This is solved by~\cite{CosmoBootstrap1}
\begin{equation} \label{eq:hom_sols_f3}
	\hat{f} (u) = A_+ \hat{f}_+ (u) + A_- \hat{f}_- (u) \qquad\text{with}\quad \hat{f}_{\pm} (u) = u^{\frac{1}{2} \pm \nu} {}_2F_1 \left[ \left. \begin{gathered} \frac{1}{4} \pm \frac{\nu}{2} \, , \,\, \frac{3}{4} \pm \frac{\nu}{2} \\ 1\pm \nu \end{gathered} \right| u^2 \right] \, .
\end{equation}
These solutions are singular in the folded limit $u\to 1$,
\begin{equation} 
	\lim_{u\to 1} \hat{f}_{\pm} (u) = \gamma_{\pm} \log (1-u) \qquad\text{with}\quad \gamma_{\pm} \equiv - \frac{ \Gamma (1\pm \nu) }{ \Gamma \left( \frac{1}{4} \pm \frac{\nu}{2} \right) \Gamma \left( \frac{3}{4} \pm \frac{\nu}{2} \right) } \, ,
\end{equation}
so requiring the absence of this singularity in the full solution (i.e. imposing the Bunch-Davies vacuum condition) fixes one of the constants,
\begin{equation}
	\lim_{u\to 1} \hat{f} (u) = \big( A_+ \gamma_+ + A_- \gamma_- \big) \log (1-u) \overset{!}{=} 0 \qquad\Rightarrow\qquad \hat{f} (u) = A_+ \left( \hat{f}_+ (u) - \frac{\gamma_+}{\gamma_-} \, \hat{f}_- (u) \right) \, .
\end{equation}
The solutions~\eqref{eq:hom_sols_f3} are also singular as $u\to -1$, which can be used to fix the remaining constant. They have a branch cut along the negative real axis so the direction in which the singularity is approached matters. Eq.~\eqref{eq:analyticity_dom_f3} instructs us to approach the limit from the upper half complex plane, $u\to -1+i\epsilon$, which yields
\begin{equation} 
	\lim_{u\to -1+i\epsilon} \hat{f}_{\pm} (u) = i \, e^{\pm i\pi\nu} \, \gamma_{\pm} \log (1+u) \, .
\end{equation}
Matching to the corresponding limit of the time integral, given by Eq.~\eqref{eq:lim-1f3}, we get
\begin{equation}
	\lim_{u\to -1+i\epsilon} \hat{f} (u) = - 2 \, \gamma_+ A_+ \sin (\pi\nu) \log (1+u) \overset{!}{=} \frac{-i}{\sqrt{2}} \log (1+u) \qquad\Rightarrow\qquad A_+ = i\, \frac{\csc (\pi\nu)}{2\sqrt{2} \, \gamma_+} \, .
\end{equation}
The three-point function is then finally
\begin{equation}
	\hat{f} (u) = i \, \frac{\csc (\pi\nu)}{2\sqrt{2}} \left( \frac{\hat{f}_+ (u)}{\gamma_+} - \frac{\hat{f}_- (u)}{\gamma_-} \right) \, .
\end{equation}

\paragraph{Full solution} From~\eqref{eq:Fpm_f_rel} we then have
\begin{equation} \label{eq:FpmFmp}
\begin{aligned}
	2 \big( \hat{F}_{+-} + \hat{F}_{-+} \big) = \hat{f} (u) \, \hat{f} (v)^* + \hat{f} (u)^* \hat{f} (v) = \frac{\csc^2 (\pi\nu)}{4} \left( \frac{\hat{f}_+ (u)}{\gamma_+} - \frac{\hat{f}_- (u)}{\gamma_-} \right) \left( \frac{\hat{f}_+ (v)}{\gamma_+} - \frac{\hat{f}_- (v)}{\gamma_-} \right) \, ,
\end{aligned}
\end{equation}
where we have used $(\hat{f}_{\pm})^*=\hat{f}_{\mp}$ and $(\gamma_{\pm})^*=\gamma_{\mp}$. The seed $\hat{F}$ was computed in~\cite{CosmoBootstrap1}, and is
\begin{equation} \label{eq:sol_F}
	\hat{F} (u,v) = \begin{dcases}
		\sum_{m,n=0}^\infty c_{mn} u^{2m+1} \left( \frac{u}{v} \right)^n + \frac{\pi}{2\cos (\pi\nu)} \, \hat{g} (u,v) & \qquad u\leq 1 \text{ and } u\leq v \, , \\ 
		\sum_{m,n=0}^\infty c_{mn} v^{2m+1} \left( \frac{v}{u} \right)^n + \frac{\pi}{2\cos (\pi\nu)} \, \hat{g} (v,u) & \qquad v\leq 1 \text{ and } v\leq u \, , 
	\end{dcases}
\end{equation}
where\footnote{Notice that~\cite{CosmoBootstrap1} used a basis $\hat{F}_{\pm}(u)$ of solutions to~\eqref{eq:DiffEqf} that have a branch cut along the positive imaginary axis, which makes the analytical continuation to negative $u$ cumbersome. Here, just like in~\cite{Jazayeri:2022kjy,Goodhew:2021oqg}, we use a different basis~\eqref{eq:hom_sols_f3} where the branch cuts lay along the negative real axis.}
\begin{align}
	c_{mn} & = \frac{(-1)^n (n+1)(n+2)\cdots (n+2m)}{ \big[ (n+\frac{1}{2})^2 - \nu^2 \big] \big[ (n+\frac{5}{2})^2 - \nu^2 \big] \cdots \big[ (n+\frac{1}{2}+2m)^2 - \nu^2 \big] } \, , \\[7pt]
	\hat{g} (u,v) & = \frac{\cot(\pi\nu)}{4\pi} \left[ \frac{\hat{f}_+ (u)}{\gamma_+} + \frac{\hat{f}_- (u)}{\gamma_-} + \frac{1}{\sin(\pi\nu)} \left( \frac{\hat{f}_+ (u)}{\gamma_+} - \frac{\hat{f}_- (u)}{\gamma_-} \right) \right] \left( \frac{\hat{f}_+ (v)}{\gamma_+} - \frac{\hat{f}_- (v)}{\gamma_-} \right) \, .
\end{align}
Then, substracting~\eqref{eq:FpmFmp} from~\eqref{eq:sol_F} we finally get our trispectrum scalar seed
\begin{keyeqn}
\begin{align} \label{eq:sol_Ftilde}
	\hat{\tilde{F}} (u,v) = \begin{dcases}
		\sum_{m,n=0}^\infty c_{mn} u^{2m+1} \left( \frac{u}{v} \right)^n + \frac{\pi}{2\cos (\pi\nu)} \, \hat{\tilde{g}} (u,v) & \qquad u\leq 1 \text{ and } u\leq v \, , \\ 
		\sum_{m,n=0}^\infty c_{mn} v^{2m+1} \left( \frac{v}{u} \right)^n + \frac{\pi}{2\cos (\pi\nu)} \, \hat{\tilde{g}} (v,u) & \qquad v\leq 1 \text{ and } v\leq u \, , 
	\end{dcases}
\end{align}
with
\begin{equation}
	\hat{\tilde{g}} (u,v) \equiv \frac{\cot(\pi\nu)}{4\pi} \left[ \frac{\hat{f}_+ (u)}{\gamma_+} + \frac{\hat{f}_- (u)}{\gamma_-} - \frac{1}{\sin(\pi\nu)} \left( \frac{\hat{f}_+ (u)}{\gamma_+} - \frac{\hat{f}_- (u)}{\gamma_-} \right) \right] \left( \frac{\hat{f}_+ (v)}{\gamma_+} - \frac{\hat{f}_- (v)}{\gamma_-} \right) \, .
\end{equation}
\end{keyeqn}
This result is valid when at least one of the variables $u$ or $v$ lays within the unit circle, but our physical set-up involves a generic sound speed $c_{\pi}$ for the inflaton so we can have $|u|,|v|>1$ (see Eq.~\eqref{eq:uv_defs}). The solution for $\hat{F}$ when $|u|,|v|>1$ was derived in~\cite{Jazayeri:2022kjy}, and reads 
\begin{align}
	\hat{F} (u,v) = & \,\, 2 \sum_{m,n=0}^\infty a_{mn} u^{-m} \left( \frac{u}{v} \right)^n + \frac{1}{2} \left( \hat{h}_+ (u) \hat{h}_- (v) + \hat{h}_- (u) \hat{h}_+ (v) \right) \log u \label{eq:sol_F_uvlarge} \\
    & - \frac{1}{4} \left( \log 2 -1 +H_{-\frac{1}{4}+\frac{\nu}{2}} +H_{-\frac{1}{4}-\frac{\nu}{2}} \right) \hat{h}_- (u) \hat{h}_+ (v) - \frac{1}{4} \left( \log 2 +H_{-\frac{1}{4}+\frac{\nu}{2}} +H_{-\frac{1}{4}-\frac{\nu}{2}} \right) \hat{h}_+ (u) \hat{h}_- (v) \nonumber \\ 
	& + \frac{1}{4\pi} \Gamma \left( \frac{3}{4}+\frac{\nu}{2} \right)^2 \Gamma \left( \frac{3}{4}-\frac{\nu}{2} \right)^2 \hat{h}_- (u) \hat{h}_- (v) \, , \qquad\qquad\qquad\qquad\qquad\qquad\qquad 1<u<v \, , \nonumber
\end{align}
where $H_x$ is the Harmonic number and the basis of solutions to the homogeneous equation is now conveniently chosen to be
\begin{equation} \label{eq:hom_sols_h}
\begin{aligned}
	\hat{h}_+ (u) = {}_2F_1 \left[ \left. \begin{gathered} \frac{1}{4} - \frac{\nu}{2} \, , \,\, \frac{1}{4} + \frac{\nu}{2} \\ \frac{1}{2} \end{gathered} \right| \frac{1}{u^2} \right] \, , \qquad \hat{h}_- (u) = \frac{2}{u} \, {}_2F_1 \left[ \left. \begin{gathered} \frac{3}{4} - \frac{\nu}{2} \, , \,\, \frac{3}{4} + \frac{\nu}{2} \\ \frac{3}{2} \end{gathered} \right| \frac{1}{u^2} \right] \, .
\end{aligned}
\end{equation}
The coefficients $a_{mn}$ can be found in~\cite{Jazayeri:2022kjy} and will not be important for us. To find our seed $\tilde{F}$ in this region of the kinematic space we can just substract~\eqref{eq:FpmFmp} from~\eqref{eq:sol_F_uvlarge}, getting
\begin{keyeqn}
\begin{align}
	\hat{\tilde{F}} (u,v) = & \,\, 2 \sum_{m,n=0}^\infty a_{mn} u^{-m} \left( \frac{u}{v} \right)^n + \frac{1}{2} \left( \hat{h}_+ (u) \hat{h}_- (v) + \hat{h}_- (u) \hat{h}_+ (v) \right) \log u \nonumber \\
    & - \frac{1}{4} \left( \log 2 -1 +H_{-\frac{1}{4}+\frac{\nu}{2}} +H_{-\frac{1}{4}-\frac{\nu}{2}} - 2\pi \sec (\pi\nu) \right) \hat{h}_- (u) \hat{h}_+ (v) \label{eq:sol_Ftilde_uvlarge} \\
    & - \frac{1}{4} \left( \log 2 +H_{-\frac{1}{4}+\frac{\nu}{2}} +H_{-\frac{1}{4}-\frac{\nu}{2}} - 2\pi \sec (\pi\nu) \right) \hat{h}_+ (u) \hat{h}_- (v) \nonumber \\ 
	& - \frac{1}{4\pi} \Gamma \left( \frac{1}{4}+\frac{\nu}{2} \right)^2 \Gamma \left( \frac{1}{4}-\frac{\nu}{2} \right)^2 \hat{h}_+ (u) \hat{h}_+ (v) \, , \qquad\qquad\qquad\qquad\qquad\qquad\qquad 1<u<v \, . \nonumber
\end{align}
\end{keyeqn}
The solution for $1<v<u$ is obtained by exchanging $u\leftrightarrow v$.

\bibliographystyle{JHEP}
\bibliography{refs}

\clearpage
\let\clearpage\relax

\end{document}